\documentclass{aa}

\usepackage{graphicx,txfonts}
\usepackage{hyperref}
\usepackage{natbib}
\usepackage[dvipsnames]{xcolor}
\usepackage[section]{placeins}

\bibpunct{(}{)}{;}{a}{}{,} % to follow the A&A style

\newcommand\teff{$T_{\rm eff}$}

\begin{document}

\title{NLTE atmospheric modelling of the ultra-hot Jupiter WASP-178b and comparison with UV and optical observations} 

\titlerunning{NLTE atmospheric modelling of the ultra-hot Jupiter WASP-178b}

\authorrunning{Fossati et al.}

\author{L. Fossati\inst{1}  \and        
        A. G. Sreejith\inst{1}  \and
        T. Koskinen\inst{2} \and
        A. Bonfanti\inst{1}   \and
        D. Shulyak\inst{3}  \and
        F. Borsa\inst{4} \and
        S. P. D. Borthakur\inst{1,5,6}\and 
        P. E. Cubillos\inst{1,7}\and
        M. E. Young\inst{8}
        }

\institute{Space Research Institute, Austrian Academy of Sciences, Schmiedlstrasse 6, 8042, Graz, Austria\\
\email{Luca.Fossati@oeaw.ac.at}
\and
Lunar and Planetary Laboratory, University of Arizona, 1629 East University Boulevard, Tucson, AZ, 85721-0092, USA
\and
Instituto de Astrof\'isica de Andaluc\'ia (CSIC), Glorieta de la Astronom\'ia s/n, 18008, Granada, Spain
\and
INAF -- Osservatorio Astronomico di Brera, Via E. Bianchi 46, 23807, Merate (LC), Italy
\and
Tartu Observatory, University of Tartu, Observatooriumi 1, T\~{o}ravere, 61602, Estonia
\and
Institute for Theoretical and Computation Physics, Graz University of Technology, Petersgasse 16, 8010 Graz, Austria
\and
INAF -- Osservatorio Astrofisico di Torino, Via Osservatorio 20, 10025 Pino Torinese, Italy
\and
Astrophysics Research Centre, Queen's University Belfast, Belfast, BT7 1NN, UK
}
\date{Received date ; Accepted date }
\abstract
{}
{We model the atmosphere of the ultra-hot Jupiter (UHJ) WASP-178b accounting for non-local thermodynamical equilibrium (NLTE) effects and compare synthetic transmission spectra with near-ultraviolet (NUV) and optical observations.}
{We use the {\sc helios} code (LTE) in the lower atmosphere and the {\sc Cloudy} code (LTE or NLTE) in the middle and upper atmosphere to compute the temperature-pressure (TP) and abundance profiles. We further use {\sc Cloudy} to compute the theoretical planetary transmission spectrum both in LTE and NLTE for comparison with observations.}
{We find an isothermal TP profile at pressures higher than 10\,mbar and lower than 10$^{-8}$\,bar, with an almost linear increase from $\sim$2200\,K to $\sim$8100\,K in between. The temperature structure is driven by NLTE effects, particularly in the form of increased heating resulting from the overpopulation of long-lived Fe{\sc ii} levels with strong transitions in the NUV band, where the stellar emission is strong, and of decreased cooling due to the underpopulation of Mg{\sc i} and Mg{\sc ii} levels that dominate the cooling. The planetary atmosphere is hydrostatic up to pressures of $\sim$1\,nbar, and thus accurately modelling spectral lines forming at pressures lower than $\sim$1\,nbar requires accounting for both hydrodynamics and NLTE effects. The NLTE synthetic transmission spectrum overestimates the observed H$\alpha$ and H$\beta$ absorption, while the LTE model is in good agreement, which is surprising as the opposite has been found for the other UHJs for which NLTE modelling has been performed. Instead, in the NUV we find an excellent match between the NLTE transmission spectrum and the HST/UVIS data, contrary to the LTE model. This contrasts previous LTE results requiring SiO absorption to fit the observations.}
{The accurate characterisation of the atmosphere of UHJs is possible only accounting for NLTE effects, and particularly for the level population of Fe and Mg, which dominate heating and cooling, respectively.}
\maketitle
\section{Introduction}\label{sec:intro}
Ultra-hot Jupiters (UHJs) are close-in gas giant planets with equilibrium temperatures $T_{\rm eq}\gtrsim$2000\,K. The high temperature favours metal ionisation in the planetary atmosphere that leads to the production of a significant amount of H$^-$, which dominates the continuum of the planetary transmission and emission spectra \citep[e.g.][]{arcangeli2018,parmentier2018_UHJs}. Thanks to the relatively large atmospheric pressure scale heights, driven by the high equilibrium temperatures, and the brightness of their host stars, UHJs have become prime targets for atmospheric observations conducted at both transmission and emission geometry. This has fostered a significant modelling effort, aiming particularly at interpreting the observations.

Several UHJs orbit around relatively bright F- and A-type stars, which has driven the community to study their characteristics in greater detail \citep[e.g.][]{ahlers2020_kelt9b,kama2023_kelt9}. Among the properties of planet-hosting stars, the high-energy emission (X-ray and extreme ultraviolet, EUV; together XUV) is one of the most important, because it has a profound impact on the planetary atmospheric properties and evolution \citep[e.g.][]{lammer2003,yelle2004,owen2017,kubyshkina2018_grid}. Observations suggested that intermediate-mass stars cooler than about 8250\,K have strong XUV emission, possibly significantly stronger than solar, while the XUV emission of hotter stars can be approximated as black-body emission, and is thus relatively low \citep[e.g.][]{fossati2018_AstarPlanets,gunther_activityAstars2022}. Therefore, we expect that UHJs orbiting intermediate-mass stars cooler than about 8250\,K have a very hot and extended upper atmosphere with hydrogen ionisation driven by the incident XUV stellar irradiation, while the atmospheres of planets orbiting hotter intermediate-mass stars are mostly controlled by the stellar ultraviolet (UV) and optical incident radiation \citep[e.g.][]{fossati2018_AstarPlanets,fossati2021_kelt9cloudy,fossati2023_kelt20mascara2,lothringer_and_barman2019,garcia2019_UHJs}.

Among the currently known UHJs, KELT-9b has the highest equilibrium temperature \citep[$T_{\rm eq}\approx4000$\,K;][]{gaudi2017,jones2022_kelt9b_cheops}, and thus it is one of the most studied UHJs to date. KELT-9b has been observed from the ground using different high-resolution spectrographs to study the emission and transmission spectra at optical wavelengths. Both hydrogen and a range of neutral and ionised metal species have been detected in the planetary atmosphere through cross-correlation and/or single line profile analyses \citep[e.g.][]{yan_and_henning2018_HalphaKELT9b,hoeijmakers2018,turner2020_kelt9b,Wyttenbach_2020,pino2020,sanchez2022_paschenBeta_kelt9b,borsato2023,darpa2024_kelt9b,stangret2024}. Eclipse and phase curve observations of this planet have enabled one to place constraints on albedo, lower atmospheric temperature, and atmospheric dynamics \citep[e.g.][]{hooton2018_kelt9b_albedo,wong2020_dynamics_kelt9b,jones2022_kelt9b_cheops}.

Atmospheric modelling of UHJs, further supported by observations, have indicated that the atmospheres of these planets are characterised by a strong temperature inversion typically occurring between the mbar and $\mu$bar level \citep[e.g.][]{lothringer2018_UHJs,kitzmann2018_kelt9b,gandhi2019_Tinversion,lothringer_and_barman2019,borsa2022_mascara2b,yan2022_mascara2b,fossati2020_KELT9datadriven,fossati2021_kelt9cloudy,fossati2023_kelt20mascara2}. Several studies aimed to identify the primary cause of the temperature inversion. Hydrogen Balmer line heating has been proposed as a key agent responsible for the temperature inversion \citep{garcia2019_UHJs}. Several studies also searched for molecules (e.g. TiO, VO, FeH) that are believed to cause temperature inversions, but with little-to-no success \citep[e.g.][]{nugroho2020,kesseli2020,johnson2023_kelt20b_inversion}. Different models of UHJ atmospheres predict that the temperature inversion is mostly driven by metal-line absorption of stellar UV radiation, more specifically of the near-UV (NUV) radiation, which is particularly strong for intermediate-mass stars \citep[e.g.][]{lothringer2018_UHJs,lothringer_and_barman2019,fossati2021_kelt9cloudy,fossati2023_kelt20mascara2}. This is supported by numerous detections of neutral and singly-ionised species in the atmospheres of UHJs.

\citet{fossati2021_kelt9cloudy,fossati2023_kelt20mascara2} showed that non-local thermodynamical equilibrium (NLTE) effects enhance the temperature inversion driven by metal-line absorption of stellar UV radiation. As a matter of fact, they have shown that Fe{\sc ii} and Mg{\sc ii} are, respectively, the species mostly responsible for heating and cooling the planetary atmosphere, and that NLTE leads to significant Fe{\sc ii} over-population and Mg{\sc ii} under-population, increasing the magnitude of the inversion. This occurs because the NLTE-driven over-population of excited Fe{\sc ii} significantly increases the absorption of stellar NUV radiation, further increasing the heating rate. Therefore, accurate characterisation of the atmosphere is only possible if NLTE effects are accounted for. This means that transmission spectra computed accounting for NLTE effects have been the only reliable forward models capable of fitting the observed hydrogen Balmer line profiles of the UHJs KELT-9b and MASCARA-2b/KELT-20b \citep[see also][]{stangret2024,darpa2024_kelt9b}.

WASP-178b is an UHJ that orbits an intermediate-mass star hotter than 8250\,K, though there are contrasting results in the literature on the actual value of the stellar effective temperature \citep{hellier2019,rodriguez-martinez2020,fouesneau2022,lothringer2025_wasp178}. \citet{lothringer2022_wasp178} presented optical and NUV low-resolution (R$\sim$70) observations conducted with the WFC3/UVIS instrument on-board of HST. They used the NUV observations to retrieve the temperature-pressure (TP) profile assuming local thermodynamical equilibrium (LTE), obtaining an inversion characterised by a temperature in the upper atmosphere that is about 3000\,K higher than the temperature in the lower atmosphere. Furthermore, they interpreted the strong NUV absorption as indicative of the presence of SiO, which is considered to be the main Si-bearing species in high-temperature environments.

\citet{cont2024} reported the results obtained from high-resolution emission spectroscopy observations carried out with CRIRES$^+$ at the Very Large Telescope (VLT). The emission spectrum shows clear indications of CO and H$_2$O emission, indicative of a temperature inversion. \citet{cont2024} used retrievals (based on LTE models), including also optical eclipse measurements from TESS\footnote{Transiting Exoplanet Survey Satellite} \citep{ricker2015_tess} and CHEOPS\footnote{CHaracterising ExOPlanets Satellite} \citep{benz2021_cheops}, to constrain the inversion obtaining an upper limit on the lower atmospheric ($\sim$1\,bar) temperature of about 2000\,K and a lower limit on the middle atmospheric ($\sim$10$^{-6}$\,bar) temperature of about 3200\,K. The retrievals also constrain the atmospheric metallicity to be solar, or possibly slightly super-solar, with a solar C/O ratio. The presence of CO and H$_2$O in the atmosphere of WASP-178b was further confirmed by \citet{lothringer2025_wasp178} on the basis of infrared transmission spectroscopy observations conducted with JWST. In particular, they run retrievals, assuming LTE, on the HST (optical and NUV)\,$+$\,JWST transmission spectra finally obtaining a solar C/O ratio as best fit, in agreement with \citet{cont2024}, though about a third of the retrieval scenarios find very low C/O ratios, which have been excluded by \citet{cont2024}. They also did not detect SiO in the JWST spectrum, but the covered infrared SiO band is significantly weaker than the NUV one, indicating that there is no discrepancy between the NUV detection by \citet{lothringer2022_wasp178} and non-detection in the JWST data.

High-resolution transmission spectroscopy observations collected with ESPRESSO \citep{pepe2021_espresso} at the VLT led to the detection of H$\alpha$, H$\beta$, and \ion{Na}{i}\,D lines with the tentative detection of one of the \ion{Mg}{i}\,b lines \citep{damasceno2024_wasp178}. The transmission spectrum has also been analysed using cross correlation, leading to the detection of \ion{Mg}{i}, \ion{Fe}{i}, and \ion{Fe}{ii} \citep{damasceno2024_wasp178}. Furthermore, from the analysis of the Balmer lines, \citet{damasceno2024_wasp178} found significant broadening of the H$\alpha$ and H$\beta$ lines with widths of 39.6$\pm$2.1\,km\,s$^{-1}$ and 27.6$\pm$4.6\,km\,s$^{-1}$, respectively, that they interpret as a signature of atmospheric escape.

WASP-178b is an ideal candidate for NLTE atmospheric modeling. It provides us with the opportunity to further explore the impact of NLTE effects in the atmospheres of UHJs and start comparing results obtained for different planets. In addition, it allows us to compare a synthetic NLTE transmission spectrum with NUV observations. This enables us to test if NLTE effects could be an alternative explanation to SiO in reproducing the NUV observations. 

This paper is organised as follows. Section~\ref{sec:star} presents the methodology and results of a stellar spectroscopic analysis aiming to independently measure the stellar effective temperature to constrain the stellar XUV emission. Section~\ref{sec:modelling} describes the planetary atmospheric modelling and transmission spectra calculation, the results of which are presented in Section~\ref{sec:results}. We discuss our interpretation of the observations in Section~\ref{sec:discussion}, focusing on the properties of the planetary upper atmosphere and inference of the mass-loss rate (Section~\ref{sec:discussion:Cs_lambda}), on the comparison of the obtained NLTE temperature-pressure profile with those of KELT-9b and MASCARA-2b/KELT-20b (Section~\ref{sec:discussion:comparisonsTP_w178_m2_k9}), and on the comparison of our model with the NUV (Section~\ref{sec:discussion:observations_HST}) and optical (Section~\ref{sec:discussion:observations_optical}) observations. Finally, Section~\ref{sec:conclusions} draws the conclusions of this work.
\section{Stellar atmospheric analysis}\label{sec:star}
%
%-------------------------------------
\begin{table}[]
\caption{Stellar atmospheric parameters obtained from the spectroscopic analysis.}
\begin{tabular}{lc}
\hline
\hline
Parameter & Value \\
\hline
$T_{\rm eff}$ [K]              & 9100$\pm$100 \\
$\log g$                       &  4.1$\pm$0.1 \\
$\nu_{\rm mic}$ [km\,s$^{-1}$] &  2.5$\pm$0.2 \\
$\nu\sin i$ [km\,s$^{-1}$]     &  8.5$\pm$1.5 \\
$\nu_{\rm mac}$ [km\,s$^{-1}$] &  5.0$\pm$2.0 \\
\,[Fe/H]                         & 0.23$\pm$0.15 \\
\hline
\end{tabular}
\label{tab:stellar_parameters}
\end{table}
%-------------------------------------
An accurate value for the stellar effective temperature (\teff) plays a critical role in models of the planetary atmosphere. This is because stars cooler than about 8250\,K are believed to have strong XUV emission, similarly to Sun-like stars, while for hotter stars the high-energy emission is several orders of magnitude smaller \citep[e.g.][]{fossati2018_AstarPlanets,gunther_activityAstars2022}. For WASP-178, there are somewhat contrasting \teff\ values reported in the literature (e.g. 8640$^{+500}_{-240}$\,K, \citealt{rodriguez-martinez2020}; 9047\,K, \citealt{fouesneau2022}; 9360$\pm$150\,K, \citealt{hellier2019}; 9200$^{+200}_{-170}$\,K, \citealt{lothringer2025_wasp178}), and thus we decided to independently conduct a full stellar atmospheric parameter retrieval and abundance analysis, to also test the classification of the star as a metallic-line (Am) star, as reported by \citet{hellier2019}. To this end, we analysed an archival spectrum of the star collected with the ESPRESSO high-resolution spectrograph on March 31, 2022. The spectrum covers the 3770--7900\,\AA\ wavelength range at a resolution of R$\approx$140\,000 with a signal-to-noise ratio per pixel of $\approx$620 at $\lambda$$\approx$5300~\AA.

We computed stellar atmosphere models employing the {\sc LLmodels} stellar atmosphere code \citep{Shulyak2004} assuming local thermodynamical equilibrium (LTE), plane-parallel geometry, and the approach of \citet{canuto1991,canuto1992} to model convection. We used the VALD database \citep{piskunov1995_vald,ryabchikova2015_vald} as a source of atomic line parameters for opacity calculations. We derived the atmospheric parameters from the ESPRESSO spectrum employing a mixture of equivalent widths, calculated through either direct integration or line profile fitting using {\sc BinMag} \citep{kochukhov2018_binmag}, and spectral synthesis, through the {\sc synth3} spectral synthesis code \citep{kochukhov2007_synth3}. We converted the equivalent widths in LTE abundances using a modified version \citep{tsymbal1996_width9} of the {\sc width9} code \citep{kurucz1993_width9}. We followed an iterative procedure in which every time any of \teff, surface gravity ($\log g$), microturbulence velocity ($\nu_{\rm mic}$), or abundances changed during the iteration process, we recalculated a new model with the implementation of the last measured quantities. This ensures that the stellar model structure is consistent with the derived abundances \citep{kochukhov2009,shulyak2009,fossati2009_reference,fossati2010_w12Specpol,fossati2011_hd32115}.

We estimated \teff\ by imposing excitation equilibrium considering Ti{\sc ii}, Cr{\sc i}, Cr{\sc ii}, Fe{\sc i}, Fe{\sc ii}, and Ni{\sc i}, while we estimated $\log g$ by imposing ionisation equilibrium for Mg, Si, Ca, Cr, Mn, Fe, and Ni. Finally, we measured $\nu_{\rm mic}$ by imposing equilibrium between line abundances and equivalent widths considering C{\sc i}, Ca{\sc i}, Ti{\sc ii}, Cr{\sc i}, Cr{\sc ii}, Fe{\sc i}, Fe{\sc ii}, and Ni{\sc i}. The use of multiple elements for the determination of the stellar atmospheric parameters increases the reliability of the results compared to an analysis based on iron only \citep[e.g.][]{ryabchikova2009_hd49933}. Because of uncertainties in the spectral normalisation, we could not use the hydrogen Balmer lines to further constrain \teff\ and $\log g$, but we used them to independently check the final values. In the end, we obtained \teff\,=\,9100$\pm$100\,K, $\log g$\,=\,4.1$\pm$0.1, and $\nu_{\rm mic}$\,=\,2.5$\pm$0.2\,km\,s$^{-1}$ (Table~\ref{tab:stellar_parameters}). The \teff\ value is in excellent agreement with that of \citet{fouesneau2022} and lies within 1$\sigma$ of those of \citet{rodriguez-martinez2020}, \citet{hellier2019}, and \citet{lothringer2025_wasp178}. Our derived $\log g$ value is lower than those present in the literature, but still in agreement within 1$\sigma$, except for those derived by \citet{hellier2019} and \citet{lothringer2025_wasp178} that are higher by about 2$\sigma$. The derived \teff\ value indicates that the star has a low XUV emission.
%-------------------------------------
\begin{figure}[t!]
                \centering
                \includegraphics[width=9cm]{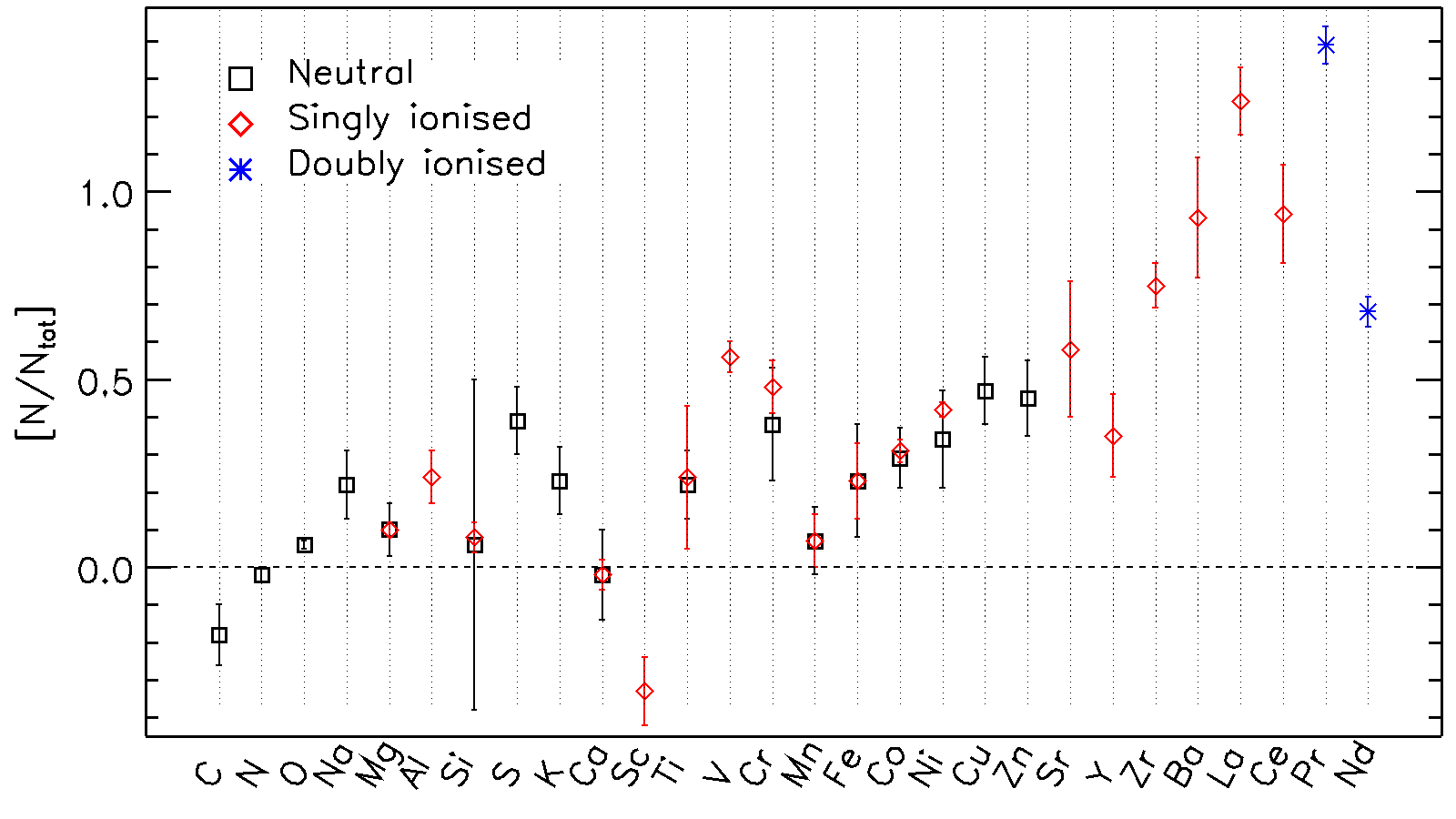}
                \caption{Ion abundance relative to the Sun \citep{asplund2009} of the WASP-178 atmosphere. On the y-axis, $N$ corresponds to the number of atoms of a given ionic species indicated on the x-axis and legend, while $N_{\rm tot}$ is the total number of atoms. Black open squares show the abundance of the neutral elements, red open diamonds are for singly ionised elements, and blue asterisks are for doubly ionised elements.} 
                \label{fig:abundances}
\end{figure}
%-------------------------------------

We estimated the projected rotational velocity ($\nu\sin i$) and macroturbulence velocity ($\nu_{\rm mac}$) values in two independent ways, namely (1) through fitting synthetic spectra to the observed spectrum of unblended and weakly blended lines and (2) following the procedure of \citet{murphy2016_vsini}. The latter involves the comparison of least-squares deconvolution \citep[LSD;][]{kochukhov2010_lsd} profiles of synthetic spectra broadened with different $\nu\sin i$ and $\nu_{\rm mac}$ values with the LSD profile of the observed spectrum to look for the best fit. Employing both techniques, we obtained compatible results and adopt those derived through LSD fitting: $\nu\sin i$\,=\,8.5$\pm$1.5\,km\,s$^{-1}$, which is in agreement with \citet{hellier2019} and \citet{damasceno2024_wasp178}, and $\nu_{\rm mac}$\,=\,5.0$\pm$2.0\,km\,s$^{-1}$ (Table~\ref{tab:stellar_parameters}). The tension between the $\nu\sin i$ values given by \citet{hellier2019} and \citet{damasceno2024_wasp178} is likely due to the fact that, having ignored $\nu_{\rm mac}$ broadening, the uncertainty reported by \citet{hellier2019} is underestimated\footnote{The $\nu\sin i$ measurement of \citet{damasceno2024_wasp178} is independent of $\nu_{\rm mac}$ broadening.}. The higher $\nu\sin i$ value reported by \citet[][12.28$^{+0.78}_{-0.82}$\,km\,s$^{-1}$]{rodriguez-martinez2020} might be the consequence of a too low effective temperature and/or of the fact that they did not consider macroturbulence broadening.

The obtained abundances are listed in Table~\ref{tab:abundances} and presented in Figure~\ref{fig:abundances}. The abundance uncertainties account for the line-to-line abundance scatter (i.e. the standard deviation from the average abundance) and the uncertainty due to the error bars on \teff, $\log g$, and $\nu_{\rm mic}$ \citep[see Section~4.1 of][]{fossati2010_w12Specpol}. We obtained an [Fe/H] abundance of 0.23$\pm$0.15\,dex, in excellent agreement with that obtained by \citet{hellier2019}, while it is likely that the sub-solar iron abundance obtained by \citet[][$-0.06\pm0.34$\,dex]{rodriguez-martinez2020} is driven by the too low \teff\ value. The ESPRESSO spectrum shows also the presence of weak He{\sc i} features and their strength matches that of a synthetic spectrum computed considering the finally adopted atmospheric parameters and solar He abundance. We anticipate here that in the case of WASP-178 the obtained stellar metallicity is proper only of the stellar photosphere and not of the entire star, therefore the derived abundance pattern should not be taken as reference for stellar evolution or comparisons with the planetary atmospheric abundance pattern (see below).

We independently estimated the stellar parameters and abundances by spectral fitting using the ZEEMAN spectral synthesis code \citep{1988Landstreet,2001Wadeetal,2012Folsom}. To estimate the stellar parameters, the spectrum was divided into smaller wavelength regions and then fit iteratively. The scatter in the parameter estimates between different spectral regions is reported as the uncertainty. The stellar parameters estimated from this method are \teff\,=\,9080$\pm$100\,K, $\log{g}$\,=\,4.0$\pm$0.1, $\nu\sin i$\,=\,9.6$\pm$0.1\,km\,s$^{-1}$ and $\nu_{\rm mic}$\,=\,2.5$\pm$0.3\,km\,s$^{-1}$. The stellar abundances are estimated by fitting individual lines of each element, and the line-by-line scatter is reported as the uncertainty. In the case of elements with single lines (e.g. Sr, Al and He), the uncertainty is estimated by varying \teff\ by 1$\sigma$. Finally, we found stellar parameters and abundances within 1$\sigma$ of those obtained with the previous method.

The abundance pattern is characterised by an underabundance of C and Sc and by an overabundance of the Fe-peak elements, of Sr, Y, Zr, Ba, and of the rare-Earth elements, which is typical of Am stars \citep[e.g.][]{fossati2007_praesepe,fossati2008_praesepe}. This conclusion is further supported by the somewhat enhanced microturbulence velocity compared to that of stars of similar temperature, which is also typical of Am stars \citep{landstreet1998_velocityfields,landstreet2009_velocityfields}. Chemical peculiarities in A-type stars develop as a result of diffusion processes that alter only the surface (i.e. photospheric) composition \citep[e.g.][]{michaud1970}. Therefore, the measured photospheric abundance pattern is not representative of that of the whole star.

\citet{rodriguez-martinez2020} found that the planetary orbital plane is almost perpendicular to the projected stellar equator, which led them to conclude that the star is likely seen pole-on as a consequence of A-type stars being typically fast rotators. \citet{pagano2024} derived a possible stellar rotation period of about 3\,days that, together with the stellar radius and $\nu\sin i$ value, leads to a stellar rotation rate of about 26\,km\,s$^{-1}$ and inclination angle of about 17 degrees, supporting the almost pole-on geometry. This inferred stellar rotation rate is however significantly smaller than the typical rotational velocity of A-type stars \citep[$>$100\,km\,s$^{-1}$; e.g.][]{abt1995_rotation,royer2007_rotationAstars,sun2024_rotationEvolution} and is in line with the Am chemical peculiarities that develop as a result of diffusion processes in stars rotating slower than $\sim$90\,km\,s$^{-1}$ \citep[e.g.][]{charbonneauMichaud1991_diffusionRotation,fossati2008_praesepe}. Given the rather uncertain nature of the stellar rotation period measurement, the strongest constrain on the stellar rotation comes from the Am nature of the star that implies a stellar rotational velocity lower than 90\,km\,s$^{-1}$, leaving large uncertainties on the stellar inclination angle. This rotational velocity is too low to lead to a significant pole-to-equator temperature difference, in agreement with the non-detection of gravity darkening \citep[e.g.][]{vonZeipel1924,espinosa2011_gravityDarkening} in the TESS and CHEOPS light curves \citep{pagano2024}.
\section{Planetary atmospheric modelling and synthetic transmission spectrum}\label{sec:modelling}
%
%-------------------------------------
\begin{table}[]
\caption{System parameters adopted for the atmospheric modelling of WASP-178b.}
\begin{tabular}{l|c|l}
\hline
\hline
Parameter & Value & Source \\
\hline
$T_{\rm eff}$ [K] & 9100 & This work \\
$M_{\rm s}$ [$M_{\odot}$] & 2.07 & \citet{hellier2019} \\
$R_{\rm s}$ [$R_{\odot}$] & 1.67 & \citet{hellier2019} \\
\hline
$M_{\rm p}$ [$M_{\rm J}$] & 1.66 & \citet{hellier2019} \\
$R_{\rm p}$ [$R_{\rm J}$] & 1.81 & \citet{hellier2019} \\
$T_{\rm eq}$ [K] & 2470 & \citet{hellier2019} \\
\hline
$a$ [AU] & 0.0558 & \citet{hellier2019} \\
$P$ [days] & 3.3448285 & \citet{hellier2019} \\
$b$ & 0.54 & \citet{hellier2019} \\
\hline
\end{tabular}
\label{tab:parameters}
\end{table}
%-------------------------------------
To simulate the structure of the planetary atmosphere, we followed the same procedure as in \citet{fossati2021_kelt9cloudy} and \citet{fossati2023_kelt20mascara2} that we describe briefly below. We computed the TP profile of the planetary atmosphere employing the {\sc helios} code \citep{malik2017,malik2019} at pressures higher than 0.3\,mbar and the {\sc Cloudy} NLTE radiative transfer code \citep[version 17.03;][]{ferland1998,ferland2013,ferland2017}, through the {\sc Cloudy} for Exoplanets (CfE) interface \citep{fossati2021_kelt9cloudy,young2024_wasp121b}, at lower pressures. We apply this separation, because {\sc helios} does not account for NLTE effects that are relevant in the middle and upper atmosphere, while {\sc Cloudy}, which considers NLTE effects, is unreliable at densities greater than 10$^{15}$\,cm$^{-3}$ \citep[see][for more details]{ferland2017}. The TP profiles computed by {\sc helios} and {\sc Cloudy} are then joined together to obtain single TP profiles under two different conditions (i.e. LTE {\sc helios} $+$ LTE {\sc Cloudy} and LTE {\sc helios} $+$ NLTE {\sc Cloudy}), which are used to derive the atmospheric chemical composition and transmission spectra.

For the computation of the TP profile with {\sc helios}, we considered additional opacities not present in the public version of the code \citep{fossati2021_kelt9cloudy} and included cross sections of all molecules available from the DACE\footnote{\tt https://dace.unige.ch/dashboard} database. Most of the corresponding line lists were provided by the ExoMol\footnote{https://www.exomol.com/} project \citep{tennyson2016_exomol},
with a few compiled from the databases HITRAN\footnote{https://hitran.org/} \citep[HBr, HCl, O$_2$, O$_3$;][]{gordon2022_hitran} and HITEMP\footnote{https://hitran.org/hitemp/} \citep[N$_2$O, NO$_2$;][]{rothman2010_hitemp,hargreaves2019_hitemp}. The atomic line opacity was extended to include \ion{Si}{i-ii},  \ion{Ca}{i-ii}, and \ion{Ti}{i-ii}, which we pre-tabulated using the \textsc{HELIOS-K}\footnote{https://github.com/exoclime/HELIOS-K} package \citep{grimm2015_helios}
and the original line lists produced by R.~Kurucz\footnote{http://kurucz.harvard.edu/linelists/gfall/}\citep{kurucz2018}. Our calculations of the {\sc helios} models covered the 10$^{2}$--10$^{-9}$\,bar pressure range divided into 120 layers and assumed a heat redistribution parameter ($f$), which accounts for the day-to-night side heat redistribution efficiency, equal to 0.25. In this way, the TP profile in the lower atmosphere is comparable to that obtained by \citet{lothringer2022_wasp178}, who also considered full heat redistribution.

For the planetary atmospheric modelling and transmission spectra calculation, we considered the system parameters listed in Table~\ref{tab:parameters}, which lie within 1$\sigma$ of those of \citet{damasceno2024_wasp178}, and solar atmospheric composition \citep{lodders2003}. Since we obtained a \teff\ value comparable to that given by \citet{hellier2019}, we decided to employ their system parameters, such that our results can be more directly compared to those present in the literature. Nevertheless, to double-check the system parameters are still consistent with the adoption of our stellar effective temperature, we re-derived the stellar and planetary mass and radius. To this end, we extracted the stellar mass and radius from interpolating across PARSEC~v1.2S \citep{marigo17} isochrones and evolutionary tracks \citep{bonfanti15,bonfanti16} using the updated $T_{\rm eff}$, G-band magnitude, parallax \citep{gaia2016_DR3,gaia2023_DR3}, and metallicity as input. Since the metallicity derived from the spectroscopic analysis does not represent that of the whole star, but it is proper only of the photosphere, we chose to adopt a more moderate super-solar metallicity of $+$0.1\,dex, with an uncertainty of 0.15\,dex. We then derived the planetary mass and radius using the stellar reflex velocity, orbital inclination, orbital velocity, and transit depth given by \citep{hellier2019}, further assuming no orbital eccentricity. Finally, we obtained $M_{\rm s}$\,=\,1.96$\pm$0.08\,$M_{\odot}$, $R_{\rm s}$\,=\,1.59$\pm$0.04\,$R_{\odot}$, $M_{\rm p}$\,=\,1.61$\pm$0.11\,$M_{\rm J}$, $R_{\rm p}$\,=\,1.72$\pm$0.05\,$R_{\rm J}$, which lie within 1$\sigma$ of the values given by \citet[][see Table~\ref{tab:parameters}]{hellier2019} and \citet{lothringer2025_wasp178}, and lead to a planetary reflex velocity fully consistent with that measured by \citet{cont2024}. For the {\sc helios} and {\sc Cloudy} calculations, we employed the synthetic stellar spectral energy distribution (SED) computed with {\sc LLmodels} for the atmospheric parameters reported in Section~\ref{sec:star}, and thus we did not add any extra high-energy emission to the {\sc LLmodels} photospheric fluxes \citep{fossati2018_AstarPlanets,gunther_activityAstars2022}.

The 10$^{15}$\,cm$^{-3}$ upper density limit for {\sc Cloudy} lies at a pressure of about 0.3\,mbar, while the continuum lies at a pressure of about 60\,mbar. This is the pressure at which the {\sc helios} model gives an optical depth of 2/3 around 7500\,\AA, which is the wavelength corresponding to the peak efficiency of the EulerCAM photometer \citep{lendl2012} used to measure the planetary radius from the transit light curve \citep{hellier2019}. For this reason, we run CfE setting the planetary transit radius at the reference pressure ($p_0$) of 60\,mbar, although this choice has no impact on the final TP profile \citep{fossati2021_kelt9cloudy,fossati2023_kelt20mascara2}. To save computation time, we run {\sc Cloudy} considering all elements up to Zn and only hydrogen molecules (i.e. H$_2$, H$_2^+$, H$_3^+$, HD), because for planets as hot as WASP-178b, the inclusion of all molecules present in the {\sc Cloudy} database does not affect the resulting middle and upper atmospheric temperature \citep{fossati2021_kelt9cloudy,fossati2023_kelt20mascara2}. In particular, this means that the {\sc Cloudy} calculations do not consider SiO. We mimicked atmospheric heat redistribution in the computation of the {\sc Cloudy} TP profiles by scaling the input stellar SED multiplying it by a factor $f_1$\,$\leq$\,1, looking for the one leading to the TP profile best matching the one given by {\sc helios} around the 10$^{-4}$\,bar level. We finally adopted $f_1$\,=\,1.0, but note that the value employed for $f_1$ has no significant ($>$200\,K) impact on the TP profile at pressures lower than about 10$^{-5}$\,bar \citep{fossati2021_kelt9cloudy,fossati2023_kelt20mascara2}. This is likely because the heating rate is relatively high for all values of $f_1$ exceeding 0.25, but also substantially offset by thermostatic radiative cooling rates.

Finally, we used atmospheric structure (i.e. TP and abundance profiles) both in LTE and NLTE to compute the LTE and NLTE transmission spectra using {\sc Cloudy} ranging from the far-ultraviolet to the near-infrared at a spectral resolution of 100,000. To this end, we followed the procedure described by \citet{young2020} and \citet{fossati2020_KELT9datadriven}, in which the atmosphere is divided into 100 layers equally spaced in $\log{p}$. We computed the NLTE transmission spectrum employing the NLTE TP profile and enabling NLTE in the {\sc Cloudy} calculations. Similarly, we derived the LTE transmission spectrum using the LTE TP profile and imposing the LTE assumption for the {\sc Cloudy} transmission spectrum calculation. Within {\sc Cloudy}, the difference between LTE and NLTE computations lies in the fact that in LTE level populations are computed using the Boltzmann equation with the exception of the first two levels of hydrogen, which are always computed in NLTE. Furthermore, by default {\sc Cloudy} include photoionisation, also in LTE. Instead, {\sc helios} does not account for photoionisation, and thus thermal ionisation and level populations are computed using the Saha and Boltzmann equations, respectively. 

The TP profile is for the sub-stellar point, but the computation of transmission spectra implies a different geometry (i.e. from emission geometry to transmission geometry), which then calls for a new calibration of the reference pressure, $p_0$, that corresponds to the reference radius, $R_0$, that is, the measured transit radius, $R_{\rm p}$. We performed this calibration following \citet{fossati2023_kelt20mascara2}, which gives more accurate results compared to that followed by \citet{fossati2021_kelt9cloudy}. In short, we computed transmission spectra at different $p_0$ values in the 0.001--0.1\,bar range, then convolved each synthetic transmission spectrum with the EulerCAM efficiency curve to obtain $R_{\rm p}/R_{\rm s}$ as a function of $p_0$. The final transmission spectrum is that computed with the $p_0$ value leading to best match the observed $R_{\rm p}/R_{\rm s}$ of 0.111 \citep{hellier2019}. For both LTE and NLTE theoretical transmission spectra, we find a best-fitting reference pressure value of 0.002\,bar.
\section{Results}\label{sec:results}
\subsection{Atmospheric temperature-pressure structure}\label{sec:TP}
\begin{figure}[h!]
                \centering
                \includegraphics[width=9cm]{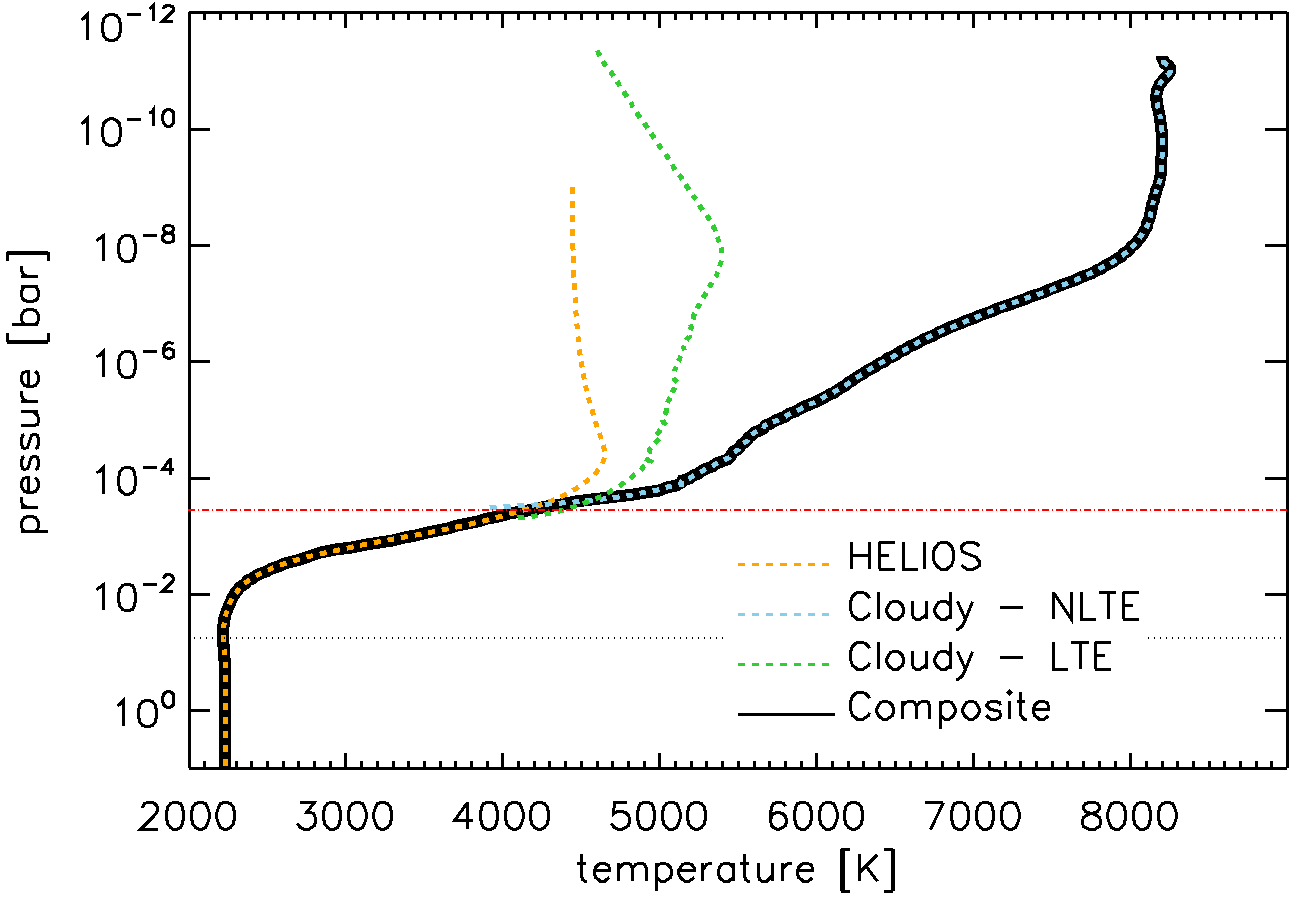}
                \includegraphics[width=9cm]{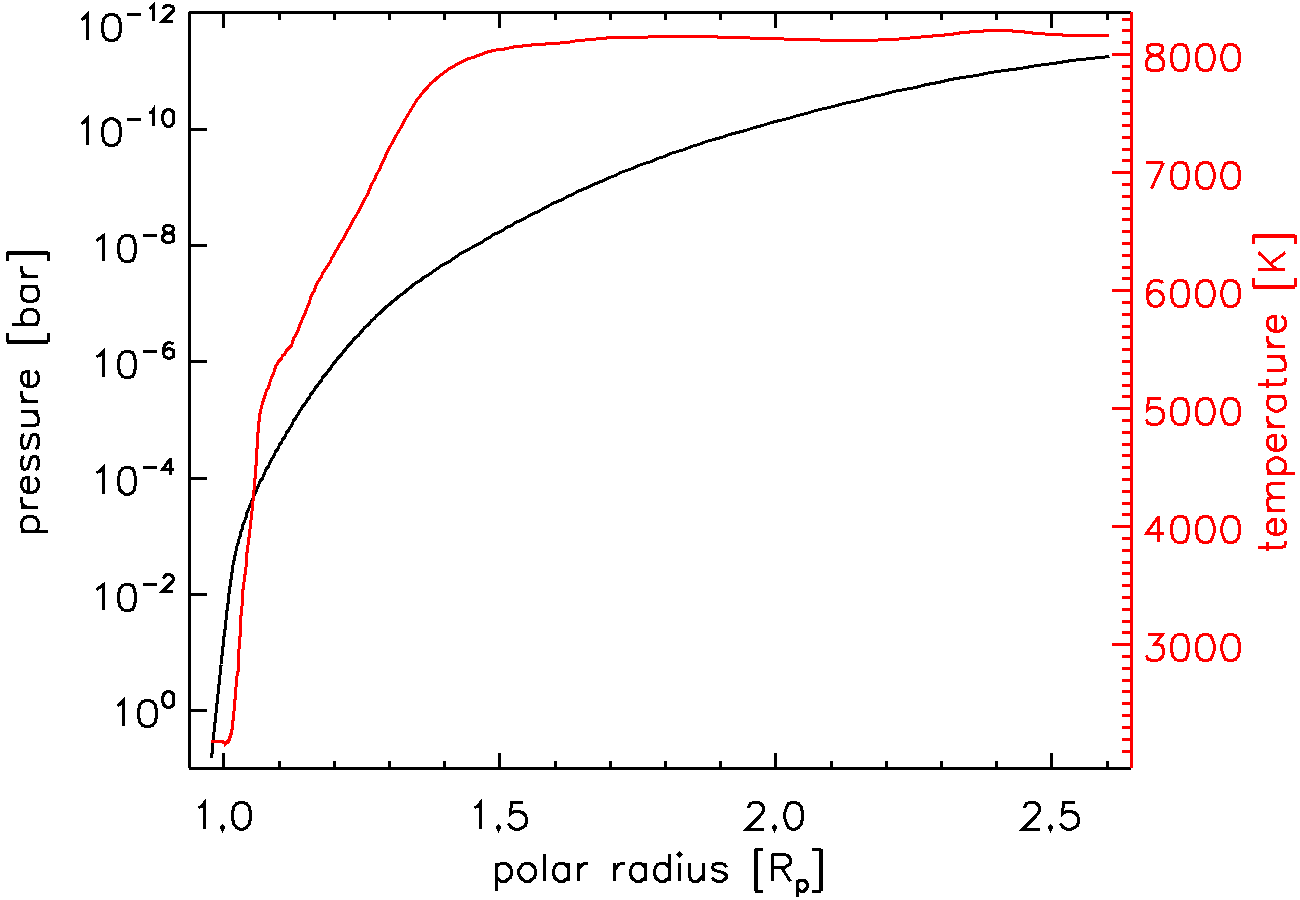}
                \caption{Theoretical atmospheric structure obtained for WASP-178b. Top: {\sc helios} (dashed orange line) and {\sc Cloudy} (NLTE; dashed cyan line) TP profiles. The solid black line shows the composite TP profile. The horizontal dotted black line indicates the location of the continuum according to the {\sc helios} model. The horizontal dash-dotted red line gives the location of {\sc Cloudy}'s upper-density limit of 10$^{15}$\,cm$^{-3}$. The dashed dark green line shows the {\sc Cloudy} TP profile computed assuming LTE. Bottom: Pressure (black; left y-axis) and temperature (red; right y-axis) composite theoretical profiles as a function of the planetary polar radius. For context, the L1 point lies at about 4.00\,$R_{\rm p}$, while the Roche radius in the limb direction lies at about 2.67\,$R_{\rm p}$.} 
                \label{fig:TPfrankenstein}
\end{figure}

We joined the {\sc helios} and {\sc Cloudy} TP profiles at a pressure of about 4$\times$10$^{-4}$\,bar resampling the final theoretical TP profile over 200 layers equally spaced in $\log{p}$ ranging from 10\,bar to 5$\times$10$^{-12}$\,bar. This range is wide enough to cover the formation region of most UV, optical, and infrared lines considered for the computation of the theoretical transmission spectrum. Figure~\ref{fig:TPfrankenstein} shows the composite theoretical TP profile in comparison to the original {\sc helios} and {\sc Cloudy} theoretical profiles. The TP profile is very similar to that obtained for MASCARA-2b/KELT-20b using the same modelling scheme \citep{fossati2023_kelt20mascara2}: isothermal at pressures higher than about 10\,mbar and lower than about 10$^{-8}$\,bar, with a linear temperature rise from about 2200\,K to about 8100\,K in between.
%-------------------------------------
%-------------------------------------

The comparison of the TP profiles obtained in LTE and NLTE (Figure~\ref{fig:TPfrankenstein}) indicates that the LTE assumption appears to be appropriate at pressures higher than about 10$^{-4}$\,bar, but it underestimates the temperature by about 3000\,K at lower pressures. This difference in upper atmospheric temperature between LTE and NLTE TP profiles is similar to that obtained for MASCARA-2b/KELT-20b \citep[$T_{\rm eq}$ $\approx$ 2260\,K;][]{talens2018}, which has an equilibrium temperature comparable to that of WASP-178b \citep{fossati2023_kelt20mascara2}. Interestingly, this temperature difference in the upper atmosphere is about 1000\,K larger than that obtained for KELT-9b \citep[$T_{\rm eq}$ $\approx$ 4050\,K;][]{gaudi2017}, which has instead a significantly higher equilibrium temperature (see Section~\ref{sec:discussion:comparisonsTP_w178_m2_k9}, for a more thorough comparison of the TP profiles of the three planets).
\subsubsection{Heating and cooling}\label{sec:heatingVScooling}
\begin{figure}[ht!]
                \centering
                \includegraphics[width=9cm]{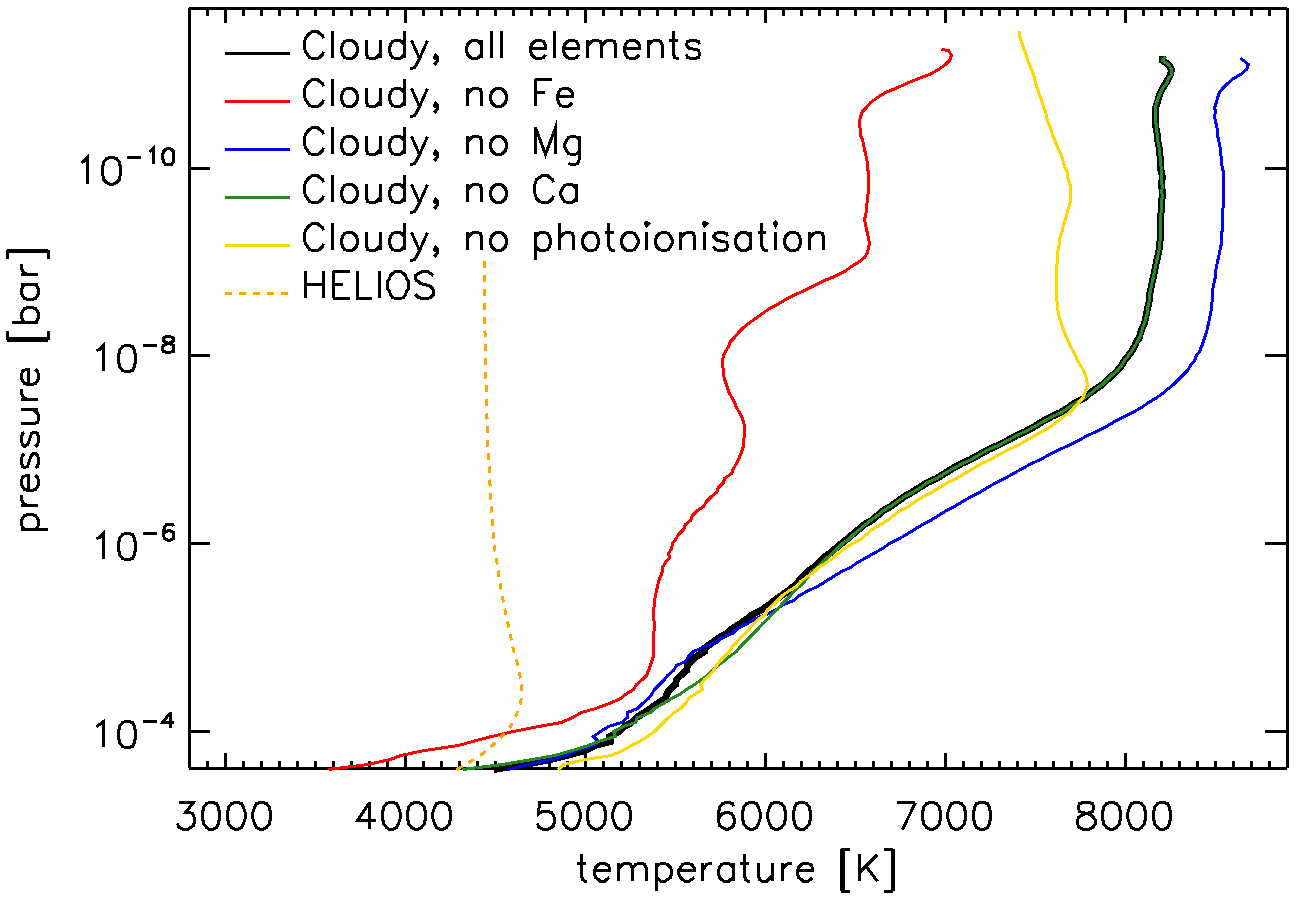}
                \caption{Comparison between the NLTE {\sc Cloudy} TP profiles obtained considering all elements up to Zn (solid black line) and excluding Fe (solid red line) or Mg (solid blue line) or Ca (solid green line that is almost overlapping with the black line) or photoionisation (yellow). The dashed orange line shows the {\sc helios} (i.e. LTE) TP profile for reference.} 
                \label{fig:NLTEnoFeMgCa}
\end{figure}

Following \citet{fossati2021_kelt9cloudy} and \citet{fossati2023_kelt20mascara2}, we identify the elements primarily responsible for the difference between the LTE and NLTE TP profiles by computing {\sc Cloudy} NLTE TP profiles excluding one of the elements at a time. Similarly to previous results, we find that Fe dominates heating and Mg dominates cooling in the middle and upper atmosphere (Figure~\ref{fig:NLTEnoFeMgCa}). Removing Fe leads to a 1000--2500\,K cooler temperature in the middle and upper atmosphere, while excluding Mg leads to an about 300\,K hotter TP profile, particularly in the upper atmosphere, at pressures lower than the $\mu$bar. The other elements, instead, contribute less than 50\,K to the atmospheric heating or cooling, with the exception of Ca for which we find that removing it from the calculation of the TP profile leads to an about 200\,K hotter temperature around the 10\,$\mu$bar level. 
%-------------------------------------
%-------------------------------------

\begin{figure}[ht!]
                \centering
                \includegraphics[width=9cm]{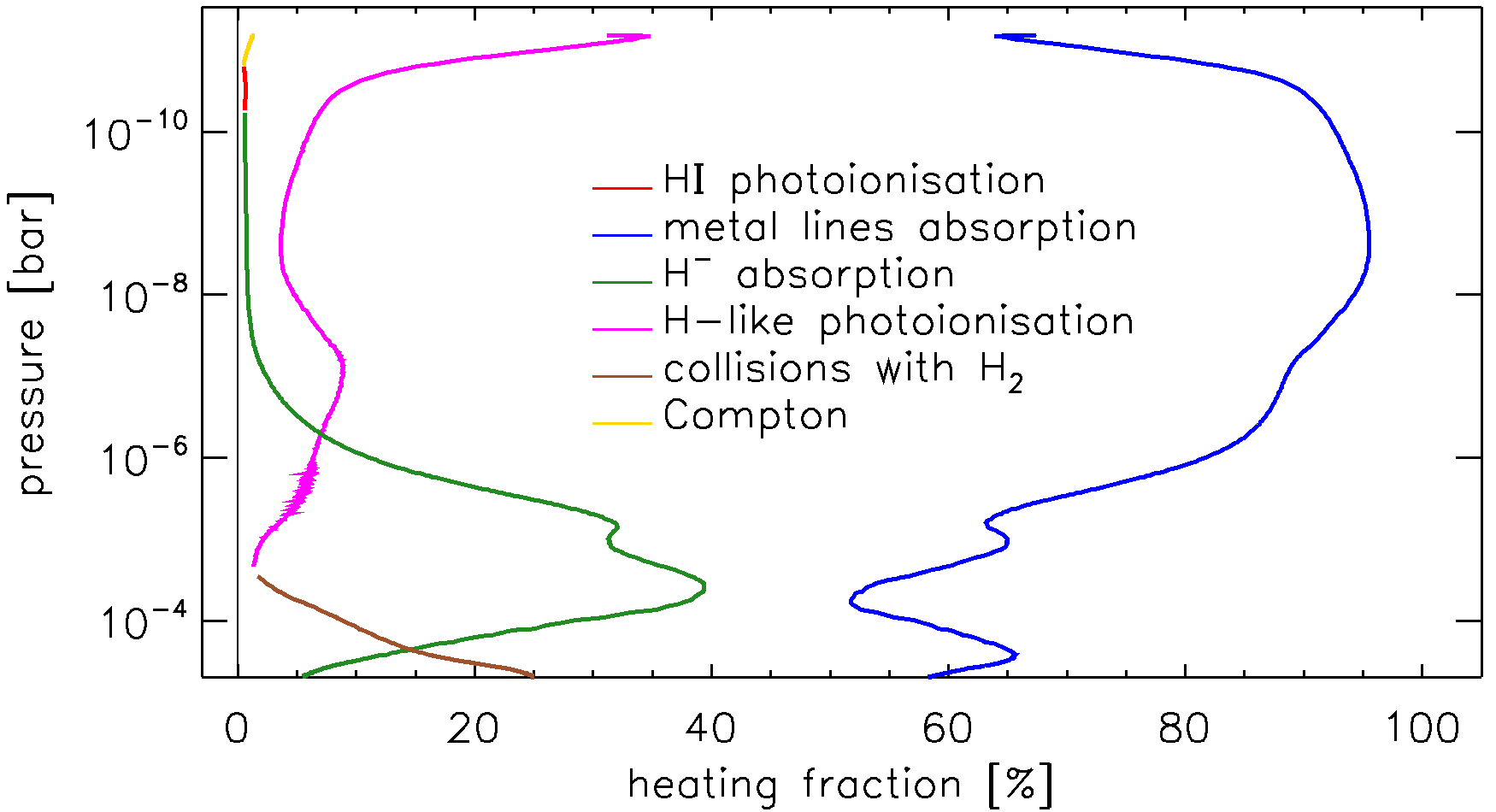}
                \includegraphics[width=9cm]{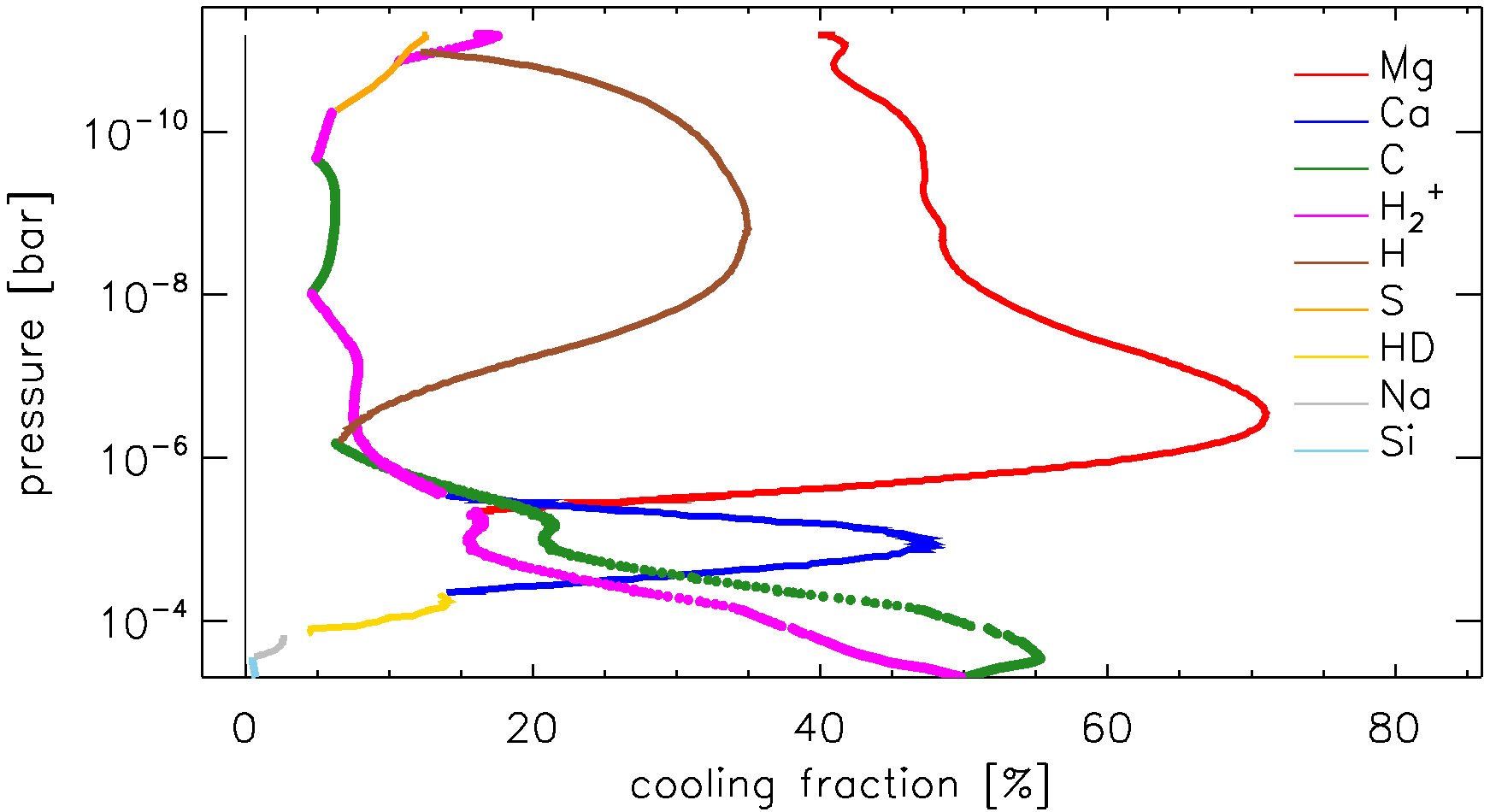}
                \caption{Heating and cooling contributions in the middle and upper atmosphere (i.e. at pressures lower than {\sc Cloudy}'s density limit) of WASP-178b. Top: Contribution to the total heating as a function of pressure. At each pressure bin, the plot shows the three most important heating processes. The main heating processes are hydrogen photoionisation (red; photoionisation of H{\sc i} lying in the ground state), metal line absorption (blue), H$^-$ absorption (green), photoionisation of hydrogenic species (magenta), collisions with H$_2$ (brown), and Compton heating (i.e. electron absorption; yellow). Bottom: Contribution to the total cooling as a function of pressure. At each pressure bin, the plot shows the three most important cooling agents. The main cooling agents are Mg (red), Ca (blue), C (green), H$_2^+$ (magenta), H (brown), S (orange), HD (i.e. hydrogen deuteride; yellow), Na (gray), and Si (light blue). Metal line absorption (primarily from Fe{\sc ii}) and Mg line emission are the primary heating and cooling mechanisms, respectively.} 
                \label{fig:heatingCooling}
\end{figure}

Figure~\ref{fig:NLTEnoFeMgCa} suggests that metals play a pivotal role in the heating and cooling of the middle and upper atmosphere. To gather more insights, we extracted from the {\sc Cloudy} run the three main heating and cooling agents that we show in Figure~\ref{fig:heatingCooling}. We find that metal line absorption (particularly of Fe; Figure~\ref{fig:NLTEnoFeMgCa}) is the main heating mechanism, with H{\sc i} photoionisation playing a minor role, while numerous species contribute to the cooling of the middle and upper atmosphere, with Mg being the dominant coolant in our steady state model. We further explore the general impact of photoionisation on the atmospheric heating by running {\sc Cloudy} without photoionisation, which leads to an upper atmospheric temperature that is just a few hundred Kelvin cooler than that obtained with photoionisation, while the temperature remains the same in the middle atmosphere (see Figure~\ref{fig:NLTEnoFeMgCa}). This indicates that photoionisation heating does not play a major role in the atmospheric thermal balance (see also the magenta line in the top panel of Figure~\ref{fig:heatingCooling}).
%-------------------------------------
%-------------------------------------

\begin{figure}[t!]
                \centering
                \includegraphics[width=9cm]{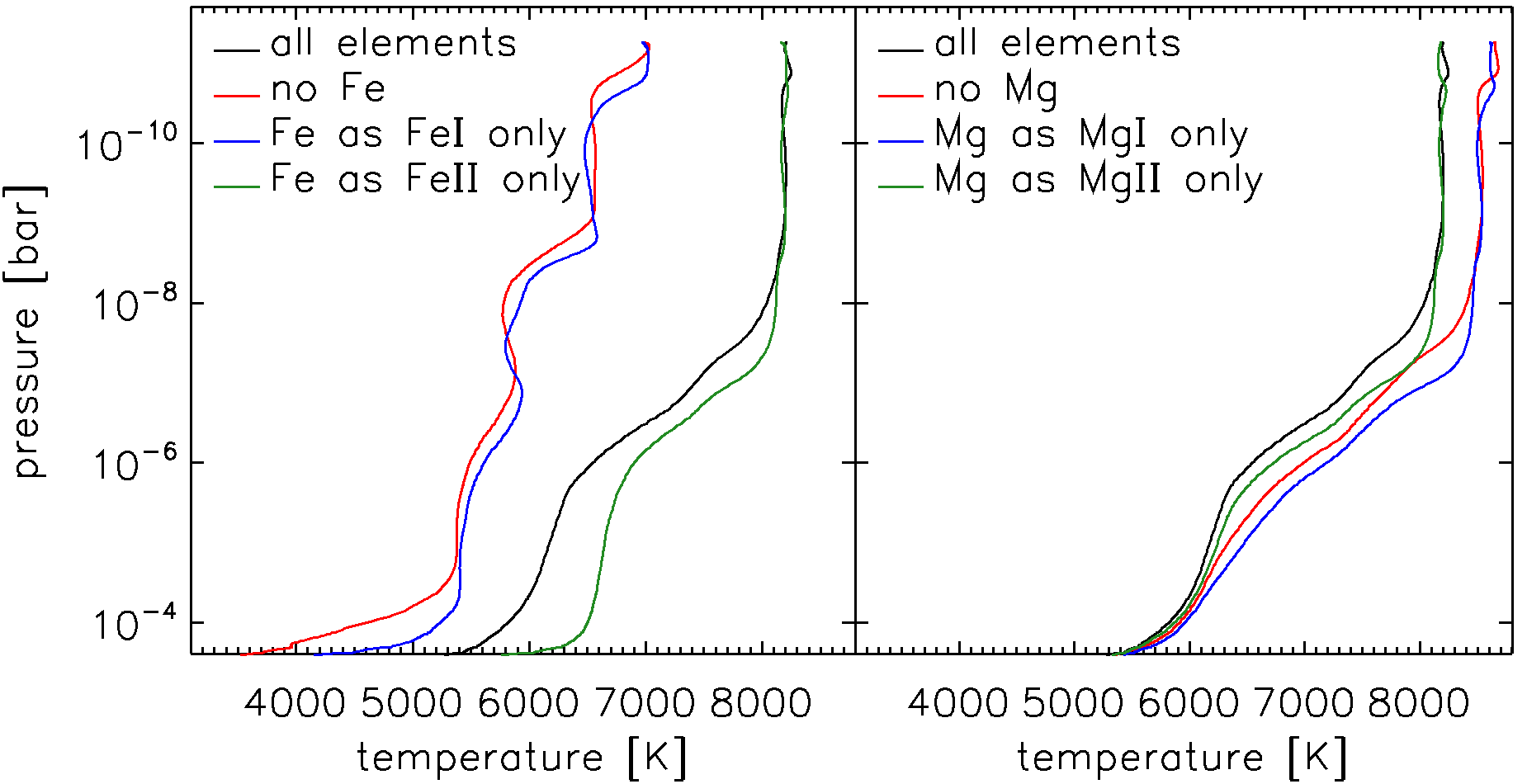}
                \caption{TP profiles obtained from isolating the impacts of Fe (left) and Mg (right) ions. The black line is for accounting for all elements (same as in Figures~\ref{fig:TPfrankenstein} and \ref{fig:NLTEnoFeMgCa}), while the other lines have been obtained considering that in the planetary atmosphere Fe or Mg exist only in the form of neutral species (blue), singly ionised species (green), or is absent (red).} 
                \label{fig:fe1fe2mg1mg2}
\end{figure}

To better understand the relative roles of Mg and Fe in modifying the TP profile, we took the output obtained after the last NLTE CfE iteration and used it as input for a further {\sc Cloudy} run, but considering only Fe{\sc i} (i.e. Fe{\sc i} is not allowed to ionise) or Fe{\sc ii} (i.e. Fe{\sc ii} is not allowed to ionise or recombine), fixing the Fe{\sc i} or Fe{\sc ii} density profile to that obtained accounting for all elements and ions. We then applied the same procedure for Mg. Figure~\ref{fig:fe1fe2mg1mg2} shows the temperature profiles obtained in all these cases. As previously found for KELT-9b and MASCARA-2b/KELT-20b, Fe{\sc ii} dominates the heating in the middle and upper atmosphere and Fe{\sc i} contributes to the cooling in the middle atmosphere, as shown by the fact that without Fe{\sc i} we obtain a higher temperature by a few hundred Kelvin at pressures higher than about 10$^{-8}$\,bar. As also shown above, removing Mg increases the atmospheric temperature across the entire middle and upper atmosphere. Considering just Mg{\sc i} being present in the planetary atmosphere leads to a TP profile that is similar to that obtained without Mg, which suggests that Mg ions are responsible for the cooling at altitudes corresponding to pressures lower than 0.1\,mbar. Considering only Mg{\sc ii} leads to a slightly cooler TP profile within the 10$^{-4}$ and 10$^{-8}$\,bar pressure range, which is where Mg cooling is the strongest (see Figure~\ref{fig:heatingCooling}), while at higher altitudes the TP profile is almost identical to that obtained considering all elements and ions. This indicates that both Mg{\sc i} and Mg{\sc ii} contribute to the cooling in the 10$^{-4}$--10$^{-8}$ bar pressure range, with Mg{\sc i} being more important that Mg{\sc ii}, while Mg{\sc ii} is the dominant Mg cooling species at lower pressures. 

Removing Fe from the atmosphere decreases the temperature significantly more than removing photoionisation, bringing it close to that obtained in LTE. Therefore, the temperature difference between the LTE and NLTE TP profiles is driven by the differences in atomic level populations of Fe, particularly of Fe{\sc ii}, driven by NLTE effects, which increase the level population of long-lived states presenting transitions mainly in the NUV band, where the stellar emission is strong. Instead, the fact that by removing Mg cooling the temperature increases by only 300\,K indicates that other cooling mechanisms replace it and are nearly as efficient as Mg cooling in the new steady state. The bottom panel of Figure~\ref{fig:heatingCooling} shows that this other cooling agent is hydrogen. Therefore, in general atomic line cooling is important and it is hard to heat the atmosphere above 9000\,K without removing all major cooling agents.
%-------------------------------------
%-------------------------------------
%
\subsection{Atmospheric abundance profiles}\label{sec:chemistry}
%
%-------------------------------------
\begin{figure}[ht!]
                \centering
                \includegraphics[width=9cm]{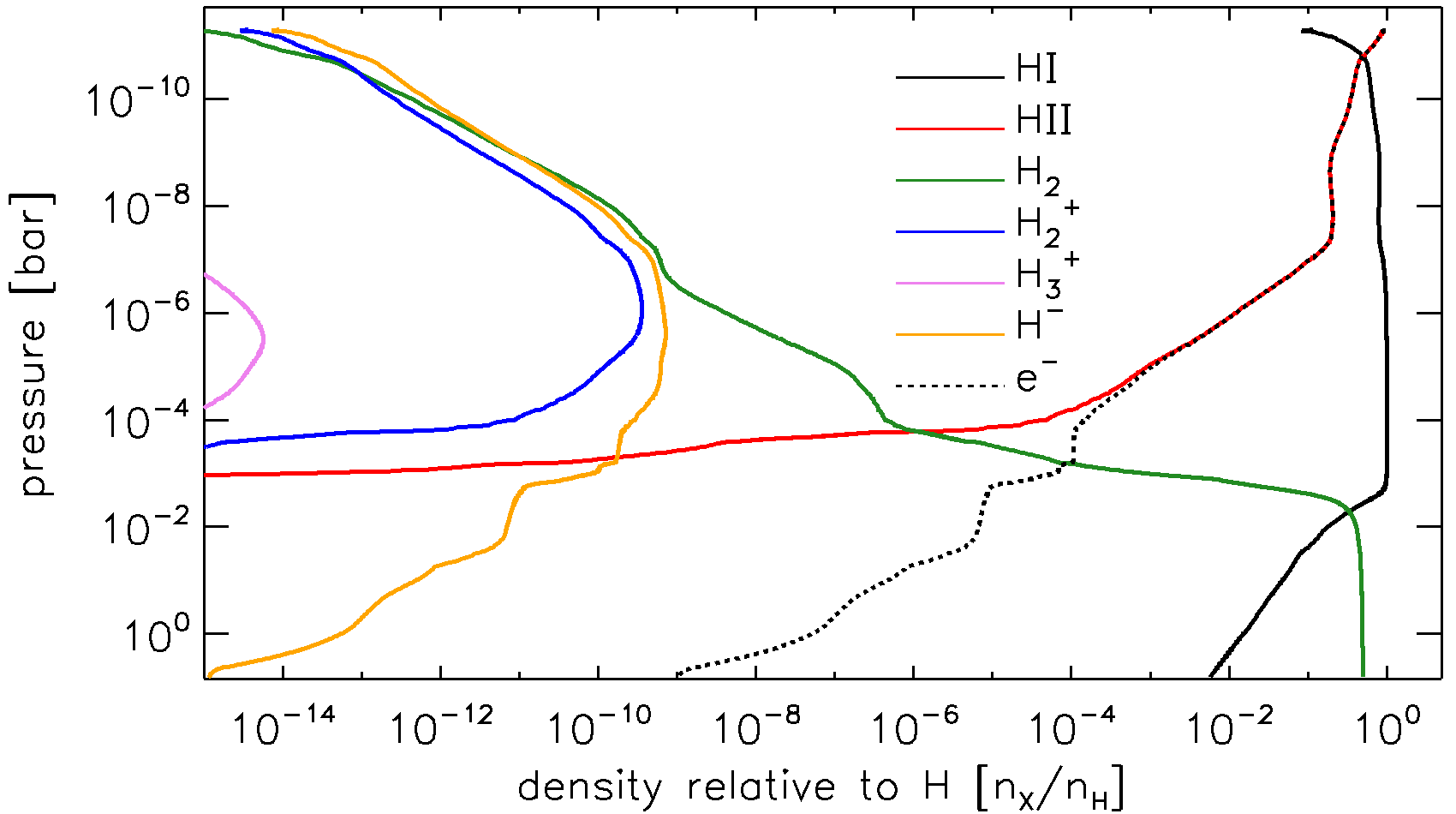}
                \caption{Density (relative to the total density of hydrogen)-pressure profiles for neutral hydrogen (H{\sc i}; solid black line), protons (H{\sc ii}; red), molecular hydrogen (H$_2$; dark green), H$_2^+$ (blue), H$_3^+$ (magenta), H$^-$ (orange), and electrons (e$^-$; dashed black).} 
                \label{fig:NLTEchemistryHydrogen}
\end{figure}
%-------------------------------------
To obtain a homogeneous synthetic physical and chemical structure throughout the entire atmosphere, we passed the composite TP profile to {\sc Cloudy} \citep[see][for more details]{fossati2021_kelt9cloudy}. We used this final {\sc Cloudy} model to extract the abundance profiles of the most important species. Figure~\ref{fig:NLTEchemistryHydrogen} shows the {\sc Cloudy} density profiles with respect to the total hydrogen density of the main hydrogen-bearing species, that is neutral hydrogen (H{\sc i}), protons (H{\sc ii}), H$^-$, molecular hydrogen (H$_2$), H$_2^+$, and H$_3^+$, plus electrons (e$^-$). The abundance profiles are very similar to those previously obtained for MASCARA-2b/KELT-20b. Neutral hydrogen is the most abundant species in the middle and upper atmosphere ($<$5$\times$10$^{-3}$\,bar), with ionised hydrogen taking over just at the very top of the atmosphere, at pressures lower than 10$^{-11}$\,bar. Instead, molecular hydrogen dominates in the lower atmosphere, but its abundance decreases rapidly at pressure lower than 5$\times$10$^{-3}$\,bar as a result of thermal dissociation. As previously found for other UHJs \citep[e.g.][]{arcangeli2018,fossati2021_kelt9cloudy,fossati2023_kelt20mascara2}, H$^-$ is relatively abundant with its density first increasing with decreasing pressure up to the 1\,$\mu$bar level and then decreasing at lower pressures.

\begin{figure}[t!]
                \centering
                \includegraphics[width=9cm]{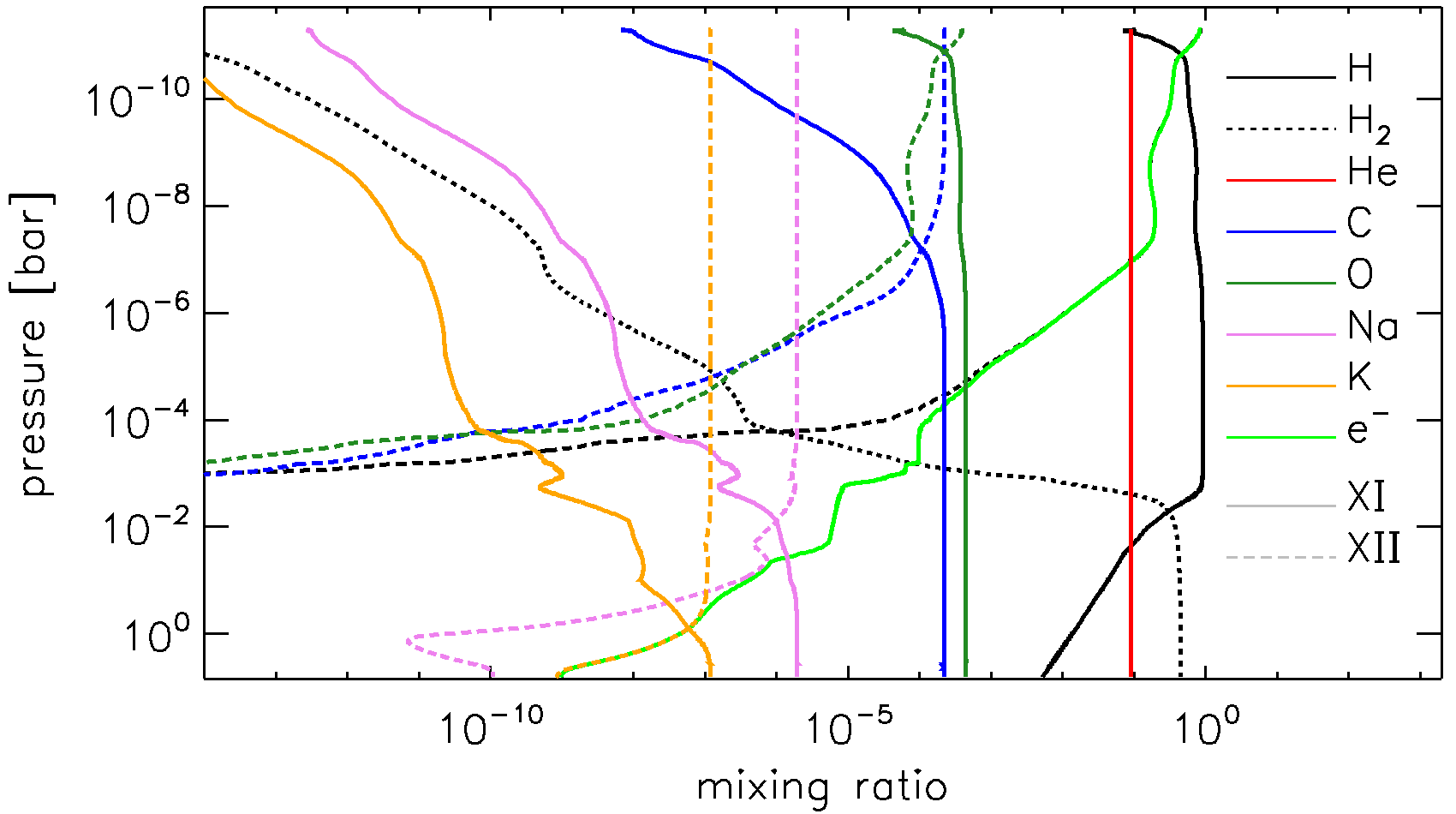}
                \includegraphics[width=9cm]{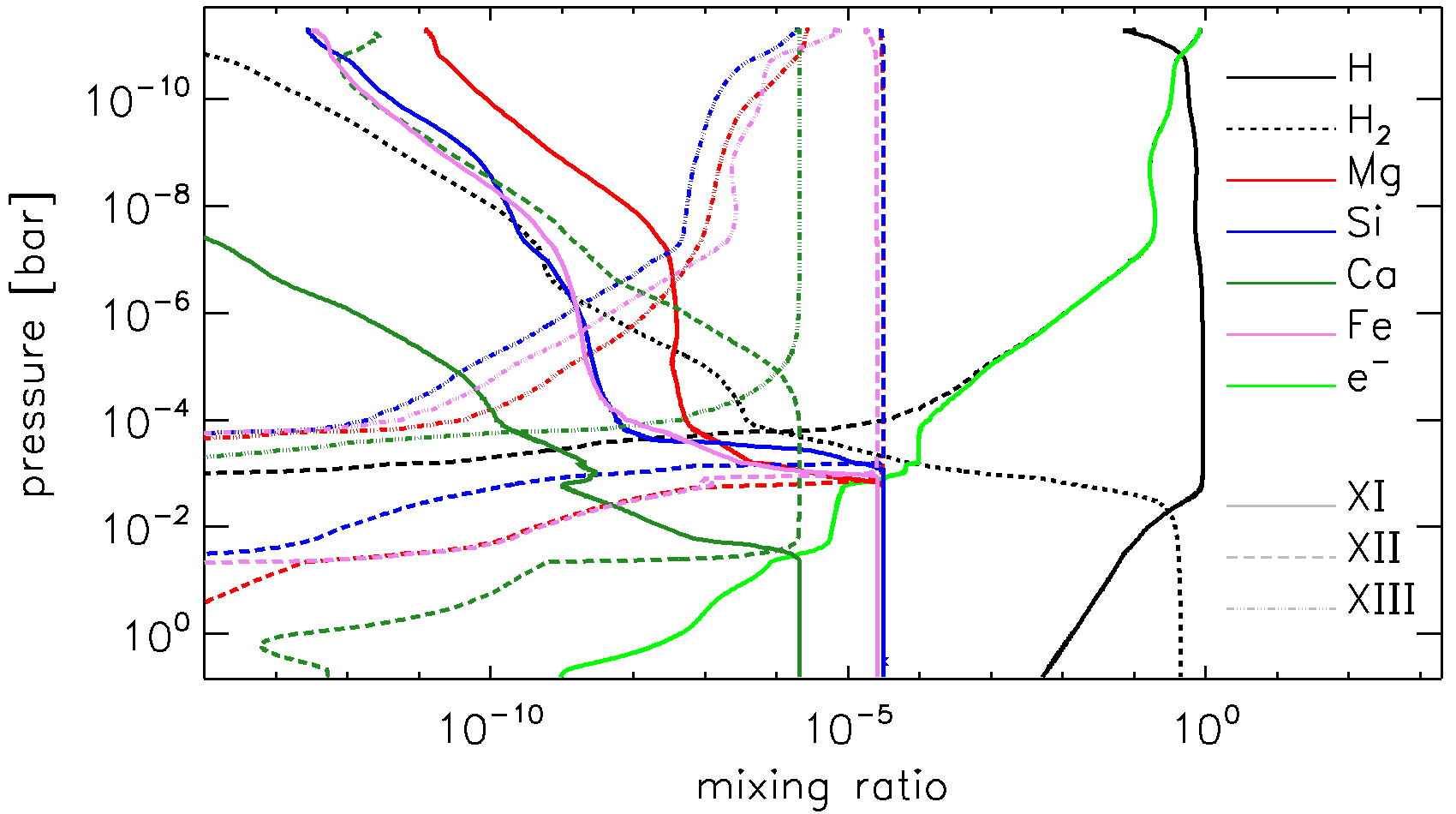}
                \caption{Atmospheric mixing ratios obtained for metals. Top: Mixing ratios for hydrogen (solid black line), H$_2$ (dotted black), He (red), C (blue), O (dark green), Na (magenta), K (orange), and electrons (bright green). Bottom: Same as the top panel, but for Mg (red), Si (blue), Ca (dark green), and Fe (magenta). The hydrogen, H$_2$, and e$^-$ mixing ratios are shown in both panels for reference. Neutral (XI), singly ionised (XII), and doubly ionised (XIII; present just in the bottom panel) species are shown as solid, dashed, and dash-dotted lines, respectively.} 
                \label{fig:NLTEchemistryMetals}
\end{figure}

Figure~\ref{fig:NLTEchemistryMetals} shows the mixing ratio as a function of pressure given by {\sc Cloudy} for some of the most abundant elements. Oxygen remains neutral up to the top of the atmosphere, where ionisation becomes dominant around 10$^{-11}$\,bar, while C{\sc ii} becomes the dominant carbon species above the 10$^{-7}$\,bar level. In the lower atmosphere ($>$10\,mbar), Ca, Na, and K are the main contributors to free electrons, while around the mbar level Fe, Mg, and Si ionisation contribute significantly to the electron population, which then starts following closely the proton abundance at lower pressures as a result of the increasing importance of hydrogen thermal ionisation. Among the species shown in Figure~\ref{fig:NLTEchemistryMetals}, Ca is the only element for which the second ionised species become dominant within the simulated pressure range.
%-------------------------------------
%-------------------------------------
%
\subsection{Synthetic transmission spectrum}\label{sec:transmission}
To enable comparisons with observations, we computed the synthetic NLTE and LTE transmission spectra ranging from the far-UV to the near-infrared (i.e. from 912\,\AA\ to 2.85\,$\mu$m). To this end, we followed the procedure described by \citet{young2020} and \citet{fossati2020_KELT9datadriven}, which consists in first mapping the TP profile onto concentric circles and calculating the lengths through successive layers of atmosphere along line-of-sight transmission chords. These lengths, along with the atmospheric properties of their respective layers, are then stacked and entered into {\sc Cloudy} as the line of sight transmission medium. The final transmission spectrum of the planet is then computed by adding up the single layer spectra, weighted by their relative area projected on the stellar disc and accounting for the planetary impact parameter.

\begin{figure}[ht!]
                \centering
                \includegraphics[width=9cm]{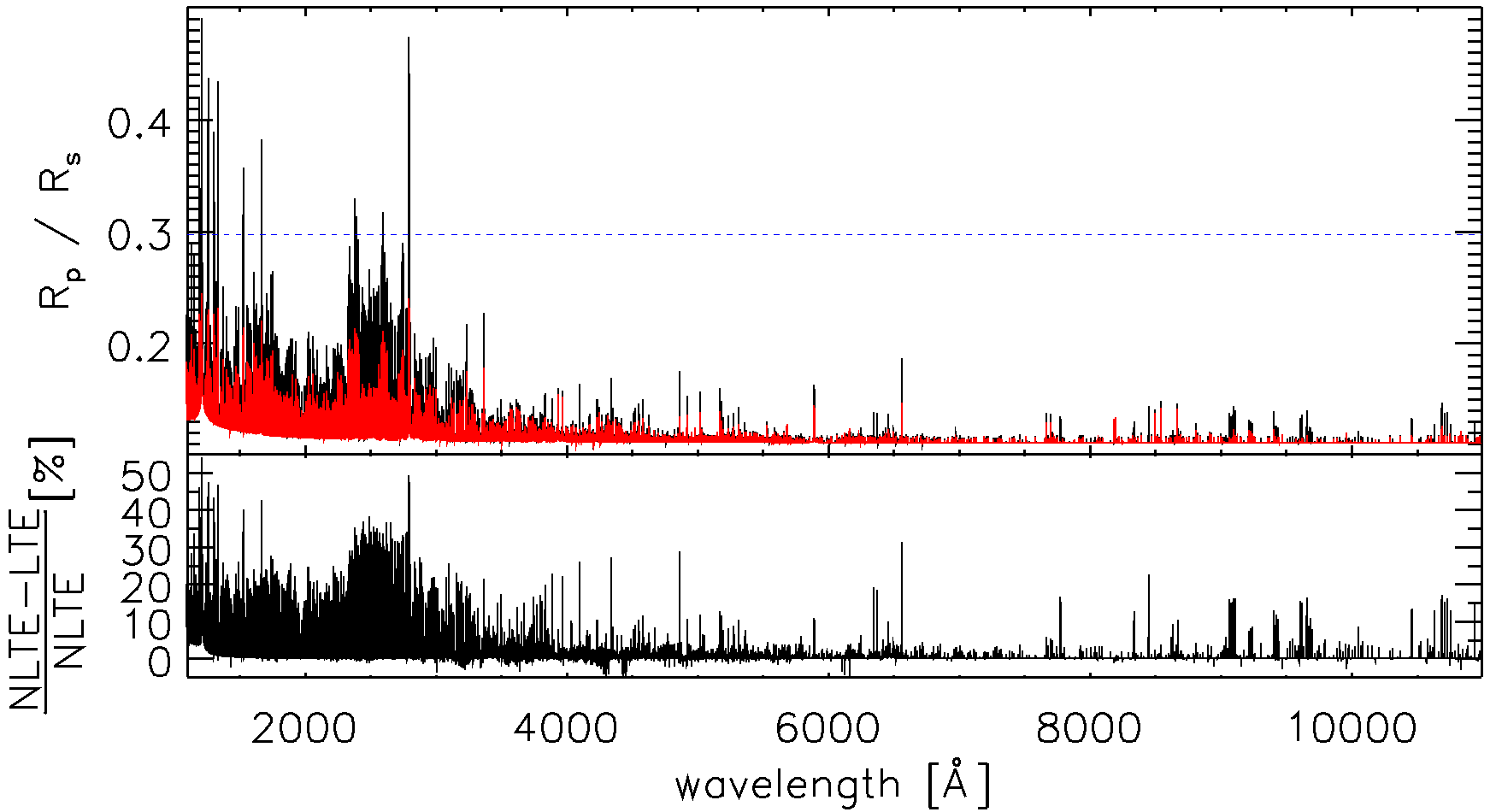}
                \includegraphics[width=9cm]{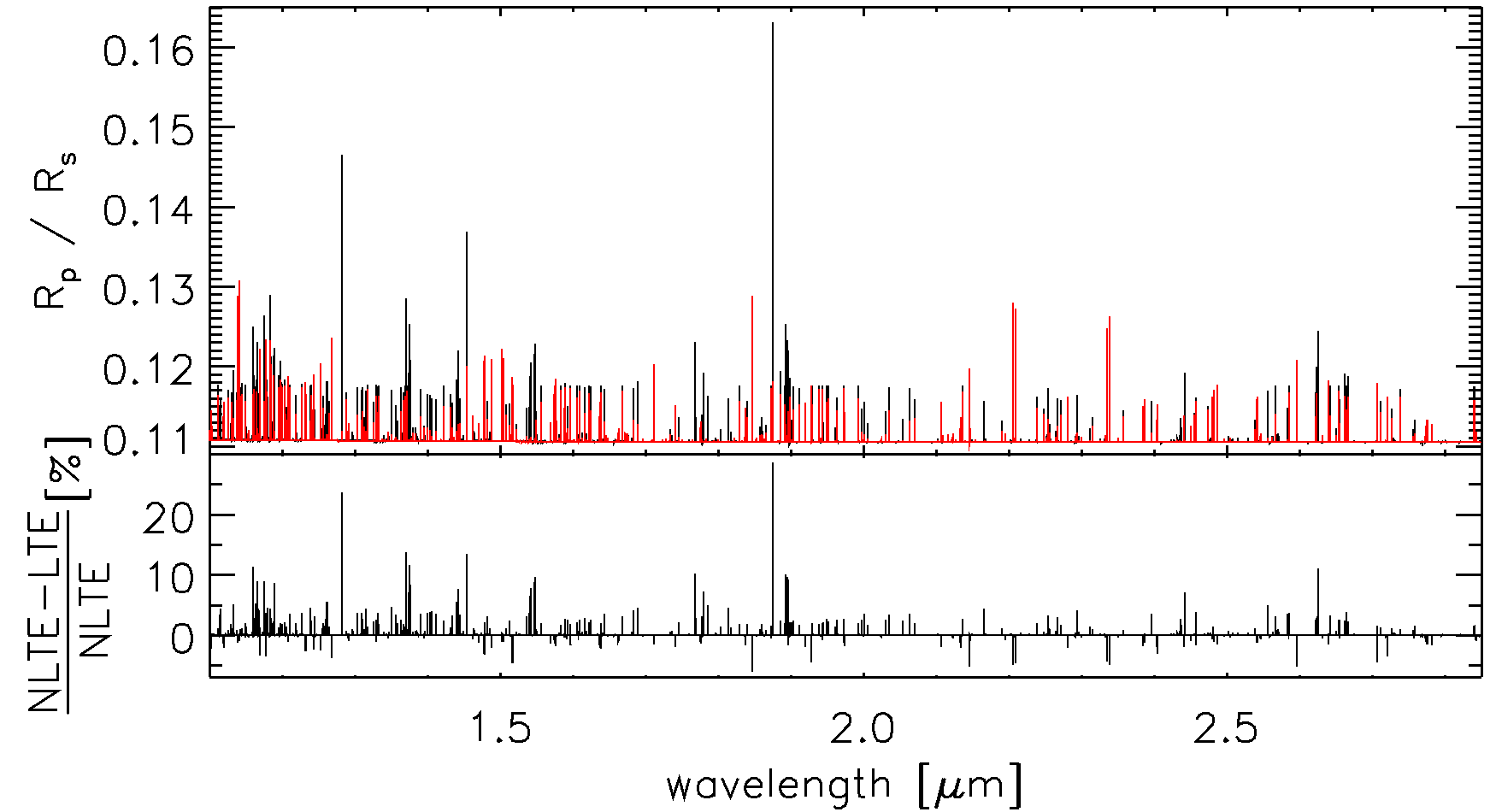}
                \caption{Comparison between the theoretical LTE (red) and NLTE (black) transmission spectra computed at a resolution of 100\,000 in the UV and optical range (top) and in the near infrared (bottom). Within each plot, the bottom panel shows the relative difference between LTE and NLTE (in percent). In the top panel, the blue horizontal dashed line indicates the location of the Roche radius in the planetary limb direction (i.e. perpendicular to the star-planet connecting line).}
                \label{fig:transmission_spectra}
\end{figure}

Figure~\ref{fig:transmission_spectra} shows the LTE and NLTE synthetic transmission spectra covering the entire wavelength range, while similar plots of specific wavelength ranges can be found in Appendix (Figures~\ref{fig:transmission_spectra_1100-1550}, \ref{fig:transmission_spectra_1500-2350}, \ref{fig:transmission_spectra_2300-3050}, \ref{fig:transmission_spectra_3000-4050}, \ref{fig:transmission_spectra_4000-6100}, and \ref{fig:transmission_spectra_6000-11000}). For completeness and easier comparison with observations, we show in Figure~\ref{fig:NLTEvsLTE_transit_depth_difference} the transit depth difference between the theoretical LTE and NLTE transmission spectra.
%-------------------------------------
%-------------------------------------

Figure~\ref{fig:transmission_spectra} indicates that, in general, the LTE transmission spectrum tends to underestimate absorption, mostly due to the fact that the LTE model is generally cooler than the NLTE model, though there are some typically weak features showing the opposite effect. The obtained behaviour is similar to that previously found for MASCARA-2b/KELT-20b and KELT-9b \citep{fossati2021_kelt9cloudy,fossati2023_kelt20mascara2}, but the amplitude of the difference between the LTE and NLTE transmission spectra is significantly larger in the case of WASP-178b. For example, in the case of the H$\alpha$ line the impact of NLTE effects on the transmission spectrum is of about 15\% for KELT-9b and about 11\% for MASCARA-2b/KELT-20b, while we find it to be about 32\% for WASP-178b, that is more than twice that found for KELT-9b. This is likely due to the more inflated nature of the atmosphere of WASP-178b resulting from the differences in planetary mass. Figure~\ref{fig:transmission_spectra} also shows that the core of the strongest UV absorption lines predicted by the NLTE model lies beyond the Roche lobe. Given that our simulations assume one-dimensional geometry, the actual strength of these lines should be taken with caution as significant differences can be expected when compared with observations.

As in previous analyses, we find the strongest deviation from LTE in the UV band, with deviations around the 10--40\%, except for strong resonance lines (e.g. Ly$\alpha$, Mg{\sc ii}\,H\&K) where the deviation reaches 50\%. In the optical and near-infrared bands, we find deviations from LTE typically below 10\%, except for specific lines, such as the hydrogen lines (Balmer and Paschen) for which we find deviations larger than 20\%. Similarly to the case of MASCARA-2b/KELT-20b, the absorption of the O{\sc i} triplet at about 7780\,\AA, which shows a prominent deviation from LTE, appears to be weaker than that of the H$\alpha$ line, while for KELT-9b the two lines had comparable strengths \citep{fossati2021_kelt9cloudy,fossati2023_kelt20mascara2,borsa2022_kelt9b_OI}. From an observational point of view, this result suggests that it will require data of significantly higher quality than that of the data analysed by \citet{borsa2022_kelt9b_OI} to detect the O{\sc i} triplet in the atmosphere of both MASCARA-2b/KELT-20b and WASP-178b. 

Also in the case of WASP-178b, the difference between the LTE and NLTE transmission spectra, particularly in the UV band, is partly due to the difference in underlying TP profiles, in addition to the different physical assumptions in the radiative transfer calculations. Most of the spectral lines lying in the UV belong to ionised species, which are more abundant in the NLTE model as a consequence of the higher temperature of the NLTE TP profile compared to the LTE TP profile, particularly in the line forming region. Furthermore, the hotter temperature of the NLTE model leads to a higher pressure scale height compared to the LTE model, which also affects the strength of the lines in the transmission spectrum.

\begin{figure}[ht!]
                \centering
                \includegraphics[width=9cm]{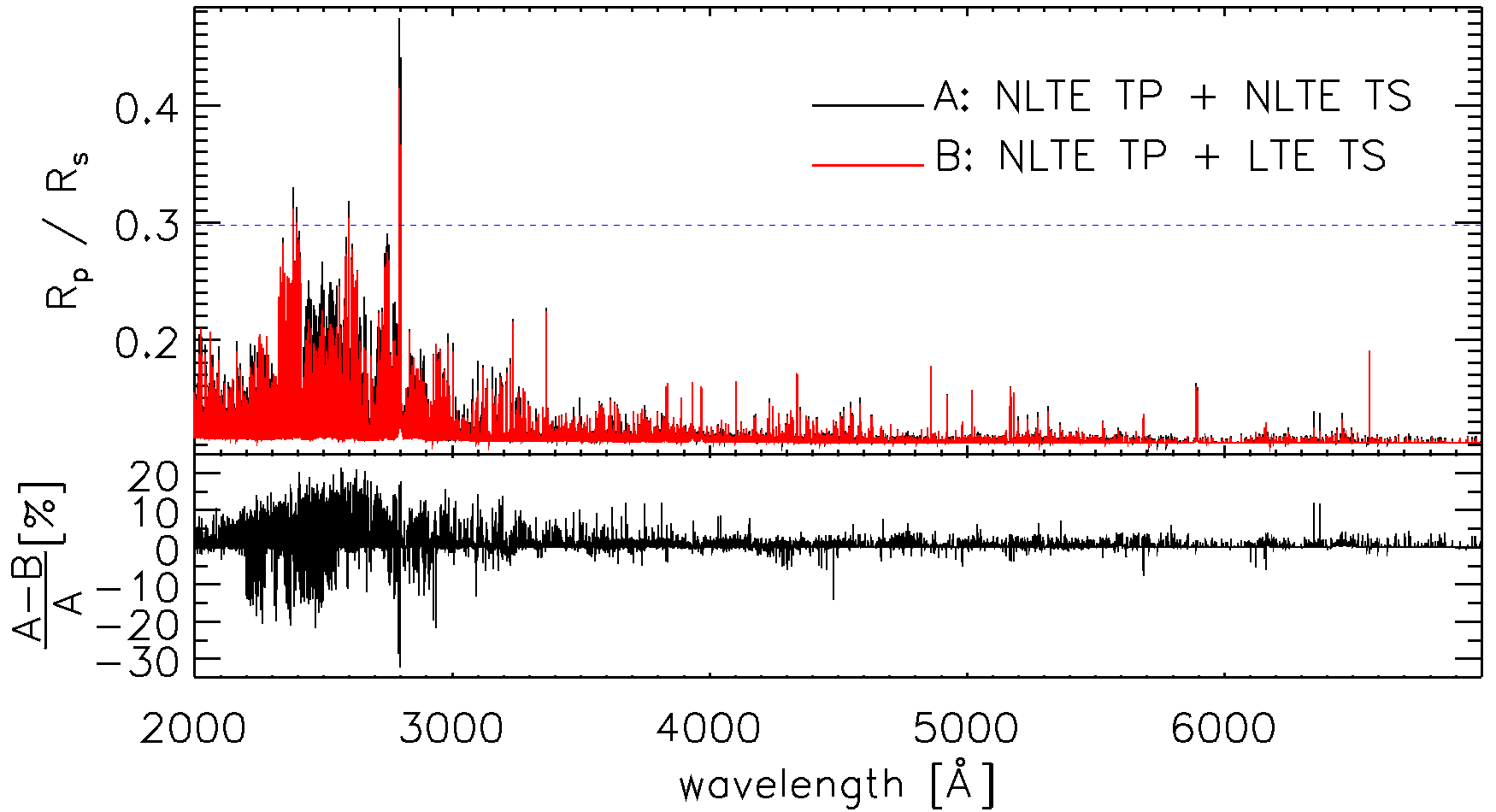}
                \caption{Comparison between theoretical transmission spectra computed with full NLTE (A; black; same as black line in Figure~\ref{fig:transmission_spectra}), namely NLTE TP profile and accounting for NLTE effects for the computation of the transmission spectrum (TS), and with partial NLTE (B, red), namely NLTE TP profile and assuming LTE for the computation of the transmission spectrum, at a resolution of 100\,000 in the NUV and optical range. The bottom panel shows the relative difference between the two spectra, in percent. In the top panel, the blue horizontal dashed line indicates the location of the Roche radius in the planetary limb direction.}
                \label{fig:transmission_spectra_FakeNLTE}
\end{figure}

Computationally, it is not yet possible to run full retrievals accounting for NLTE effects, however it might be interesting to explore the possibility of deriving abundances and/or abundance profiles on the basis of a fixed TP profile computed using forward modelling that accounts for NLTE effects. To this end, we used the NLTE TP profile as input to {\sc Cloudy} to compute a transmission spectrum assuming LTE. Figure~\ref{fig:transmission_spectra_FakeNLTE} shows a comparison between the NUV and optical transmission spectrum computed with partial NLTE (in red) and the one computed with full NLTE (in black). In the optical, the difference is typically small and below 10\%, however we remark that {\sc Cloudy} in LTE still assumes NLTE level populations for the first two hydrogen states, which is most likely why there is no difference between the two H$\alpha$ lines. In the NUV, the difference increases significantly up to around 20\%. Nevertheless, a comparison between Figure~\ref{fig:transmission_spectra} and \ref{fig:transmission_spectra_FakeNLTE} shows that the partial NLTE transmission spectrum is on average closer than the LTE one to the full NLTE transmission spectrum. This indicates that retrievals based on the partial NLTE approach might lead to more reliable results than retrievals fully based on the LTE assumption.
%-------------------------------------
%-------------------------------------
%
\subsubsection{Line formation region}\label{sec:line_formation}
We used the {\sc Cloudy} outputs obtained for each atmospheric layer in the synthetic transmission spectra calculations to estimate the line formation region in transmission geometry for a few of the strongest lines detected in the optical transmission spectra of UHJs: H$\alpha$, the reddest of the Mg{\sc i}\,b lines at $\sim$5183\,\AA, the Ca{\sc i} line at $\sim$6439\,\AA, the Ca{\sc ii}\,K line at $\sim$3933\,\AA, the Fe{\sc i} line at $\sim$4071\,\AA, and the Fe{\sc ii} line at $\sim$5169\,\AA. In particular, we started from the top of the atmosphere (at 5$\times$10$^{-12}$\,bar) looking at the fraction of light absorbed through increasingly thick atmospheric layers at the center of each line. In this scheme, the top boundary of the formation region of a given line (i.e. at low pressure) is set where, with increasing pressure of the lower considered layer, the line strength in the transmitted spectrum starts deviating from zero, while the bottom boundary of the line formation region (i.e. at high pressure) is set at the pressure for which the line strength stops increasing with increasing pressure.

\begin{figure}[t!]
                \centering
                \includegraphics[width=9cm]{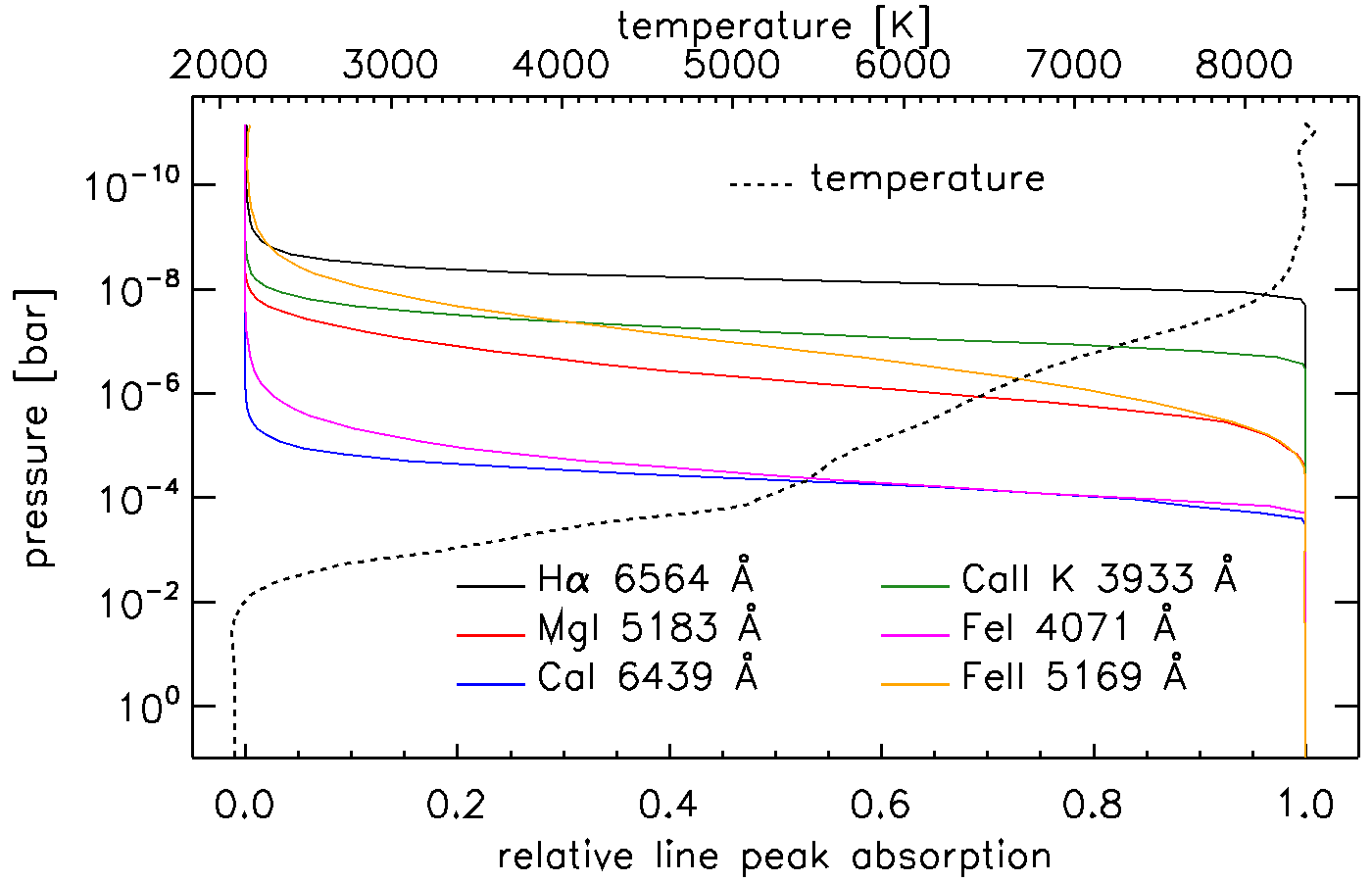}
                \includegraphics[width=9cm]{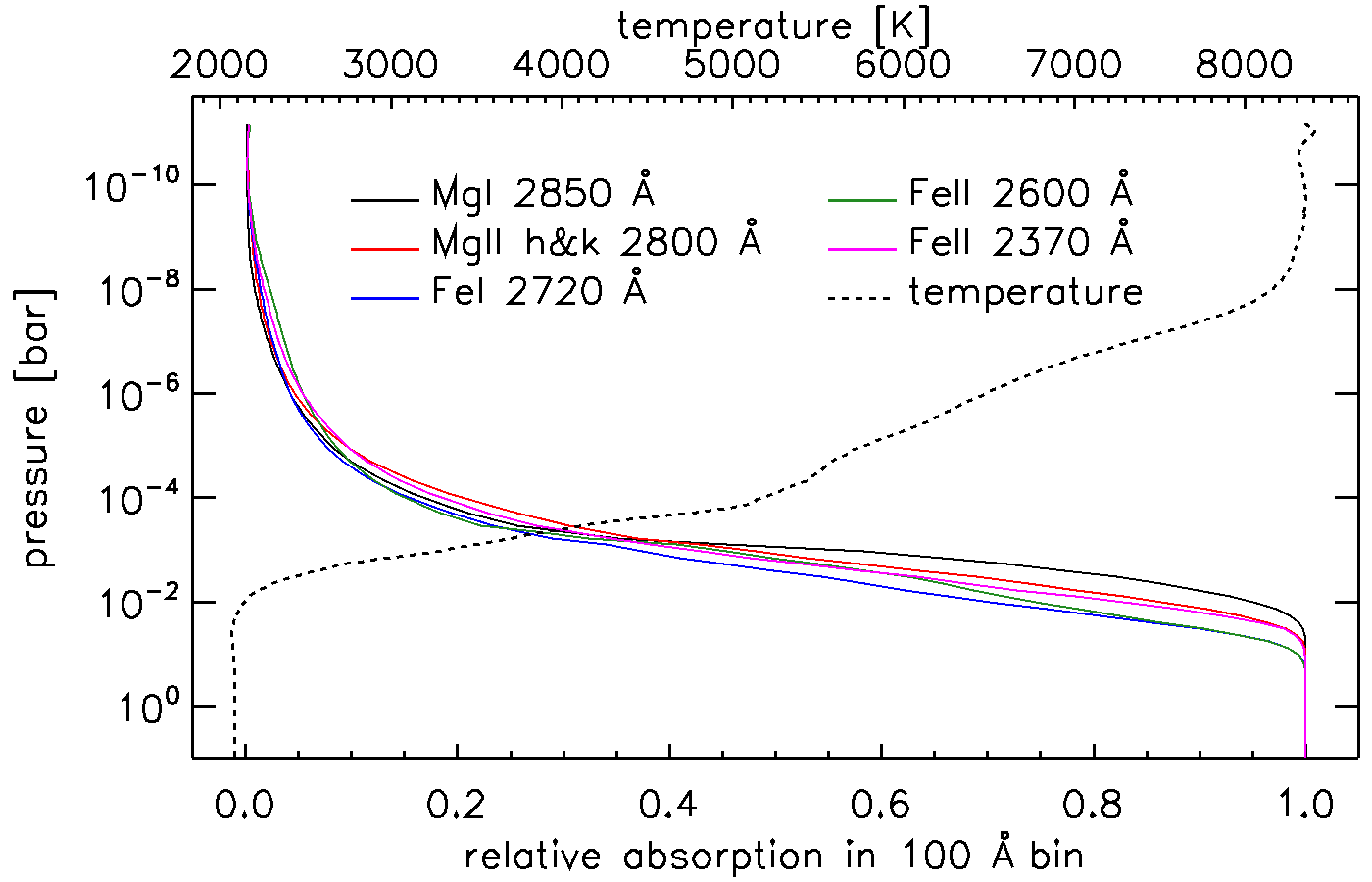}
                \caption{Line formation analysis in transmission geometry. Top: relative peak absorption for H$\alpha$ (black), the Mg{\sc i}\,b line at $\sim$5183\,\AA\ (red), the Ca{\sc i} line at $\sim$6439\,\AA\ (blue), the Ca{\sc ii}\,K line (green), the Fe{\sc i} line at $\sim$4071\,\AA\ (magenta), and the Fe{\sc ii} line at $\sim$5169\,\AA\ (orange). Bottom: relative absorption in a 100\,\AA\ spectral bin centered at the Mg{\sc i} resonance line at $\sim$2850\,\AA\ (black), the Mg{\sc ii}\,h\&k resonance lines at $\sim$2800\,\AA\ (red), the Fe{\sc i} NUV absorption band at $\sim$2720\,\AA\ (blue), the Fe{\sc ii} resonance line at $\sim$2600\,\AA\ (green), and the Fe{\sc ii} NUV absorption band at $\sim$2370\,\AA\ (magenta) as a function of pressure. In both panels, the dashed line shows the TP profile (top x-axis). The line formation region corresponds to the pressure values in which the relative absorption ranges between zero and one.} 
                \label{fig:line_formation}
\end{figure}

The top panel of Figure~\ref{fig:line_formation} shows the relative peak absorption of the considered lines as a function of pressure and thus the line formation region corresponds to the pressure range in which the relative line peak absorption lies between zero and one. Among those considered, H$\alpha$ is the line forming higher up in the atmosphere, in the 1--10\,nbar range, while the Ca{\sc i} line at $\sim$6439\,\AA\ and the Fe{\sc i} line at $\sim$4071\,\AA\ are those forming at the higher pressures. In general, the lines considered here have a rather narrow formation region, spanning 1--2 orders of magnitude in pressure, except for the Fe{\sc ii} line at $\sim$5169\,\AA\ that has a line formation region spanning about five orders of magnitude in pressure (10$^{-5}$--10$^{-10}$\,bar). The line formation region shown in Figure~\ref{fig:line_formation} reflects the abundance profile of each ion. 
%-------------------------------------
%-------------------------------------

Figure~\ref{fig:TPfrankenstein} and the top panel of Figure~\ref{fig:line_formation} indicate that in the optical band lines form at a temperature higher than 5000\,K and that the H$\alpha$ line forms in the hottest part of the atmosphere, where the temperature is higher than 8000\,K. Similarly to the cases of KELT-9b and MASCARA-2b/KELT-20b, the complete lack of Ly\,$\alpha$ emission from the host star implies that the excitation of H{\sc i} to the $n$\,=\,2 level, which leads to the formation of the Balmer lines, has to occur mostly thermally. Furthermore, because of the very low EUV emission of the host star, photoionisation and subsequent recombination rates to the $n$\,=\,2 level are low compared to thermal excitation \citep{fossati2023_kelt20mascara2}. Figure~\ref{fig:line_formation} also shows the pressure range probed by the lines considered here and thus the atmospheric layers constrained by observations through both forward modelling and retrieval analyses.

We carried out the same line formation analysis described above, but computing {\sc Cloudy} models at a much lower spectral resolution of 100. In particular, we computed the average relative absorption in a 100\,\AA\ range centered on the strongest spectral features in the NUV band, namely the Mg{\sc i} resonance line at $\sim$2850\,\AA, the Mg{\sc ii}\,h\&k resonance lines at $\sim$2800\,\AA, the Fe{\sc i} NUV absorption band at $\sim$2720\,\AA, the Fe{\sc ii} resonance line at $\sim$2600\,\AA, and the Fe{\sc ii} NUV absorption band at $\sim$2370\,\AA. These are the approximate spectral resolution and wavelength bin width covered by the WFC3/UVIS observations presented by \citet{lothringer2022_wasp178}. The bottom panel of Figure~\ref{fig:line_formation} shows the average relative absorption as a function of pressure, indicating that at the resolution of WFC3/UVIS the observations probe mainly the $\approx$10$^{-5}$--10$^{-1}$\,bar pressure range. 

Figure~\ref{fig:TPliteratureCompare} shows a comparison of the LTE and NLTE TP profiles obtained joining {\sc helios} and {\sc Cloudy} with those presented by \citet{lothringer2022_wasp178}, \citet{cont2024}, and \citet{lothringer2025_wasp178} assuming LTE. All results support the presence of a temperature inversion, as also evinced by the observations, but the magnitude of the inversion is significantly larger when accounting for NLTE effects. In the lower atmosphere ($>$1\,mbar), all TP profiles are comparable (particularly when considering that at high pressures the temperature of \citealt{cont2024} is a lower limit), which is partly driven by the fact that the {\sc helios} calculations have been set up to produce a profile comparable to that of {\sc phoenix}. The main difference lies in the middle and upper atmosphere, where the profiles of \citet{lothringer2022_wasp178} and \citet{cont2024}, as well as our LTE profile, predict a temperature that is a few 1000 K cooler than our NLTE profile. In the 10$^{-4}$--10$^{-5}$\,bar range, the temperature retrieved by \citet{lothringer2025_wasp178} is particularly low compared to all other profiles, although they obtained it including in the retrieval the observed NUV transmission spectrum.

In the case of the TP profiles obtained through forward modelling, the difference between our result and previous results can be ascribed mainly to the impact of accounting for NLTE effects. Instead, in the case of the retrieved profiles the difference with our NLTE profile is due not only to the different assumptions in the radiative transfer modelling (i.e. LTE vs NLTE), but also to the fact that the observations probe mainly the lower part of the atmosphere, at pressures higher than 10$^{-5}$\,bar.
\section{Discussion}\label{sec:discussion}
\subsection{Upper atmosphere and mass-loss rate}\label{sec:discussion:Cs_lambda}
\begin{figure}[t!]
                \centering
                \includegraphics[width=9cm]{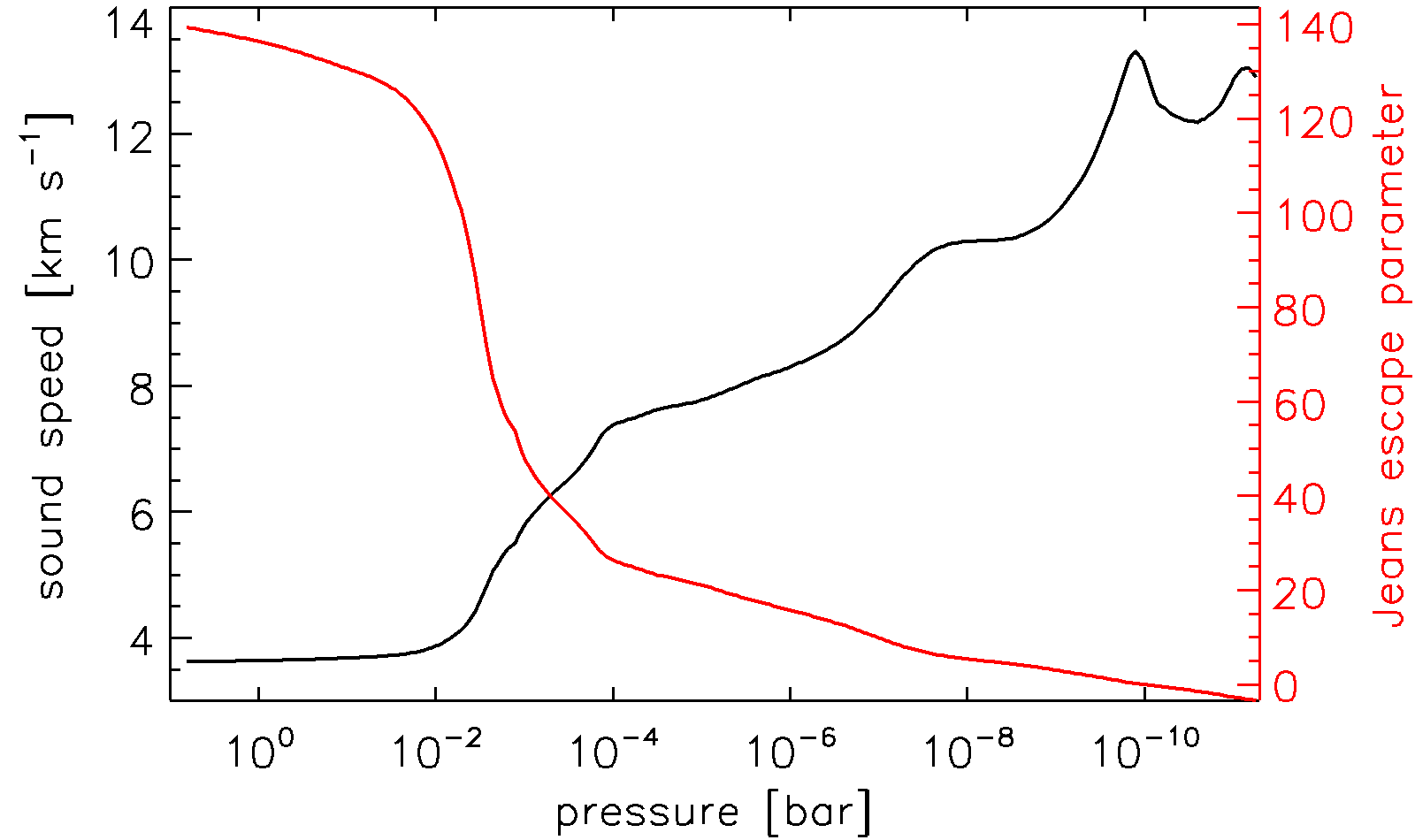}
                \caption{Atmospheric sound speed (black; left y-axis) and Jeans escape parameter (red; right y-axis) profiles as a function of pressure computed in the direction of the sub-stellar point (i.e. the Roche lobe lies at the L1 point).}
                \label{fig:Cs_lambda}
\end{figure}

{\sc Cloudy} is a hydrostatic code, but part of the modelled atmosphere lies beyond the Roche lobe (Figure~\ref{fig:transmission_spectra}), which calls for checking the validity of the hydrostatic approximation throughout the considered pressure range. Figure~\ref{fig:Cs_lambda} shows the profiles of the Jeans escape parameter and of the sound speed as a function of pressure. For the computation of the Jeans escape parameter, we used Equation (7) of \citet[][see also \citealt{volkov2011}]{fossati2017}, setting the bulk flow velocity equal to zero. In this way, the Jeans escape parameter accounts for the gravitational potential difference between every point in the atmosphere and the Roche lobe, where the Jeans escape parameter goes to zero. The strong metal lines in the optical band form roughly in the 10$^{-4}$--10$^{-8}$\,bar range, where the Jeans escape parameter is larger than five, and thus the atmosphere is hydrostatic. This is also the case of the atmospheric regions probed by the WFC3/UVIS and JWST observations. However, the H$\alpha$ line forms in the 1--10\,nbar range, where the Jeans escape parameter is between two and five, and thus the bulk outward velocity starts becoming non-negligible, though the velocity remains subsonic. Therefore, in the case of WASP-178b, in contrast to KELT-9b and MASCARA-2b/KELT-20b, high-resolution transmission spectroscopy observations of the \ion{H}{i} Balmer lines probe the upper atmosphere, and thus can provide information on the on-going hydrodynamic escape and eventually constrain the mass-loss rate. Our NLTE model supports the suggestion of \citet{damasceno2024_wasp178} that the significant broadening they found characterising the hydrogen Balmer line profiles can be related to atmospheric escape. However, the velocities \citet{damasceno2024_wasp178} inferred from the line broadening (39.6$\pm$2.1\,km\,s$^{-1}$ for H$\alpha$ and 27.6$\pm$4.6\,km\,s$^{-1}$ for H$\beta$) are significantly higher than the sound speed in the line formation region, indicating that only part of the broadening might be related to the motion of the escaping material, while other processes (e.g. thermal and non-thermal broadening and winds) must contribute to it.
%-------------------------------------
%-------------------------------------

We estimate the mass-loss rate from WASP-178b as
\begin{equation}
\dot{M} \approx 4 \pi x_R z_R \frac{\rho_R}{2 \sqrt{\pi}} \sqrt{\frac{2 k T_R}{m_R}}\,,
\end{equation}
where $x_R$ is the radial distance to the L1 point, $z_R$ is the radial distance to pole, $\rho_R$ is the mass density at the Roche lobe boundary, $T_R$ is the temperature at the Roche lobe boundary, and $m_R$ is the mean molecular weight at the Roche lobe boundary. Detailed simulations of atmospheric escape show that this approach is reasonably accurate, to a factor of a few, as long as the exobase is below the Roche lobe boundary \citep{koskinen2022}. Based on our model of the atmosphere, we obtain a mass-loss rate of 4.5$\times$10$^{10}$\,g\,s$^{-1}$, which is comparable to that obtained for the archetypal hot Jupiter HD209458b. This mass-loss rate is likely high enough to affect the structure of the atmosphere above the heating peak, by causing temperatures to decrease with altitude above the $\sim$1 nbar level. As we indicate above, this means that self-consistent models of escape and thermal structure are required to simulate absorption signatures of strong atomic lines probing lower pressures at high radial distances from the planet.   

\subsection{WASP-178b vs MASCARA-2b/KELT-20b vs KELT-9b}\label{sec:discussion:comparisonsTP_w178_m2_k9}
\begin{figure}[t!]
                \centering
                \includegraphics[width=9cm]{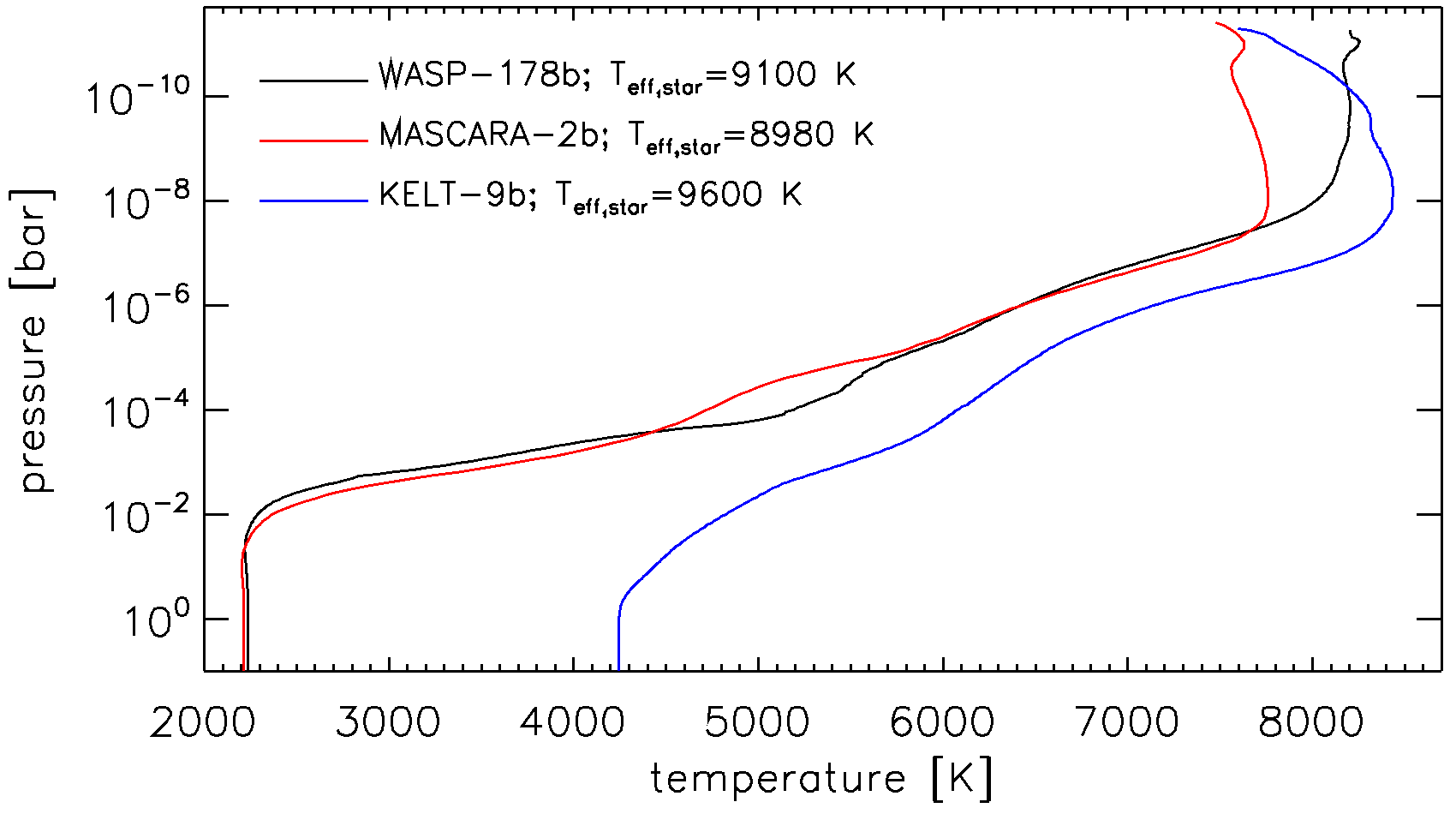}
                \caption{Comparison between the theoretical NLTE TP profiles of WASP-178b (black; this work), MASCARA-2b/KELT-20b \citep[red; from][]{fossati2023_kelt20mascara2}, and KELT-9b \citep[blue; from][]{fossati2021_kelt9cloudy}. For context, the legend gives also the effective temperature of the hosts stars from Section~\ref{sec:star} (for WASP-178b), \citet[][for MASCARA-2b/KELT-20b]{talens2018}, and \citet[][for KELT-9b]{borsa2019}.}
                \label{fig:comparisonsTP_w178_m2_k9}
\end{figure}

The {\sc helios} and {\sc Cloudy} models have been used to construct the theoretical TP profiles, accounting for NLTE effects in the middle and upper atmosphere, of three UHJs: WASP-178b, MASCARA-2b/KELT-20b, and KELT-9b. Figure~\ref{fig:comparisonsTP_w178_m2_k9} shows a direct comparison between the theoretical TP profiles of the three planets\footnote{There is a slight difference between the KELT-9b TP profile shown in Figure~\ref{fig:comparisonsTP_w178_m2_k9} and that presented by \citet{fossati2021_kelt9cloudy}. This is due to the fact that we recomputed the TP profile of KELT-9b in the middle and upper atmosphere (i.e. with {\sc Cloudy}) using a spectral resolution for the input stellar flux and opacity sampling that is equal to those used to compute the TP profiles of WASP-178b and MASCARA-2b/KELT-20b, which is significantly higher than that used by \citet{fossati2021_kelt9cloudy} that was constrained by the computing power available at the time.}.
%-------------------------------------
%-------------------------------------

As a result of the hotter host star, shorter orbital separation and therefore a higher planetary equilibrium temperature, the lower atmosphere of KELT-9b is significantly hotter than that of the other two planets, which have similar equilibrium temperatures of about 2400\,K \citep{lund2017,talens2018,hellier2019}. All three planets have a significant temperature inversion in the middle atmosphere and the three TP profiles get closer to each other with decreasing pressure. The temperature inversion is more significant for the two cooler planets, and in particular for WASP-178b, which shows the strongest inversion with a temperature difference of about 6000\,K between the top and bottom of the atmosphere. Remarkably, at the top of the atmosphere, the temperature of WASP-178b is comparable to that of KELT-9b, while that of MASCARA-2b/KELT-20b is about 500\,K lower. 

The characteristics of the TP profiles in the upper atmosphere do not appear to be directly related to either the temperature of the host star or the orbital separation. Furthermore, the host stars are very similar in terms of high-energy emission as all have low XUV emission. One of the main parameters distinguishing the three planets is the bulk density, which impacts the atmospheric pressure scale height, thus the penetration depth of the stellar radiation in the planetary atmosphere, and in turn the shape of the TP profile. As a matter of fact, among the three planets WASP-178b has the lowest bulk density \citep[$\approx$0.35\,g\,cm$^{-3}$;][]{hellier2019}, while MASCARA-2b/KELT-20b has the highest bulk density  \citep[$\approx$0.71\,g\,cm$^{-3}$;][]{talens2018}, with KELT-9b lying roughly in between \citep[$\approx$0.49\,g\,cm$^{-3}$;][]{borsa2019}. Furthermore, WASP-178b and KELT-9b have Roche radii that are significantly closer to the planetary radius compared to MASCARA-2b/KELT-20b. At the same time, atomic line cooling could act as a thermostat, such that independently of energy input the heating rate is offset by a cooling rate that increases with temperature. Therefore, a parameter space study could lead to the explanation of the origin of the behaviour shown in Figure~\ref{fig:comparisonsTP_w178_m2_k9}.
\subsection{Comparison with HST observations}\label{sec:discussion:observations_HST}
\citet{lothringer2022_wasp178} presented NUV transmission spectroscopy observations performed at low-resolution ($R$\,$\approx$\,70) with the WFC3/UVIS instrument on-board HST, which they interpreted using both forward and retrieval modelling based on the {\sc phoenix} code assuming LTE. They fitted the data by adding SiO opacity to the model comprising Mg and Fe, among other species, though the data could have also been fitted by increasing the atmospheric metallicity. In the end, they preferred the first option. This choice was based on a NUV medium-resolution ($R$\,$\approx$\,30\,000) transmission spectrum collected with the STIS spectrograph on-board HST, which, despite the higher spectral resolution, did not show extra-absorption at the position of strong Fe and Mg resonance lines compared to the UVIS spectrum. 

Below, we present a comparison of the LTE and NLTE synthetic transmission spectra with the HST observations. We extracted the low-resolution WFC3/UVIS transmission spectrum from \citet{lothringer2022_wasp178}, but we could not do the same for the STIS mid-resolution spectrum because \citet{lothringer2022_wasp178} did not provide it in machine readable format and their extended data Figure~8 shows just short portions ($\approx$90\,\AA\ wide) around the Mg{\sc ii}\,h\&k lines and the strongest Fe{\sc ii} NUV resonance lines. Therefore, to enable a comparison of the synthetic transmission spectra with the STIS observations, we downloaded the data from the HST archive and analysed them independently. The details of the STIS data analysis are given in Appendix~\ref{appendix:hst_stis_analysis}.
\subsubsection{WFC3/UVIS}
As shown in Figure~\ref{fig:TPliteratureCompare}, by considering NLTE effects, we calculate a significantly higher temperature in the middle and upper atmosphere compared to those obtained by \citet{lothringer2022_wasp178} and \citet{lothringer2025_wasp178} using forward modelling or retrievals. This leads to larger pressure scale heights, and thus a larger extent of the atmosphere, which, together with the stronger absorption features, might allow us to reproduce the UVIS data without the need of SiO absorption or super-solar abundances. 

\begin{figure}[t!]
                \centering
                \includegraphics[width=9cm]{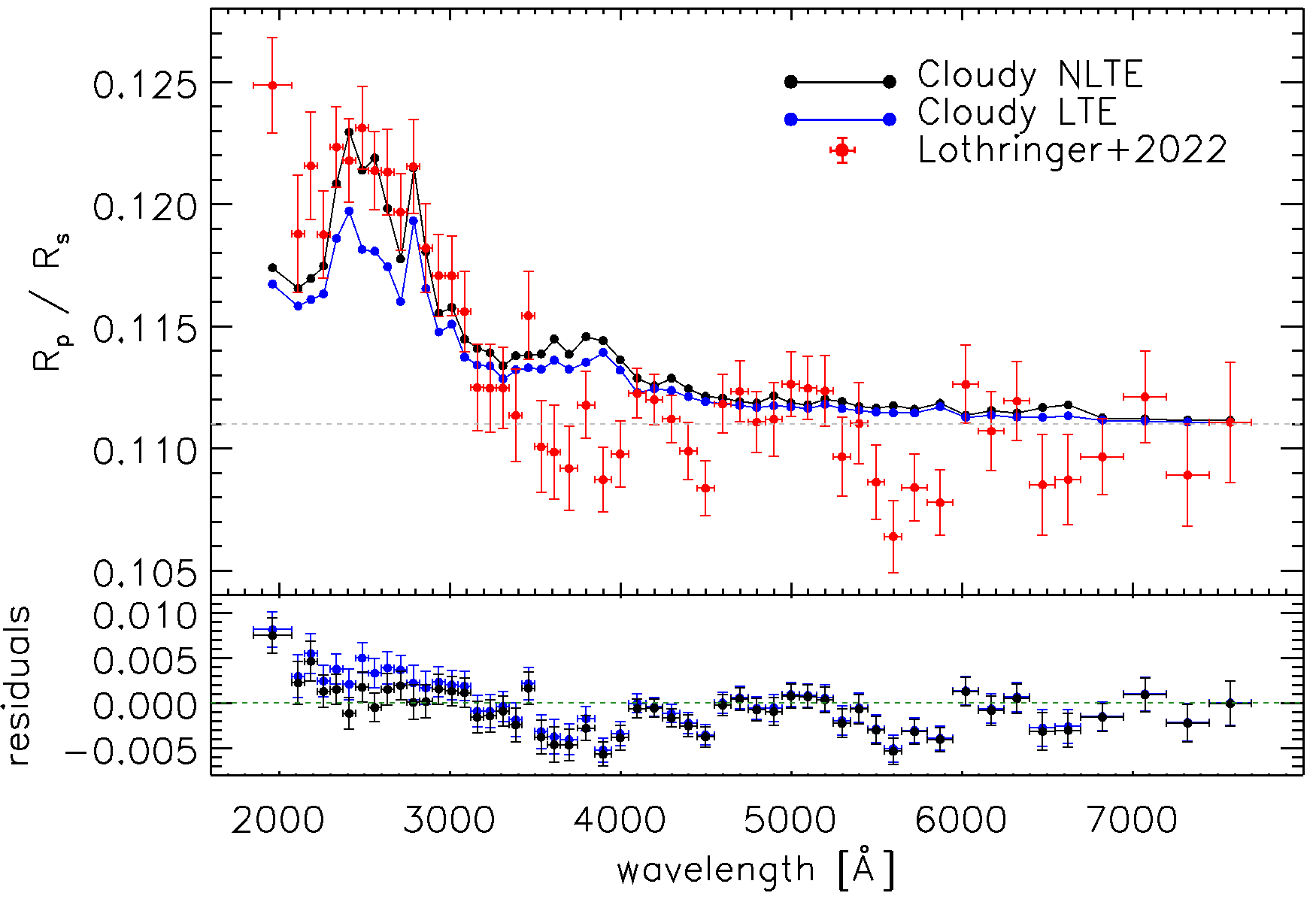}
                \includegraphics[width=9cm]{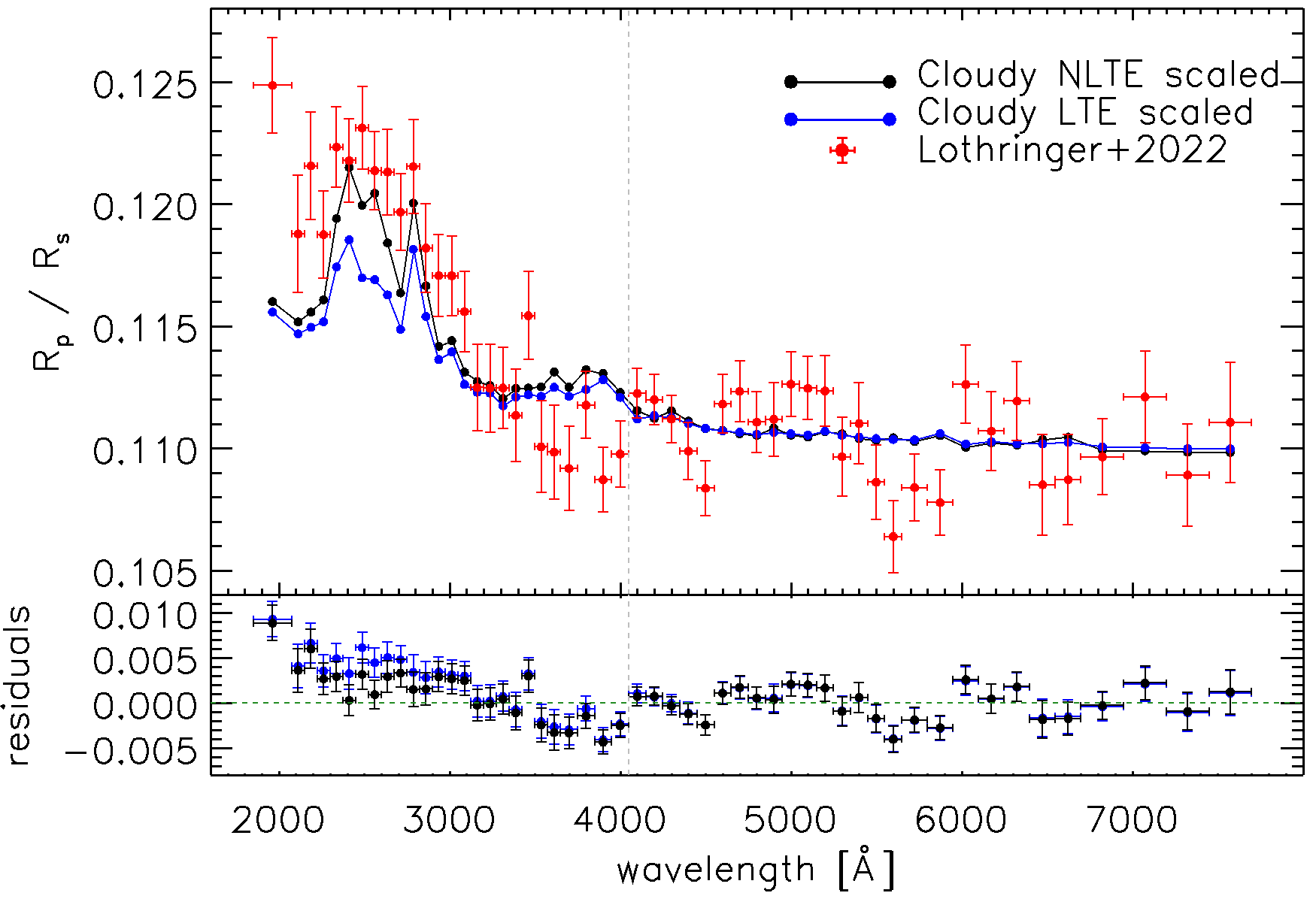}
                \caption{Comparison of the NLTE (black) and LTE (blue) theoretical transmission spectra with the WFC3/UVIS observations published by \citet[][ red]{lothringer2022_wasp178}. The horizontal error bars shown for the red points correspond to the size of the wavelength bins. The top plot is without scaling of the models, while in the bottom plot the NLTE and LTE models have been scaled to match the average of the UVIS observations at wavelengths longer than 4050\,\AA, marked by the vertical gray dashed line. In the top plot, the gray horizontal dashed line indicates the broad-band $R_{\rm p}/R_{\rm s}$ obtained from the EulerCAM optical transits \citep{hellier2019}. In each plot, the bottom panel shows the residuals between the observation and each of the models, with a green dashed line at zero for reference.}
                \label{fig:comparison_UVIS}
\end{figure}

The top panel of Figure~\ref{fig:comparison_UVIS} shows the comparison between the theoretical transmission spectra we obtained in LTE and NLTE with the UVIS data. The theoretical transmission spectra have been obtained by averaging the high-resolution spectra shown in Figure~\ref{fig:transmission_spectra} for each of the bins used to obtain the UVIS transmission spectrum. We find that in the NUV, the NLTE transmission spectrum provides an excellent fit to the data with a reduced $\chi_{\nu}^2$ of 0.95 in the 2000--3000\,\AA\ range, with 11 degrees of freedom, while the LTE transmission spectrum lies significantly below the observations returning a higher reduced $\chi_{\nu}^2$ of 3.77. Indeed, accounting for NLTE effects increases NUV absorption compared to what we obtained assuming LTE and this enables us to fit the UVIS observation without adding SiO absorption or increasing metallicity. 
%-------------------------------------
%-------------------------------------

However, in the optical band, the UVIS transmission spectrum lies slightly below the models, indicative of the presence of a small offset between the UVIS data and the broad-band radius ratio. Given the magnitude of the residual structured red noise present in the data, this offset might not be significant. However, to be conservative, we shifted the LTE and NLTE models to match the UVIS data at wavelengths longer than 4000\,\AA\ (bottom panel of Figure~\ref{fig:comparison_UVIS}). This leads to a worse fit to the data in the NUV band as indicated by the reduced $\chi_{\nu}^2$ values of 2.59 and 6.69 obtained for the NLTE and LTE models, respectively. Therefore, taking into account the possible offset, we find that to match the UVIS data one might require additional absorption across the entire NUV band. This could be achieved indeed by including some SiO absorption, or by a slight increase in the metallicity, which would agree with the results of \citet{cont2024} and \citet{damasceno2024_wasp178}. In fact, for example, a slight increase in Fe abundance would not only lead to an increase in the Fe opacity and thus NUV absorption, but also to increase the atmospheric temperature in the middle and upper atmosphere, which would further favour Fe thermal ionisation and increase the pressure scale height, with the net effect of increased NUV absorption. Also, we cannot rule out the possibility that the middle and upper atmosphere is even hotter than predicted by the NLTE model, which would further strengthen NUV absorption. In fact, low-resolution NUV observations of the UHJ WASP-189b conducted with the Colorado Ultraviolet Transit Experiment (CUTE) CubeSat suggest that the temperature in the upper atmosphere may be significantly higher than that predicted by state-of-the-art hydrodynamic models \citep{sreejith2023_cute_wasp189b}. Finally, it is important to note that our model is one-dimensional while three-dimensional effects are likely to be important particularly at the terminator region. This might impact the TP profile, and in turn the shape of the transmission spectrum. 
\subsubsection{STIS}
\begin{figure}[t!]
                \centering
                \includegraphics[width=9cm]{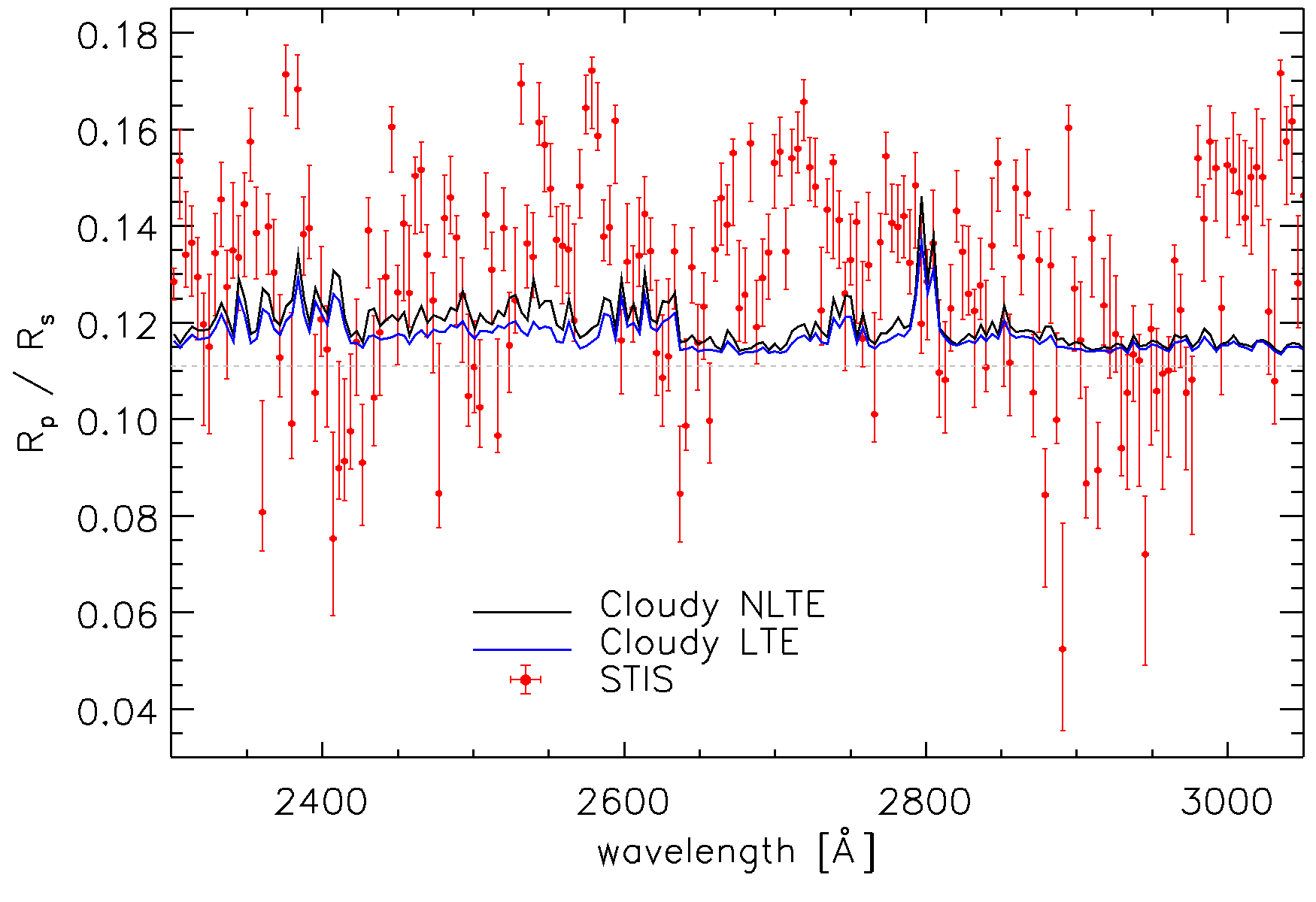}
                \includegraphics[width=9cm]{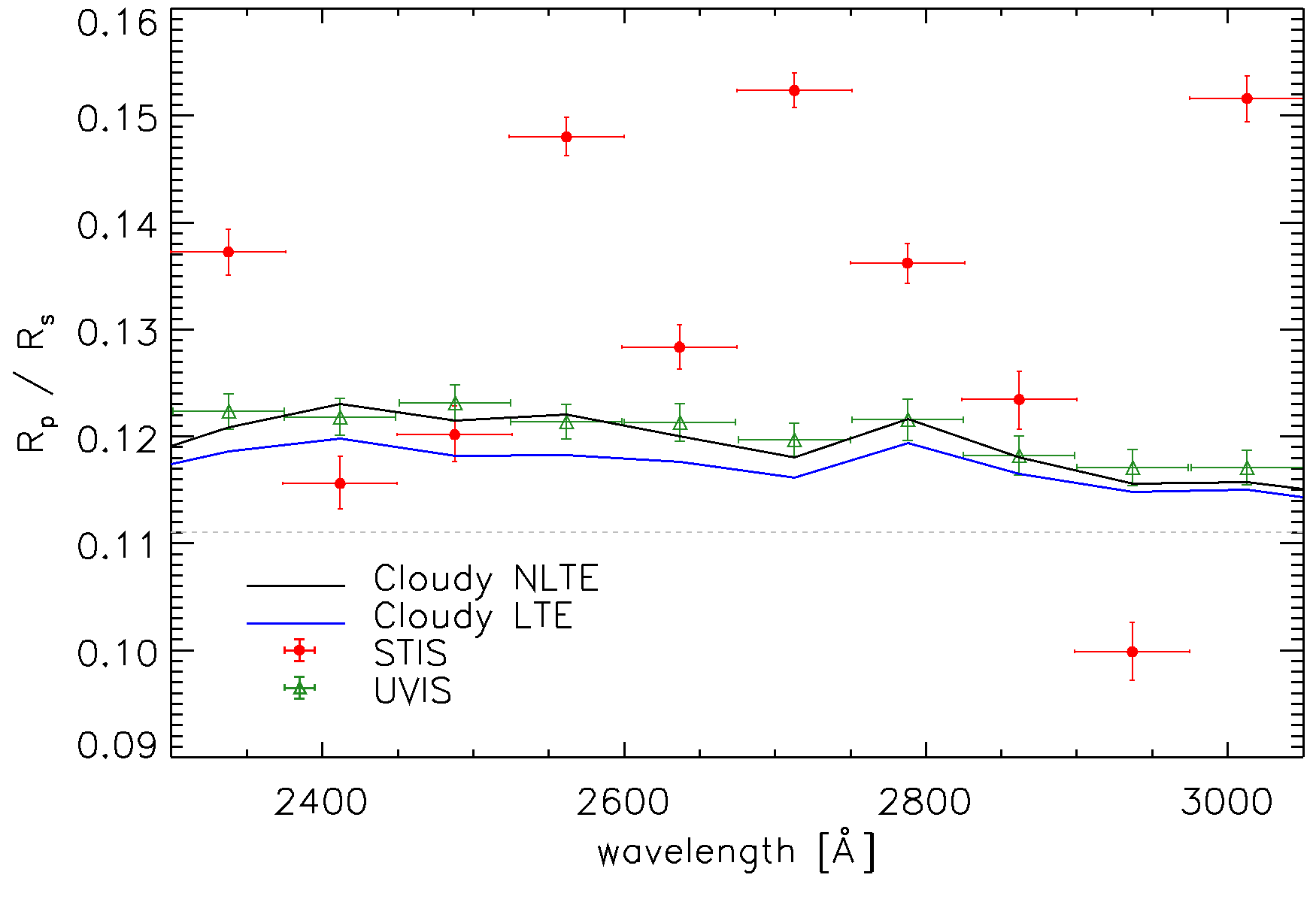}
                \caption{Comparison between observed and synthetic transmission spectra in the region covered by the STIS data. Top: the mid-resolution ($\approx$9\,\AA\ binning) STIS transmission spectrum is shown by red points. Bottom: the low-resolution ($\approx$74\,\AA\ binning) STIS and UVIS transmission spectra are shown by the red circles and green triangles, respectively. In both panels, the blue and black lines show the LTE and NLTE synthetic transmission spectra, respectively, while the gray horizontal dashed line indicates the broad-band $R_{\rm p}/R_{\rm s}$ obtained from the EulerCAM optical transits \citep{hellier2019}. The horizontal error bars shown for the red and green points correspond to the size of the wavelength bins.}
                \label{fig:comparison_STIS}
\end{figure}

The top panel of Figure~\ref{fig:comparison_STIS} shows the comparison of the LTE and NLTE synthetic transmission spectra with the mid-resolution ($\approx$9\,\AA\ binning) STIS transmission spectrum obtained using the `multiple jitter parameter' analysis method to remove systematic noise, as detailed in Appendix~\ref{appendix:hst_stis_analysis}. The data present a significant scatter around the expected transit depth and the shape of the transmission spectrum does not follow what is predicted by any of the two models. Therefore, to check the quality of the STIS data and systematic removal we compare the STIS transmission spectrum with the UVIS one at the same resolution. To this end, we re-extracted the STIS transmission spectrum from the data, but using the same binning as that of the UVIS data ($\approx$74\,\AA\ binning). The bottom panel of Figure~\ref{fig:comparison_STIS} shows the comparison of the LTE and NLTE synthetic transmission spectra with the UVIS and STIS transmission spectra, all at the same binning. The low-resolution STIS transmission spectrum does not match the UVIS transmission spectrum. Notably, the STIS transmission spectrum we extracted from the data following the procedure described in Appendix~\ref{appendix:hst_stis_analysis} matches well what presented by \citet[][see Figure~\ref{fig:tp_comparison}]{lothringer2022_wasp178}. Furthermore, Appendix~\ref{appendix:hst_stis_analysis} shows that the results obtained from the analysis of the STIS data is model-dependent. Therefore, we conclude that the STIS dataset of WASP-178b may not be reliably used to gather information about the atmosphere of WASP-178b. The residual noise present in the STIS data might result from slit losses caused by the use of a too narrow slit. Therefore, we recommend using a wider slit for future STIS transmission spectroscopy observations\footnote{For context, the UVIS observations have been obtained slitless, while the STIS observations have been obtained using a slit with an aperture of 0.2\arcsec.}.
%-------------------------------------
%-------------------------------------

\citet{lothringer2022_wasp178} fitted the UVIS transmission spectrum using an LTE model and required additional opacity to be able to fit the deeper NUV transit depths shown by the data. They identified in SiO the source of this opacity and used the lack of strong absorption at the position of the Mg{\sc ii}\,h\&k and of Fe{\sc ii} resonance lines in the STIS data to exclude the scenario in which hydrodynamic escape is responsible for the increased NUV absorption. However, our results indicate that SiO absorption is not required to fit the data. Nevertheless, even if the shape of the UVIS transmission spectrum in the NUV is the result of NLTE effects and not of SiO absorption, our findings support the conclusion of \citet{lothringer2022_wasp178} that Fe and Mg have not rained out.
\subsection{Comparison with optical observations}\label{sec:discussion:observations_optical}
\begin{figure}[t!]
                \includegraphics[width=9cm]{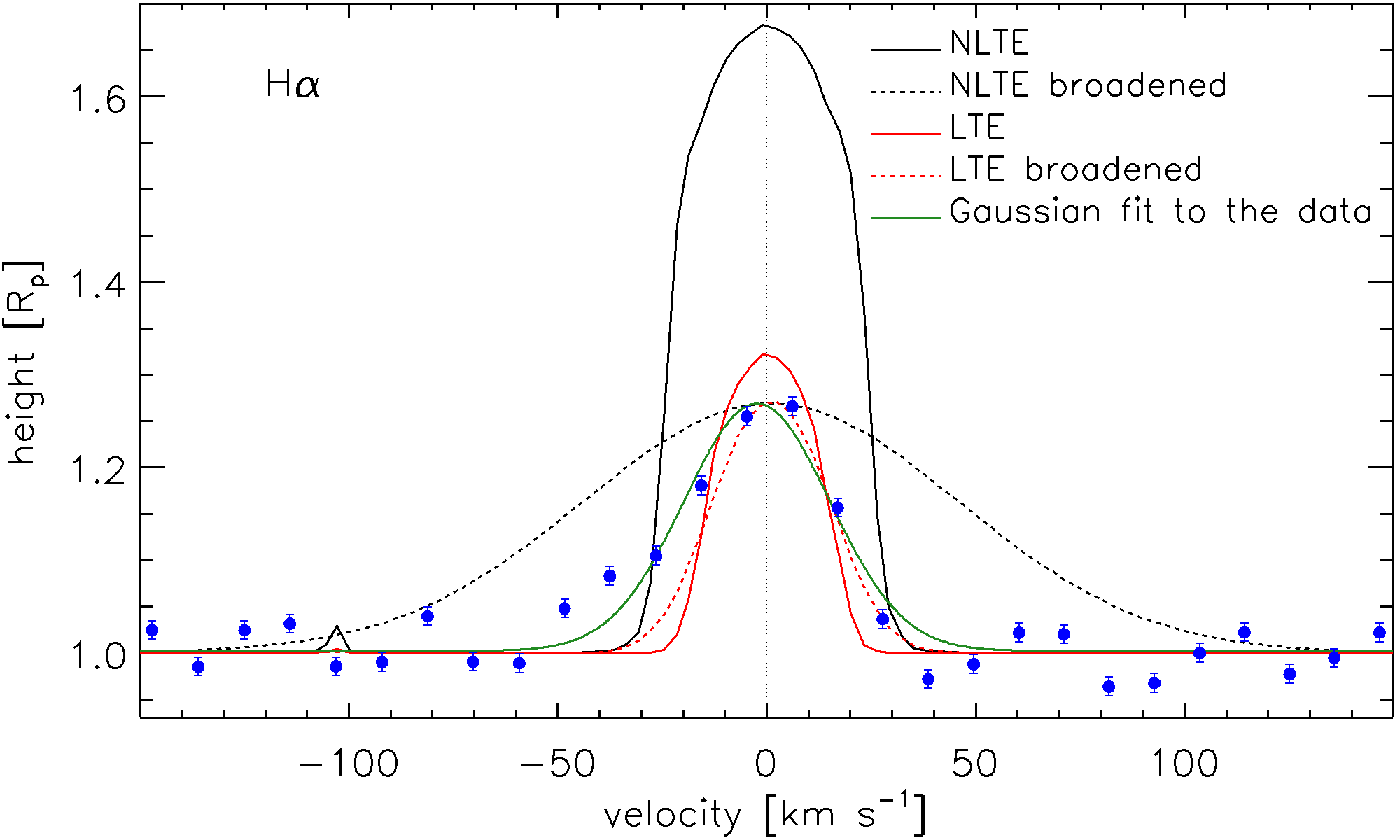}
                \includegraphics[width=9cm]{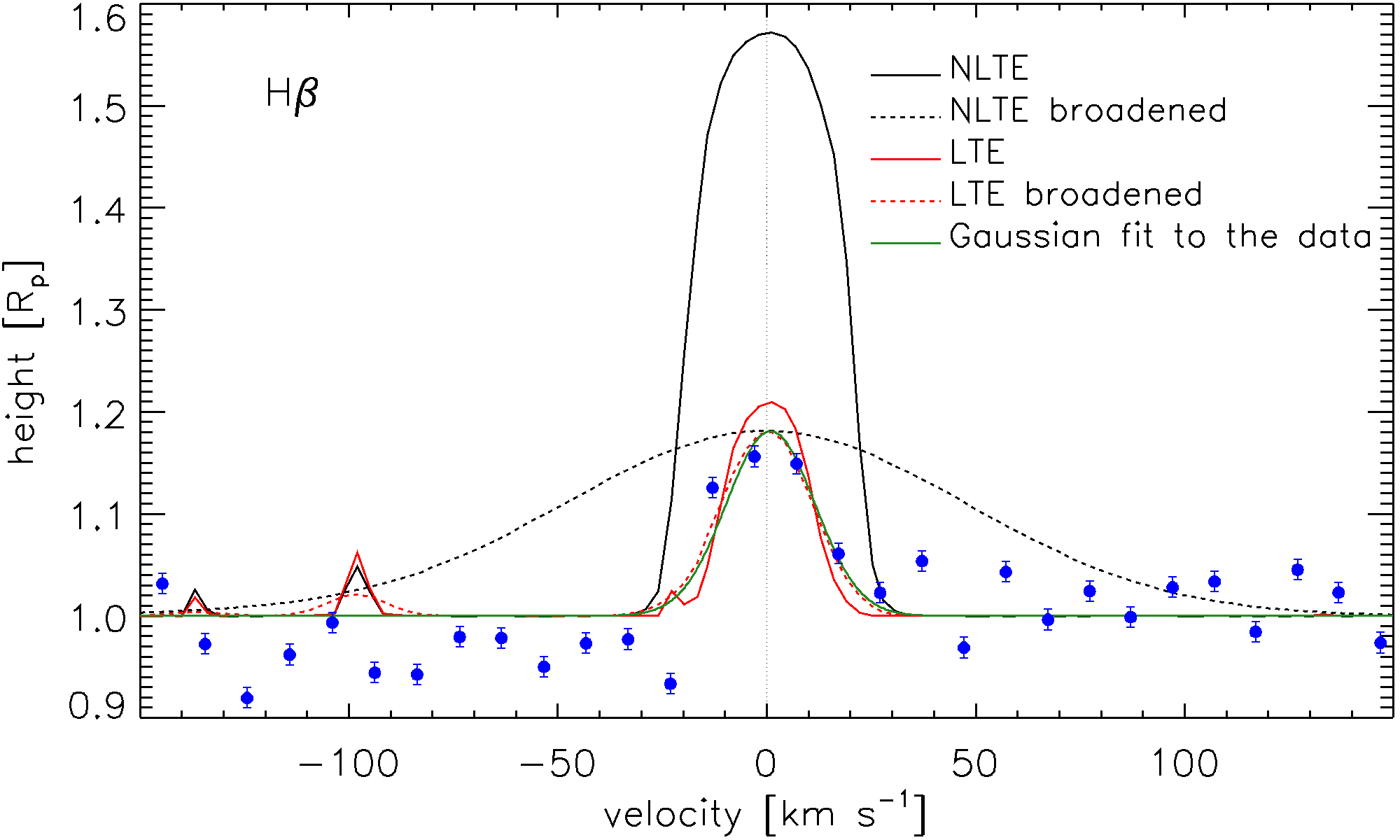}
                \includegraphics[width=9cm]{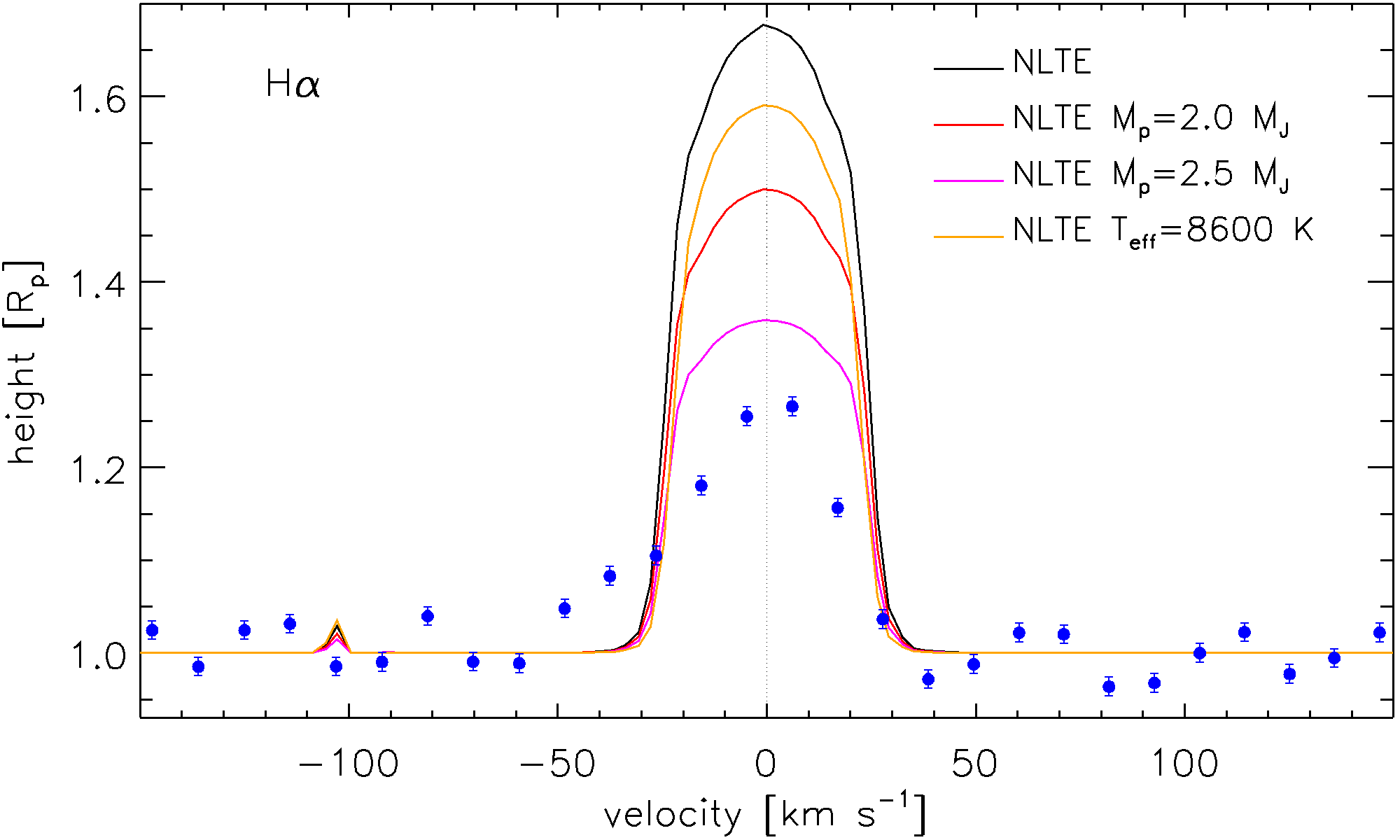}
                \caption{Comparison between synthetic and observed hydrogen Balmer line profiles. Top: NLTE (black) and LTE (red) theoretical transmission spectra compared to the observed H$\alpha$ line profile (blue dots) presented by \citet{damasceno2024_wasp178}. The green line is a Gaussian fit to the observed line profiles. The black and red dashed lines correspond, respectively, to the NLTE and LTE line profiles convolved with a Gaussian profile having a width such that the height of the lines equals the height of the Gaussian fit to the observations. Middle: same as top panel, but for the H$\beta$ line. Bottom: comparison of the observed H$\alpha$ line (blue dots) with NLTE synthetic spectra computed with the system parameters listed in Table~\ref{tab:parameters} (black; same as in the top panel), with $M_{\rm p}$\,=\,2.0\,$M_{\rm J}$ (red) and 2.5\,$M_{\rm J}$ (magenta), and with $T_{\rm eff}$\,=\,8600\,K (orange).}
                \label{fig:comparison_optical}
\end{figure}

Figure~\ref{fig:comparison_optical} shows the comparison of the synthetic NLTE and LTE transmission spectra with the high-resolution observations of the H$\alpha$ and H$\beta$ lines presented by \citet{damasceno2024_wasp178}. The NLTE synthetic line profiles are significantly stronger than the observations. Instead, the line profiles obtained assuming LTE appear to be a better match to the data, particularly following the inclusion of some additional broadening.
%-------------------------------------
%-------------------------------------

Previous works have shown that for the UHJs KELT-9b and MASCARA-2b/KELT-20b the NLTE synthetic transmission spectra provide a significantly better match to the observed \ion{H}{i} Balmer lines compared to the LTE models \citep{fossati2021_kelt9cloudy,fossati2023_kelt20mascara2}. In terms of system properties, WASP-178b shares several similarities with KELT-9b and MASCARA-2b/KELT-20b, and thus the results shown in Figure~\ref{fig:comparison_optical} are unexpected and call for a deeper investigation.

We first attempted to increase planetary mass or decrease stellar $T_{\rm eff}$ as both decrease the strength of the H$\alpha$ line \citep[see e.g.][]{fossati2023_kelt20mascara2}. The bottom panel of Figure~\ref{fig:comparison_optical} shows a comparison between the observed H$\alpha$ line and NLTE synthetic spectra computed for $M_{\rm p}$\,=\,2.0 and 2.5\,M$_{\rm J}$, and for $T_{\rm eff}$\,=\,8600\,K. As expected, a higher planetary mass decreases the strength of the H$\alpha$ line, but fitting the observation would require a mass higher than 2.5\,M$_{\rm J}$ that is inconsistent at more than 8$\sigma$ with the measured stellar and planetary reflex motions \citep{hellier2019,cont2024}. Similarly, considering a  500\,K cooler input stellar SED decreases the strength of the H$\alpha$ line, but fitting the observation would require reducing $T_{\rm eff}$ to values inconsistent with the observed spectrum.

Having discarded uncertainties on  the basic system parameters as origin of the discrepancy, we turn our attention to the planetary atmosphere as the weaker observed Balmer lines compared to the NLTE model suggest a cooler TP profile, particularly around the line formation region. Considering that by dropping the LTE assumption the NLTE model can be deemed being physically more accurate than the LTE one, the explanation for this mismatch can be sought by looking for planetary atmospheric properties or physical processes unaccounted for in the model that would lead to a decrease in the atmospheric temperature.

As can be inferred from Figures~\ref{fig:NLTEnoFeMgCa}, \ref{fig:heatingCooling}, and \ref{fig:fe1fe2mg1mg2}, the Mg and/or Fe abundance can significantly impact the shape of the TP profile. Therefore, a lower middle and upper atmospheric temperature can be obtained by decreasing the Fe abundance, which decreases the heating, and/or by increasing the Mg abundance, which increases the cooling. A low Fe abundance would be in contrast with the clear detection of both \ion{Fe}{i} and \ion{Fe}{ii} by \citet{damasceno2024_wasp178}. However, a high Mg abundance would be in line with the detection of \ion{Mg}{i} by \citep{damasceno2024_wasp178}. Nevertheless, our NLTE transmission spectrum computed assuming solar abundances fits the UVIS observations well. Therefore, varying the atmospheric abundance pattern might not significantly improve the fit of the NLTE synthetic spectrum to the available NUV and optical observations. In relation to the planetary abundance pattern, we remark that in the case of the WASP-178 system the stellar photospheric abundances cannot be taken as reference for inferring planetary properties. This is because in Am stars (and generally in all intermediate-mass chemically peculiar stars), the peculiar abundance pattern is a characteristic of the stellar atmospheric layers, while the bulk stellar abundance pattern is different and can hardly be accurately constrained for field stars \citep[see e.g.][]{fossati2008_praesepe,fossati2011_ngc5460}. 

The most notable difference between WASP-178b and the other two UHJs is the lower planetary mass of WASP-178b, and thus lower planetary gravity. As already mentioned in Section~\ref{sec:discussion:comparisonsTP_w178_m2_k9}, the lower gravity moves towards higher pressures the level at which hydrodynamic escape (not accounted for in our modelling) becomes important. This might significantly impact the TP profile and the strength of the absorption lines, especially of the \ion{H}{i} Balmer lines that form in the 1--10\,nbar range, that is the pressure level at which hydrodynamic effects are expected to become important (see Section~\ref{sec:discussion:Cs_lambda}). In particular, with the inclusion of hydrodynamic effects, the temperature in the upper atmosphere would decrease as a consequence of adiabatic cooling, leading to a decrease in the strength of the Balmer lines. However, the inclusion of hydrodynamics in the modelling might also lead to strengthen the lines that (partly) form in the upper atmosphere as a result of advection. Therefore, although the presence of hydrodynamic effects in the upper atmosphere might explain the discrepancy between the NLTE synthetic spectrum and the observed Balmer line profiles, dedicated hydrodynamic modelling properly accounting for metal line heating and cooling processes as well as advection \citep[e.g.][]{huang2023_wasp121b,kubyshkina2024_chain} is needed to confirm or reject this hypothesis. Furthermore, dedicated hydrodynamic modelling would clarify whether the significant line broadening measured by \citet{damasceno2024_wasp178} is indeed related to atmospheric escape or not.
\section{Conclusions}\label{sec:conclusions}
We performed forward NLTE atmospheric modelling of the UHJ WASP-178b \citep{hellier2019}. To identify the most adequate stellar SED to use as input to the model, we conducted a stellar spectroscopic analysis from which we derived an effective temperature of 9100$\pm$100\,K and confirmed that the stellar atmosphere is characterised by Am chemical peculiarities. Therefore, the star does not present a strong XUV emission (i.e. only photospheric emission) and the metallicity, and more in general the abundance pattern, cannot be taken as reference for the system, because Am peculiarities are the result of chemical diffusion processes localised to the stellar atmosphere.

We conducted planetary atmospheric modelling combining TP profiles computed with {\sc helios} (in LTE) for the lower atmosphere and {\sc Cloudy} (in both LTE or NLTE) for the middle and upper atmosphere, following the same approach previously used to model the atmospheres of the UHJs KELT-9b and MASCARA-2b/KELT-20b \citep{fossati2021_kelt9cloudy,fossati2023_kelt20mascara2}. 
We finally obtained an isothermal TP profile at pressures higher than 10\,mbar and lower than 10$^{-8}$\,bar, with a roughly linear rise from $\sim$2200\,K to $\sim$8100\,K in between. 
A detailed analysis of the atmospheric thermal balance shows that accounting for NLTE effects increases significantly the middle and upper atmospheric temperature. This is due to a combination of increased heating due to the overpopulation of long-lived Fe{\sc ii} levels presenting transitions mainly in the NUV band, where the stellar emission is strong, and of decreased cooling due to the underpopulation of Mg{\sc i} and Mg{\sc ii} levels that dominate radiative cooling. We also identified that hydrogen line cooling, which is the second most important coolant after Mg, acts as a thermostat, because it does not allow the atmospheric temperature to go above 9000\,K independently of the Mg level underpopulation and abundance. This shows that NLTE effects, and particularly the resulting increased Fe line heating and decreased Mg line cooling, cannot be ignored when modelling hot planetary atmospheres.

Despite the hot upper atmosphere and the rather low planetary mass, the lack of strong stellar XUV emission implies that the atmosphere remains mostly hydrostatic, up to pressures of $\sim$1\,nbar. We used the NLTE model to constrain the planetary mass-loss rate for which we obtained a value of 4.5$\times$10$^{10}$\,g\,s$^{-1}$, which is comparable to that of classical hot Jupiters such as HD209458b. We computed the planetary synthetic transmission spectrum covering from the far-UV to the near infrared obtaining that a number of strong UV lines extend to pressures lower than 1\,nbar and even beyond the Roche lobe. Therefore, more sophisticated models accounting also for hydrodynamics \citep[e.g.][]{huang2023_wasp121b,kubyshkina2024_chain} are required to more adequately constrain the mass-loss rate and model the strongest lines, particularly at high spectral resolution.

We compared the synthetic transmission spectra computed in LTE and NLTE with the available observations at optical (H$\alpha$ and H$\beta$, from \citealt{damasceno2024_wasp178}; high spectral resolution) and NUV (HST STIS and WFC3/UVIS from \citealt{lothringer2022_wasp178}; medium and low spectral resolution, respectively) wavelengths. We find that the NLTE transmission spectrum overestimates the observed H$\alpha$ and H$\beta$ absorption, while the LTE model is more in line with the data. This is surprising given that the NLTE models of the hydrogen Balmer lines computed for the UHJs KELT-9b and MASCARA-2b/KELT-20b match the observations remarkably well (and much better than the LTE models). We could not identify the origin of this discrepancy and suggest that, because of the hydrodynamic nature of the upper atmosphere where the H$\alpha$ and H$\beta$ lines form, particularly at high spectral resolution, future attempts to reproduce the observations should account for hydrodynamics in addition to NLTE effects. We find that the NLTE transmission spectrum is an excellent match to the low resolution UVIS observation, particularly in the NUV band, without the need to consider SiO absorption or super-solar metallicity, as required instead when assuming LTE \citep{lothringer2022_wasp178,lothringer2025_wasp178}. This is mainly the outcome of the hotter NLTE TP profile compared to the LTE one. We performed an independent analysis of the STIS observations aiming at minimising the impact of systematic noise, but obtained that red noise is too strong and the resulting transmission spectrum is not consistent with the UVIS one. Furthermore, the resulting transmission spectrum is significantly dependent on the applied data analysis method. Therefore, the STIS spectrum cannot be used to support the detection of SiO absorption presented by \citet{lothringer2022_wasp178}, and thus SiO may have condensed in the atmosphere of WASP-178b \citep{moran2024,lothringer2025_wasp178}, but Fe- and Mg-bearing silicates did not. Furthermore, our results indicate that SiO absorption might not be present at all in the NUV transmission spectrum of WASP-178b, which would indicate a significantly lower Si abundance compared to that derived by \citet{lothringer2025_wasp178} with consequences on inferences with respect to comparisons with other UHJs and possible formation scenarios. Therefore, accounting for NLTE effects in the computation of the TP profile and transmission spectrum might be critical for making best use of the available observations to advance our understanding of the hottest planets.
\begin{acknowledgements}
This work has made use of the VALD database, operated at Uppsala University, the Institute of Astronomy RAS in Moscow, and the University of Vienna. This work has made use of data from the European Space Agency (ESA) mission {\it Gaia} (\url{https://www.cosmos.esa.int/gaia}), processed by the {\it Gaia} Data Processing and Analysis Consortium (DPAC,
\url{https://www.cosmos.esa.int/web/gaia/dpac/consortium}). Funding for the DPAC has been provided by national institutions, in particular the institutions participating in the {\it Gaia} Multilateral Agreement. LF and SB gratefully acknowledge funding from the European Union's Horizon Europe research and innovation programme under grant agreement No. 101079231 (EXOHOST), and from UK Research and Innovation (UKRI) under the UK government’s Horizon Europe funding guarantee (grant number 10051045). AGS gratefully acknowledge funding from the Austrian Science Fund (FWF) grant J4596-N. D.S. acknowledges financial support from the project PID2021-126365NB-C21(MCI/AEI/FEDER, UE) and from the Severo Ochoa grant CEX2021-001131-S funded by MCIN/AEI/ 10.13039/501100011033. M.E.Y. acknowledges support from the UK Science and Technology Facilities Council (STFC, grant number ST/X00094X/1). We thank the anonymous referee for the useful comments that led to improve the paper.
\end{acknowledgements}
\bibliographystyle{aa}
\bibliography{main}      

\begin{thebibliography}{121}
\expandafter\ifx\csname natexlab\endcsname\relax\def\natexlab#1{#1}\fi

\bibitem[{{Abt} \& {Morrell}(1995)}]{abt1995_rotation}
{Abt}, H.~A. \& {Morrell}, N.~I. 1995, \apjs, 99, 135

\bibitem[{{Ahlers} {et~al.}(2020){Ahlers}, {Johnson}, {Stassun}, {Col{\'o}n},
  {Barnes}, {Stevens}, {Beatty}, {Gaudi}, {Collins}, {Rodriguez}, {Ricker},
  {Vanderspek}, {Latham}, {Seager}, {Winn}, {Jenkins}, {Caldwell}, {Goeke},
  {Osborn}, {Paegert}, {Rowden}, \& {Tenenbaum}}]{ahlers2020_kelt9b}
{Ahlers}, J.~P., {Johnson}, M.~C., {Stassun}, K.~G., {et~al.} 2020, \aj, 160, 4

\bibitem[{{Arcangeli} {et~al.}(2018){Arcangeli}, {D{\'e}sert}, {Line}, {Bean},
  {Parmentier}, {Stevenson}, {Kreidberg}, {Fortney}, {Mansfield}, \&
  {Showman}}]{arcangeli2018}
{Arcangeli}, J., {D{\'e}sert}, J.-M., {Line}, M.~R., {et~al.} 2018, \apjl, 855,
  L30

\bibitem[{{Asplund} {et~al.}(2009){Asplund}, {Grevesse}, {Sauval}, \&
  {Scott}}]{asplund2009}
{Asplund}, M., {Grevesse}, N., {Sauval}, A.~J., \& {Scott}, P. 2009, \araa, 47,
  481

\bibitem[{{Benz} {et~al.}(2021){Benz}, {Broeg}, {Fortier}, {Rando}, {Beck},
  {Beck}, {Queloz}, {Ehrenreich}, {Maxted}, {Isaak}, {Billot}, {Alibert},
  {Alonso}, {Ant{\'o}nio}, {Asquier}, {Bandy}, {B{\'a}rczy}, {Barrado},
  {Barros}, {Baumjohann}, {Bekkelien}, {Bergomi}, {Biondi}, {Bonfils},
  {Borsato}, {Brandeker}, {Busch}, {Cabrera}, {Cessa}, {Charnoz}, {Chazelas},
  {Collier Cameron}, {Corral Van Damme}, {Cortes}, {Davies}, {Deleuil},
  {Deline}, {Delrez}, {Demangeon}, {Demory}, {Erikson}, {Farinato}, {Fossati},
  {Fridlund}, {Futyan}, {Gandolfi}, {Garcia Munoz}, {Gillon}, {Guterman},
  {Gutierrez}, {Hasiba}, {Heng}, {Hernandez}, {Hoyer}, {Kiss}, {Kovacs},
  {Kuntzer}, {Laskar}, {Lecavelier des Etangs}, {Lendl}, {L{\'o}pez}, {Lora},
  {Lovis}, {L{\"u}ftinger}, {Magrin}, {Malvasio}, {Marafatto}, {Michaelis}, {de
  Miguel}, {Modrego}, {Munari}, {Nascimbeni}, {Olofsson}, {Ottacher},
  {Ottensamer}, {Pagano}, {Palacios}, {Pall{\'e}}, {Peter}, {Piazza}, {Piotto},
  {Pizarro}, {Pollaco}, {Ragazzoni}, {Ratti}, {Rauer}, {Ribas}, {Rieder},
  {Rohlfs}, {Safa}, {Salatti}, {Santos}, {Scandariato}, {S{\'e}gransan},
  {Simon}, {Smith}, {Sordet}, {Sousa}, {Steller}, {Szab{\'o}}, {Szoke},
  {Thomas}, {Tschentscher}, {Udry}, {Van Grootel}, {Viotto}, {Walter},
  {Walton}, {Wildi}, \& {Wolter}}]{benz2021_cheops}
{Benz}, W., {Broeg}, C., {Fortier}, A., {et~al.} 2021, Experimental Astronomy,
  51, 109

\bibitem[{{Bonfanti} \& {Gillon}(2020)}]{2020AndreaMCMCI}
{Bonfanti}, A. \& {Gillon}, M. 2020, \aap, 635, A6

\bibitem[{{Bonfanti} {et~al.}(2016){Bonfanti}, {Ortolani}, \&
  {Nascimbeni}}]{bonfanti16}
{Bonfanti}, A., {Ortolani}, S., \& {Nascimbeni}, V. 2016, \aap, 585, A5

\bibitem[{{Bonfanti} {et~al.}(2015){Bonfanti}, {Ortolani}, {Piotto}, \&
  {Nascimbeni}}]{bonfanti15}
{Bonfanti}, A., {Ortolani}, S., {Piotto}, G., \& {Nascimbeni}, V. 2015, \aap,
  575, A18

\bibitem[{{Borsa} {et~al.}(2021){Borsa}, {Fossati}, {Koskinen}, {Young}, \&
  {Shulyak}}]{borsa2022_kelt9b_OI}
{Borsa}, F., {Fossati}, L., {Koskinen}, T., {Young}, M.~E., \& {Shulyak}, D.
  2021, Nature Astronomy, 6, 226

\bibitem[{{Borsa} {et~al.}(2022){Borsa}, {Giacobbe}, {Bonomo}, {Brogi}, {Pino},
  {Fossati}, {Lanza}, {Nascimbeni}, {Sozzetti}, {Amadori}, {Benatti}, {Biazzo},
  {Bignamini}, {Boschin}, {Claudi}, {Cosentino}, {Covino}, {Desidera},
  {Fiorenzano}, {Guilluy}, {Harutyunyan}, {Maggio}, {Maldonado}, {Mancini},
  {Micela}, {Molinari}, {Molinaro}, {Pagano}, {Pedani}, {Piotto}, {Poretti},
  {Rainer}, {Scandariato}, \& {Stoev}}]{borsa2022_mascara2b}
{Borsa}, F., {Giacobbe}, P., {Bonomo}, A.~S., {et~al.} 2022, \aap, 663, A141

\bibitem[{{Borsa} {et~al.}(2019){Borsa}, {Rainer}, {Bonomo}, {Barbato},
  {Fossati}, {Malavolta}, {Nascimbeni}, {Lanza}, {Esposito}, {Affer},
  {Andreuzzi}, {Benatti}, {Biazzo}, {Bignamini}, {Brogi}, {Carleo}, {Claudi},
  {Cosentino}, {Covino}, {Damasso}, {Desidera}, {Garrido Rubio}, {Giacobbe},
  {Gonz{\'a}lez-{\'A}lvarez}, {Harutyunyan}, {Knapic}, {Leto}, {Ligi},
  {Maggio}, {Maldonado}, {Mancini}, {Fiorenzano}, {Masiero}, {Micela},
  {Molinari}, {Pagano}, {Pedani}, {Piotto}, {Pino}, {Poretti}, {Scandariato},
  {Smareglia}, \& {Sozzetti}}]{borsa2019}
{Borsa}, F., {Rainer}, M., {Bonomo}, A.~S., {et~al.} 2019, \aap, 631, A34

\bibitem[{{Borsato} {et~al.}(2023){Borsato}, {Hoeijmakers}, {Prinoth},
  {Thorsbro}, {Forsberg}, {Kitzmann}, {Jones}, \& {Heng}}]{borsato2023}
{Borsato}, N.~W., {Hoeijmakers}, H.~J., {Prinoth}, B., {et~al.} 2023, \aap,
  673, A158

\bibitem[{{Canuto} \& {Mazzitelli}(1991)}]{canuto1991}
{Canuto}, V.~M. \& {Mazzitelli}, I. 1991, \apj, 370, 295

\bibitem[{{Canuto} \& {Mazzitelli}(1992)}]{canuto1992}
{Canuto}, V.~M. \& {Mazzitelli}, I. 1992, \apj, 389, 724

\bibitem[{{Charbonneau} \&
  {Michaud}(1991)}]{charbonneauMichaud1991_diffusionRotation}
{Charbonneau}, P. \& {Michaud}, G. 1991, \apj, 370, 693

\bibitem[{{Cont} {et~al.}(2024){Cont}, {Nortmann}, {Yan}, {Lesjak}, {Czesla},
  {Lavail}, {Reiners}, {Piskunov}, {Hatzes}, {Boldt-Christmas}, {Kochukhov},
  {Marquart}, {Nagel}, {Rains}, {Rengel}, {Seemann}, \& {Shulyak}}]{cont2024}
{Cont}, D., {Nortmann}, L., {Yan}, F., {et~al.} 2024, arXiv e-prints,
  arXiv:2406.08166

\bibitem[{{Cubillos} {et~al.}(2023){Cubillos}, {Fossati}, {Koskinen}, {Huang},
  {Sreejith}, {France}, {Wilson Cauley}, \&
  {Haswell}}]{cubillos2023_nuv_stis_hd189733b}
{Cubillos}, P.~E., {Fossati}, L., {Koskinen}, T., {et~al.} 2023, \aap, 671,
  A170

\bibitem[{{Cubillos} {et~al.}(2020){Cubillos}, {Fossati}, {Koskinen}, {Young},
  {Salz}, {France}, {Sreejith}, \& {Haswell}}]{cubillos2020_nuv_stis_hd209458b}
{Cubillos}, P.~E., {Fossati}, L., {Koskinen}, T., {et~al.} 2020, \aj, 159, 111

\bibitem[{{Damasceno} {et~al.}(2024){Damasceno}, {Seidel}, {Prinoth},
  {Psaridi}, {Esparza-Borges}, {Stangret}, {Santos}, {Zapatero-Osorio},
  {Alibert}, {Allart}, {Azevedo Silva}, {Cointepas}, {Costa Silva}, {Cristo},
  {Di Marcantonio}, {Ehrenreich}, {Gonz{\'a}lez Hern{\'a}ndez},
  {Herrero-Cisneros}, {Lendl}, {Lillo-Box}, {Martins}, {Micela}, {Pall{\'e}},
  {Sousa}, {Steiner}, {Vaulato}, {Zhao}, \& {Pepe}}]{damasceno2024_wasp178}
{Damasceno}, Y.~C., {Seidel}, J.~V., {Prinoth}, B., {et~al.} 2024, arXiv
  e-prints, arXiv:2406.08348

\bibitem[{{D'Arpa} {et~al.}(2024){D'Arpa}, {Saba}, {Borsa}, {Fossati},
  {Micela}, {Di Maio}, {Stangret}, {Tripodo}, {Affer}, {Bonomo}, {Benatti},
  {Brogi}, {Fardella}, {Lanza}, {Guilluy}, {Maldonado}, {Mantovan},
  {Nascimbeni}, {Pino}, {Scandariato}, {Sicilia}, {Sozzetti}, {Spinelli},
  {Andreuzzi}, {Bignamini}, {Claudi}, {Desidera}, {Ghedina}, {Knapic}, \&
  {Lorenzi}}]{darpa2024_kelt9b}
{D'Arpa}, M.~C., {Saba}, A., {Borsa}, F., {et~al.} 2024, \aap, 690, A237

\bibitem[{{Espinosa Lara} \& {Rieutord}(2011)}]{espinosa2011_gravityDarkening}
{Espinosa Lara}, F. \& {Rieutord}, M. 2011, \aap, 533, A43

\bibitem[{{Espinoza} \&
  {Jord{\'a}n}(2015)}]{EspinozaJordan2015mnrasLimbDarkeningI}
{Espinoza}, N. \& {Jord{\'a}n}, A. 2015, \mnras, 450, 1879

\bibitem[{{Ferland} {et~al.}(2017){Ferland}, {Chatzikos}, {Guzm{\'a}n},
  {Lykins}, {van Hoof}, {Williams}, {Abel}, {Badnell}, {Keenan}, {Porter}, \&
  {Stancil}}]{ferland2017}
{Ferland}, G.~J., {Chatzikos}, M., {Guzm{\'a}n}, F., {et~al.} 2017, \rmxaa, 53,
  385

\bibitem[{{Ferland} {et~al.}(1998){Ferland}, {Korista}, {Verner}, {Ferguson},
  {Kingdon}, \& {Verner}}]{ferland1998}
{Ferland}, G.~J., {Korista}, K.~T., {Verner}, D.~A., {et~al.} 1998, \pasp, 110,
  761

\bibitem[{{Ferland} {et~al.}(2013){Ferland}, {Porter}, {van Hoof}, {Williams},
  {Abel}, {Lykins}, {Shaw}, {Henney}, \& {Stancil}}]{ferland2013}
{Ferland}, G.~J., {Porter}, R.~L., {van Hoof}, P.~A.~M., {et~al.} 2013, \rmxaa,
  49, 137

\bibitem[{{Feroz} \& {Hobson}(2008)}]{feroz2008_nested_sampling}
{Feroz}, F. \& {Hobson}, M.~P. 2008, \mnras, 384, 449

\bibitem[{{Feroz} {et~al.}(2019){Feroz}, {Hobson}, {Cameron}, \&
  {Pettitt}}]{feroz2019_nested_sampling}
{Feroz}, F., {Hobson}, M.~P., {Cameron}, E., \& {Pettitt}, A.~N. 2019, The Open
  Journal of Astrophysics, 2, 10

\bibitem[{{Folsom} {et~al.}(2012){Folsom}, {Bagnulo}, {Wade}, {Alecian},
  {Landstreet}, {Marsden}, \& {Waite}}]{2012Folsom}
{Folsom}, C.~P., {Bagnulo}, S., {Wade}, G.~A., {et~al.} 2012, \mnras, 422, 2072

\bibitem[{{Fossati} {et~al.}(2010){Fossati}, {Bagnulo}, {Elmasli}, {Haswell},
  {Holmes}, {Kochukhov}, {Shkolnik}, {Shulyak}, {Bohlender}, {Albayrak},
  {Froning}, \& {Hebb}}]{fossati2010_w12Specpol}
{Fossati}, L., {Bagnulo}, S., {Elmasli}, A., {et~al.} 2010, \apj, 720, 872

\bibitem[{{Fossati} {et~al.}(2008){Fossati}, {Bagnulo}, {Landstreet}, {Wade},
  {Kochukhov}, {Monier}, {Weiss}, \& {Gebran}}]{fossati2008_praesepe}
{Fossati}, L., {Bagnulo}, S., {Landstreet}, J., {et~al.} 2008, \aap, 483, 891

\bibitem[{{Fossati} {et~al.}(2007){Fossati}, {Bagnulo}, {Monier}, {Khan},
  {Kochukhov}, {Landstreet}, {Wade}, \& {Weiss}}]{fossati2007_praesepe}
{Fossati}, L., {Bagnulo}, S., {Monier}, R., {et~al.} 2007, \aap, 476, 911

\bibitem[{{Fossati} {et~al.}(2023){Fossati}, {Biassoni}, {Cappello}, {Borsa},
  {Shulyak}, {Bonomo}, {Gandolfi}, {Haardt}, {Koskinen}, {Lanza}, {Nascimbeni},
  {Sicilia}, {Young}, {Aresu}, {Bignamini}, {Brogi}, {Carleo}, {Claudi},
  {Cosentino}, {Guilluy}, {Knapic}, {Malavolta}, {Mancini}, {Nardiello},
  {Pinamonti}, {Pino}, {Poretti}, {Rainer}, {Rigamonti}, \&
  {Sozzetti}}]{fossati2023_kelt20mascara2}
{Fossati}, L., {Biassoni}, F., {Cappello}, G.~M., {et~al.} 2023, \aap, 676, A99

\bibitem[{{Fossati} {et~al.}(2017){Fossati}, {Erkaev}, {Lammer}, {Cubillos},
  {Odert}, {Juvan}, {Kislyakova}, {Lendl}, {Kubyshkina}, \&
  {Bauer}}]{fossati2017}
{Fossati}, L., {Erkaev}, N.~V., {Lammer}, H., {et~al.} 2017, \aap, 598, A90

\bibitem[{{Fossati} {et~al.}(2011{\natexlab{a}}){Fossati}, {Folsom}, {Bagnulo},
  {Grunhut}, {Kochukhov}, {Landstreet}, {Paladini}, \&
  {Wade}}]{fossati2011_ngc5460}
{Fossati}, L., {Folsom}, C.~P., {Bagnulo}, S., {et~al.} 2011{\natexlab{a}},
  \mnras, 413, 1132

\bibitem[{{Fossati} {et~al.}(2018){Fossati}, {Koskinen}, {Lothringer},
  {France}, {Young}, \& {Sreejith}}]{fossati2018_AstarPlanets}
{Fossati}, L., {Koskinen}, T., {Lothringer}, J.~D., {et~al.} 2018, \apjl, 868,
  L30

\bibitem[{{Fossati} {et~al.}(2009){Fossati}, {Ryabchikova}, {Bagnulo},
  {Alecian}, {Grunhut}, {Kochukhov}, \& {Wade}}]{fossati2009_reference}
{Fossati}, L., {Ryabchikova}, T., {Bagnulo}, S., {et~al.} 2009, \aap, 503, 945

\bibitem[{{Fossati} {et~al.}(2011{\natexlab{b}}){Fossati}, {Ryabchikova},
  {Shulyak}, {Haswell}, {Elmasli}, {Pandey}, {Barnes}, \&
  {Zwintz}}]{fossati2011_hd32115}
{Fossati}, L., {Ryabchikova}, T., {Shulyak}, D.~V., {et~al.}
  2011{\natexlab{b}}, \mnras, 417, 495

\bibitem[{{Fossati} {et~al.}(2020){Fossati}, {Shulyak}, {Sreejith}, {Koskinen},
  {Young}, {Cubillos}, {Lara}, {France}, {Rengel}, {Cauley}, {Turner},
  {Wyttenbach}, \& {Yan}}]{fossati2020_KELT9datadriven}
{Fossati}, L., {Shulyak}, D., {Sreejith}, A.~G., {et~al.} 2020, \aap, 643, A131

\bibitem[{{Fossati} {et~al.}(2021){Fossati}, {Young}, {Shulyak}, {Koskinen},
  {Huang}, {Cubillos}, {France}, \& {Sreejith}}]{fossati2021_kelt9cloudy}
{Fossati}, L., {Young}, M.~E., {Shulyak}, D., {et~al.} 2021, \aap, 653, A52

\bibitem[{{Fouesneau} {et~al.}(2022){Fouesneau}, {Andrae}, {Dharmawardena},
  {Rybizki}, {Bailer-Jones}, \& {Demleitner}}]{fouesneau2022}
{Fouesneau}, M., {Andrae}, R., {Dharmawardena}, T., {et~al.} 2022, \aap, 662,
  A125

\bibitem[{{Gaia Collaboration} {et~al.}(2016){Gaia Collaboration}, {Prusti},
  {de Bruijne}, {Brown}, {Vallenari}, {Babusiaux}, {Bailer-Jones}, {Bastian},
  {Biermann}, {Evans}, {Eyer}, {Jansen}, {Jordi}, {Klioner}, {Lammers},
  {Lindegren}, {Luri}, {Mignard}, {Milligan}, {Panem}, {Poinsignon},
  {Pourbaix}, {Randich}, {Sarri}, {Sartoretti}, {Siddiqui}, {Soubiran},
  {Valette}, {van Leeuwen}, {Walton}, {Aerts}, {Arenou}, {Cropper}, {Drimmel},
  {H{\o}g}, {Katz}, {Lattanzi}, {O'Mullane}, {Grebel}, {Holland}, {Huc},
  {Passot}, {Bramante}, {Cacciari}, {Casta{\~n}eda}, {Chaoul}, {Cheek}, {De
  Angeli}, {Fabricius}, {Guerra}, {Hern{\'a}ndez}, {Jean-Antoine-Piccolo},
  {Masana}, {Messineo}, {Mowlavi}, {Nienartowicz}, {Ord{\'o}{\~n}ez-Blanco},
  {Panuzzo}, {Portell}, {Richards}, {Riello}, {Seabroke}, {Tanga},
  {Th{\'e}venin}, {Torra}, {Els}, {Gracia-Abril}, {Comoretto},
  {Garcia-Reinaldos}, {Lock}, {Mercier}, {Altmann}, {Andrae}, {Astraatmadja},
  {Bellas-Velidis}, {Benson}, {Berthier}, {Blomme}, {Busso}, {Carry},
  {Cellino}, {Clementini}, {Cowell}, {Creevey}, {Cuypers}, {Davidson}, {De
  Ridder}, {de Torres}, {Delchambre}, {Dell'Oro}, {Ducourant}, {Fr{\'e}mat},
  {Garc{\'\i}a-Torres}, {Gosset}, {Halbwachs}, {Hambly}, {Harrison}, {Hauser},
  {Hestroffer}, {Hodgkin}, {Huckle}, {Hutton}, {Jasniewicz}, {Jordan},
  {Kontizas}, {Korn}, {Lanzafame}, {Manteiga}, {Moitinho}, {Muinonen},
  {Osinde}, {Pancino}, {Pauwels}, {Petit}, {Recio-Blanco}, {Robin}, {Sarro},
  {Siopis}, {Smith}, {Smith}, {Sozzetti}, {Thuillot}, {van Reeven}, {Viala},
  {Abbas}, {Abreu Aramburu}, {Accart}, {Aguado}, {Allan}, {Allasia},
  {Altavilla}, {{\'A}lvarez}, {Alves}, {Anderson}, {Andrei}, {Anglada Varela},
  {Antiche}, {Antoja}, {Ant{\'o}n}, {Arcay}, {Atzei}, {Ayache}, {Bach},
  {Baker}, {Balaguer-N{\'u}{\~n}ez}, {Barache}, {Barata}, {Barbier}, {Barblan},
  {Baroni}, {Barrado y Navascu{\'e}s}, {Barros}, {Barstow}, {Becciani},
  {Bellazzini}, {Bellei}, {Bello Garc{\'\i}a}, {Belokurov}, {Bendjoya},
  {Berihuete}, {Bianchi}, {Bienaym{\'e}}, {Billebaud}, {Blagorodnova},
  {Blanco-Cuaresma}, {Boch}, {Bombrun}, {Borrachero}, {Bouquillon}, {Bourda},
  {Bouy}, {Bragaglia}, {Breddels}, {Brouillet}, {Br{\"u}semeister},
  {Bucciarelli}, {Budnik}, {Burgess}, {Burgon}, {Burlacu}, {Busonero}, {Buzzi},
  {Caffau}, {Cambras}, {Campbell}, {Cancelliere}, {Cantat-Gaudin}, {Carlucci},
  {Carrasco}, {Castellani}, {Charlot}, {Charnas}, {Charvet}, {Chassat},
  {Chiavassa}, {Clotet}, {Cocozza}, {Collins}, {Collins}, \&
  {Costigan}}]{gaia2016_DR3}
{Gaia Collaboration}, {Prusti}, T., {de Bruijne}, J.~H.~J., {et~al.} 2016,
  \aap, 595, A1

\bibitem[{{Gaia Collaboration} {et~al.}(2023){Gaia Collaboration}, {Vallenari},
  {Brown}, {Prusti}, {de Bruijne}, {Arenou}, {Babusiaux}, {Biermann},
  {Creevey}, {Ducourant}, {Evans}, {Eyer}, {Guerra}, {Hutton}, {Jordi},
  {Klioner}, {Lammers}, {Lindegren}, {Luri}, {Mignard}, {Panem}, {Pourbaix},
  {Randich}, {Sartoretti}, {Soubiran}, {Tanga}, {Walton}, {Bailer-Jones},
  {Bastian}, {Drimmel}, {Jansen}, {Katz}, {Lattanzi}, {van Leeuwen}, {Bakker},
  {Cacciari}, {Casta{\~n}eda}, {De Angeli}, {Fabricius}, {Fouesneau},
  {Fr{\'e}mat}, {Galluccio}, {Guerrier}, {Heiter}, {Masana}, {Messineo},
  {Mowlavi}, {Nicolas}, {Nienartowicz}, {Pailler}, {Panuzzo}, {Riclet}, {Roux},
  {Seabroke}, {Sordo}, {Th{\'e}venin}, {Gracia-Abril}, {Portell}, {Teyssier},
  {Altmann}, {Andrae}, {Audard}, {Bellas-Velidis}, {Benson}, {Berthier},
  {Blomme}, {Burgess}, {Busonero}, {Busso}, {C{\'a}novas}, {Carry}, {Cellino},
  {Cheek}, {Clementini}, {Damerdji}, {Davidson}, {de Teodoro}, {Nu{\~n}ez
  Campos}, {Delchambre}, {Dell'Oro}, {Esquej}, {Fern{\'a}ndez-Hern{\'a}ndez},
  {Fraile}, {Garabato}, {Garc{\'\i}a-Lario}, {Gosset}, {Haigron}, {Halbwachs},
  {Hambly}, {Harrison}, {Hern{\'a}ndez}, {Hestroffer}, {Hodgkin}, {Holl},
  {Jan{\ss}en}, {Jevardat de Fombelle}, {Jordan}, {Krone-Martins}, {Lanzafame},
  {L{\"o}ffler}, {Marchal}, {Marrese}, {Moitinho}, {Muinonen}, {Osborne},
  {Pancino}, {Pauwels}, {Recio-Blanco}, {Reyl{\'e}}, {Riello}, {Rimoldini},
  {Roegiers}, {Rybizki}, {Sarro}, {Siopis}, {Smith}, {Sozzetti}, {Utrilla},
  {van Leeuwen}, {Abbas}, {{\'A}brah{\'a}m}, {Abreu Aramburu}, {Aerts},
  {Aguado}, {Ajaj}, {Aldea-Montero}, {Altavilla}, {{\'A}lvarez}, {Alves},
  {Anders}, {Anderson}, {Anglada Varela}, {Antoja}, {Baines}, {Baker},
  {Balaguer-N{\'u}{\~n}ez}, {Balbinot}, {Balog}, {Barache}, {Barbato},
  {Barros}, {Barstow}, {Bartolom{\'e}}, {Bassilana}, {Bauchet}, {Becciani},
  {Bellazzini}, {Berihuete}, {Bernet}, {Bertone}, {Bianchi}, {Binnenfeld},
  {Blanco-Cuaresma}, {Blazere}, {Boch}, {Bombrun}, {Bossini}, {Bouquillon},
  {Bragaglia}, {Bramante}, {Breedt}, {Bressan}, {Brouillet}, {Brugaletta},
  {Bucciarelli}, {Burlacu}, {Butkevich}, {Buzzi}, {Caffau}, {Cancelliere},
  {Cantat-Gaudin}, {Carballo}, {Carlucci}, {Carnerero}, {Carrasco},
  {Casamiquela}, {Castellani}, {Castro-Ginard}, {Chaoul}, {Charlot}, {Chemin},
  {Chiaramida}, {Chiavassa}, {Chornay}, {Comoretto}, {Contursi}, {Cooper},
  {Cornez}, {Cowell}, {Crifo}, {Cropper}, {Crosta}, {Crowley}, {Dafonte},
  {Dapergolas}, {David}, {David}, {de Laverny}, {De Luise}, \& {De
  March}}]{gaia2023_DR3}
{Gaia Collaboration}, {Vallenari}, A., {Brown}, A.~G.~A., {et~al.} 2023, \aap,
  674, A1

\bibitem[{{Gandhi} \& {Madhusudhan}(2019)}]{gandhi2019_Tinversion}
{Gandhi}, S. \& {Madhusudhan}, N. 2019, \mnras, 485, 5817

\bibitem[{{Garc{\'\i}a Mu{\~n}oz} \& {Schneider}(2019)}]{garcia2019_UHJs}
{Garc{\'\i}a Mu{\~n}oz}, A. \& {Schneider}, P.~C. 2019, \apjl, 884, L43

\bibitem[{{Gaudi} {et~al.}(2017){Gaudi}, {Stassun}, {Collins}, {Beatty},
  {Zhou}, {Latham}, {Bieryla}, {Eastman}, {Siverd}, {Crepp}, {Gonzales},
  {Stevens}, {Buchhave}, {Pepper}, {Johnson}, {Colon}, {Jensen}, {Rodriguez},
  {Bozza}, {Novati}, {D'Ago}, {Dumont}, {Ellis}, {Gaillard}, {Jang-Condell},
  {Kasper}, {Fukui}, {Gregorio}, {Ito}, {Kielkopf}, {Manner}, {Matt}, {Narita},
  {Oberst}, {Reed}, {Scarpetta}, {Stephens}, {Yeigh}, {Zambelli}, {Fulton},
  {Howard}, {James}, {Penny}, {Bayliss}, {Curtis}, {Depoy}, {Esquerdo},
  {Gould}, {Joner}, {Kuhn}, {Labadie-Bartz}, {Lund}, {Marshall}, {McLeod},
  {Pogge}, {Relles}, {Stockdale}, {Tan}, {Trueblood}, \&
  {Trueblood}}]{gaudi2017}
{Gaudi}, B.~S., {Stassun}, K.~G., {Collins}, K.~A., {et~al.} 2017, \nat, 546,
  514

\bibitem[{{Gordon} {et~al.}(2022){Gordon}, {Rothman}, {Hargreaves}, {Hashemi},
  {Karlovets}, {Skinner}, {Conway}, {Hill}, {Kochanov}, {Tan}, {Wcis{\l}o},
  {Finenko}, {Nelson}, {Bernath}, {Birk}, {Boudon}, {Campargue}, {Chance},
  {Coustenis}, {Drouin}, {Flaud}, {Gamache}, {Hodges}, {Jacquemart}, {Mlawer},
  {Nikitin}, {Perevalov}, {Rotger}, {Tennyson}, {Toon}, {Tran}, {Tyuterev},
  {Adkins}, {Baker}, {Barbe}, {Can{\`e}}, {Cs{\'a}sz{\'a}r}, {Dudaryonok},
  {Egorov}, {Fleisher}, {Fleurbaey}, {Foltynowicz}, {Furtenbacher}, {Harrison},
  {Hartmann}, {Horneman}, {Huang}, {Karman}, {Karns}, {Kassi}, {Kleiner},
  {Kofman}, {Kwabia-Tchana}, {Lavrentieva}, {Lee}, {Long}, {Lukashevskaya},
  {Lyulin}, {Makhnev}, {Matt}, {Massie}, {Melosso}, {Mikhailenko}, {Mondelain},
  {M{\"u}ller}, {Naumenko}, {Perrin}, {Polyansky}, {Raddaoui}, {Raston},
  {Reed}, {Rey}, {Richard}, {T{\'o}bi{\'a}s}, {Sadiek}, {Schwenke},
  {Starikova}, {Sung}, {Tamassia}, {Tashkun}, {Vander Auwera}, {Vasilenko},
  {Vigasin}, {Villanueva}, {Vispoel}, {Wagner}, {Yachmenev}, \&
  {Yurchenko}}]{gordon2022_hitran}
{Gordon}, I.~E., {Rothman}, L.~S., {Hargreaves}, R.~J., {et~al.} 2022, \jqsrt,
  277, 107949

\bibitem[{{Grimm} \& {Heng}(2015)}]{grimm2015_helios}
{Grimm}, S.~L. \& {Heng}, K. 2015, \apj, 808, 182

\bibitem[{{G{\"u}nther} {et~al.}(2022){G{\"u}nther}, {Melis}, {Robrade},
  {Schneider}, {Wolk}, \& {Yadav}}]{gunther_activityAstars2022}
{G{\"u}nther}, H.~M., {Melis}, C., {Robrade}, J., {et~al.} 2022, \aj, 164, 8

\bibitem[{{G{\"u}nther} \& {Daylan}(2021)}]{guenther2021_allesfitter}
{G{\"u}nther}, M.~N. \& {Daylan}, T. 2021, \apjs, 254, 13

\bibitem[{{Hargreaves} {et~al.}(2019){Hargreaves}, {Gordon}, {Rothman},
  {Tashkun}, {Perevalov}, {Lukashevskaya}, {Yurchenko}, {Tennyson}, \&
  {M{\"u}ller}}]{hargreaves2019_hitemp}
{Hargreaves}, R.~J., {Gordon}, I.~E., {Rothman}, L.~S., {et~al.} 2019, \jqsrt,
  232, 35

\bibitem[{{Hellier} {et~al.}(2019){Hellier}, {Anderson}, {Barkaoui},
  {Benkhaldoun}, {Bouchy}, {Burdanov}, {Collier Cameron}, {Delrez}, {Gillon},
  {Jehin}, {Nielsen}, {Maxted}, {Pepe}, {Pollacco}, {Pozuelos}, {Queloz},
  {S{\'e}gransan}, {Smalley}, {Triaud}, {Turner}, {Udry}, \&
  {West}}]{hellier2019}
{Hellier}, C., {Anderson}, D.~R., {Barkaoui}, K., {et~al.} 2019, \mnras, 490,
  1479

\bibitem[{{Hoeijmakers} {et~al.}(2018){Hoeijmakers}, {Ehrenreich}, {Heng},
  {Kitzmann}, {Grimm}, {Allart}, {Deitrick}, {Wyttenbach}, {Oreshenko}, {Pino},
  {Rimmer}, {Molinari}, \& {Di Fabrizio}}]{hoeijmakers2018}
{Hoeijmakers}, H.~J., {Ehrenreich}, D., {Heng}, K., {et~al.} 2018, \nat, 560,
  453

\bibitem[{{Hooton} {et~al.}(2018){Hooton}, {Watson}, {de Mooij}, {Gibson}, \&
  {Kitzmann}}]{hooton2018_kelt9b_albedo}
{Hooton}, M.~J., {Watson}, C.~A., {de Mooij}, E. J.~W., {Gibson}, N.~P., \&
  {Kitzmann}, D. 2018, \apjl, 869, L25

\bibitem[{{Huang} {et~al.}(2023){Huang}, {Koskinen}, {Lavvas}, \&
  {Fossati}}]{huang2023_wasp121b}
{Huang}, C., {Koskinen}, T., {Lavvas}, P., \& {Fossati}, L. 2023, \apj, 951,
  123

\bibitem[{{Johnson} {et~al.}(2023){Johnson}, {Wang}, {Asnodkar}, {Bonomo},
  {Gaudi}, {Henning}, {Ilyin}, {Keles}, {Malavolta}, {Mallonn},
  {Molaverdikhani}, {Nascimbeni}, {Patience}, {Poppenhaeger}, {Scandariato},
  {Schlawin}, {Shkolnik}, {Sicilia}, {Sozzetti}, {Strassmeier}, {Veillet}, \&
  {Yan}}]{johnson2023_kelt20b_inversion}
{Johnson}, M.~C., {Wang}, J., {Asnodkar}, A.~P., {et~al.} 2023, \aj, 165, 157

\bibitem[{{Jones} {et~al.}(2022){Jones}, {Morris}, {Demory}, {Heng}, {Hooton},
  {Billot}, {Ehrenreich}, {Hoyer}, {Simon}, {Lendl}, {Demangeon}, {Sousa},
  {Bonfanti}, {Wilson}, {Salmon}, {Csizmadia}, {Parviainen}, {Bruno},
  {Alibert}, {Alonso}, {Anglada}, {B{\'a}rczy}, {Barrado}, {Barros},
  {Baumjohann}, {Beck}, {Beck}, {Benz}, {Bonfils}, {Brandeker}, {Broeg},
  {Cabrera}, {Charnoz}, {Collier Cameron}, {Davies}, {Deleuil}, {Deline},
  {Delrez}, {Erikson}, {Fortier}, {Fossati}, {Fridlund}, {Gandolfi}, {Gillon},
  {G{\"u}del}, {Isaak}, {Kiss}, {Laskar}, {Lecavelier des Etangs}, {Lovis},
  {Magrin}, {Maxted}, {Nascimbeni}, {Olofsson}, {Ottensamer}, {Pagano},
  {Pall{\'e}}, {Peter}, {Piotto}, {Pollacco}, {Queloz}, {Ragazzoni}, {Rando},
  {Ratti}, {Rauer}, {Reimers}, {Ribas}, {Santos}, {Scandariato},
  {S{\'e}gransan}, {Smith}, {Steller}, {Szab{\'o}}, {Thomas}, {Udry}, {Van
  Grootel}, {Walter}, {Walton}, \& {Wang Jungo}}]{jones2022_kelt9b_cheops}
{Jones}, K., {Morris}, B.~M., {Demory}, B.~O., {et~al.} 2022, \aap, 666, A118

\bibitem[{{Kama} {et~al.}(2023){Kama}, {Folsom}, {Jermyn}, \&
  {Teske}}]{kama2023_kelt9}
{Kama}, M., {Folsom}, C.~P., {Jermyn}, A.~S., \& {Teske}, J.~K. 2023, \mnras,
  518, 3116

\bibitem[{{Kesseli} {et~al.}(2020){Kesseli}, {Snellen}, {Alonso-Floriano},
  {Molli{\`e}re}, \& {Serindag}}]{kesseli2020}
{Kesseli}, A.~Y., {Snellen}, I.~A.~G., {Alonso-Floriano}, F.~J.,
  {Molli{\`e}re}, P., \& {Serindag}, D.~B. 2020, \aj, 160, 228

\bibitem[{{Kitzmann} {et~al.}(2018){Kitzmann}, {Heng}, {Rimmer}, {Hoeijmakers},
  {Tsai}, {Malik}, {Lendl}, {Deitrick}, \& {Demory}}]{kitzmann2018_kelt9b}
{Kitzmann}, D., {Heng}, K., {Rimmer}, P.~B., {et~al.} 2018, \apj, 863, 183

\bibitem[{{Kochukhov}(2018)}]{kochukhov2018_binmag}
{Kochukhov}, O. 2018, {BinMag: Widget for comparing stellar observed with
  theoretical spectra}, Astrophysics Source Code Library, record ascl:1805.015

\bibitem[{{Kochukhov} {et~al.}(2010){Kochukhov}, {Makaganiuk}, \&
  {Piskunov}}]{kochukhov2010_lsd}
{Kochukhov}, O., {Makaganiuk}, V., \& {Piskunov}, N. 2010, \aap, 524, A5

\bibitem[{{Kochukhov} {et~al.}(2009){Kochukhov}, {Shulyak}, \&
  {Ryabchikova}}]{kochukhov2009}
{Kochukhov}, O., {Shulyak}, D., \& {Ryabchikova}, T. 2009, \aap, 499, 851

\bibitem[{{Kochukhov}(2007)}]{kochukhov2007_synth3}
{Kochukhov}, O.~P. 2007, in Physics of Magnetic Stars, ed. I.~I. {Romanyuk},
  D.~O. {Kudryavtsev}, O.~M. {Neizvestnaya}, \& V.~M. {Shapoval}, 109--118

\bibitem[{{Koskinen} {et~al.}(2022){Koskinen}, {Lavvas}, {Huang}, {Bergsten},
  {Fernandes}, \& {Young}}]{koskinen2022}
{Koskinen}, T.~T., {Lavvas}, P., {Huang}, C., {et~al.} 2022, \apj, 929, 52

\bibitem[{{Kubyshkina} {et~al.}(2024){Kubyshkina}, {Fossati}, \&
  {Erkaev}}]{kubyshkina2024_chain}
{Kubyshkina}, D., {Fossati}, L., \& {Erkaev}, N.~V. 2024, \aap, 684, A26

\bibitem[{{Kubyshkina} {et~al.}(2018){Kubyshkina}, {Fossati}, {Erkaev},
  {Johnstone}, {Cubillos}, {Kislyakova}, {Lammer}, {Lendl}, \&
  {Odert}}]{kubyshkina2018_grid}
{Kubyshkina}, D., {Fossati}, L., {Erkaev}, N.~V., {et~al.} 2018, \aap, 619,
  A151

\bibitem[{{Kurucz}(1993)}]{kurucz1993_width9}
{Kurucz}, R.~L. 1993, {SYNTHE spectrum synthesis programs and line data}

\bibitem[{{Kurucz}(2018)}]{kurucz2018}
{Kurucz}, R.~L. 2018, in Astronomical Society of the Pacific Conference Series,
  Vol. 515, Workshop on Astrophysical Opacities, 47

\bibitem[{{Lammer} {et~al.}(2003){Lammer}, {Selsis}, {Ribas}, {Guinan},
  {Bauer}, \& {Weiss}}]{lammer2003}
{Lammer}, H., {Selsis}, F., {Ribas}, I., {et~al.} 2003, \apjl, 598, L121

\bibitem[{{Landstreet}(1988)}]{1988Landstreet}
{Landstreet}, J.~D. 1988, \apj, 326, 967

\bibitem[{{Landstreet}(1998)}]{landstreet1998_velocityfields}
{Landstreet}, J.~D. 1998, \aap, 338, 1041

\bibitem[{{Landstreet} {et~al.}(2009){Landstreet}, {Kupka}, {Ford}, {Officer},
  {Sigut}, {Silaj}, {Strasser}, \& {Townshend}}]{landstreet2009_velocityfields}
{Landstreet}, J.~D., {Kupka}, F., {Ford}, H.~A., {et~al.} 2009, \aap, 503, 973

\bibitem[{{Lendl} {et~al.}(2012){Lendl}, {Anderson}, {Collier-Cameron},
  {Doyle}, {Gillon}, {Hellier}, {Jehin}, {Lister}, {Maxted}, {Pepe},
  {Pollacco}, {Queloz}, {Smalley}, {S{\'e}gransan}, {Smith}, {Triaud}, {Udry},
  {West}, \& {Wheatley}}]{lendl2012}
{Lendl}, M., {Anderson}, D.~R., {Collier-Cameron}, A., {et~al.} 2012, \aap,
  544, A72

\bibitem[{{Lodders}(2003)}]{lodders2003}
{Lodders}, K. 2003, \apj, 591, 1220

\bibitem[{{Lothringer} \& {Barman}(2019)}]{lothringer_and_barman2019}
{Lothringer}, J.~D. \& {Barman}, T. 2019, \apj, 876, 69

\bibitem[{{Lothringer} {et~al.}(2018){Lothringer}, {Barman}, \&
  {Koskinen}}]{lothringer2018_UHJs}
{Lothringer}, J.~D., {Barman}, T., \& {Koskinen}, T. 2018, \apj, 866, 27

\bibitem[{{Lothringer} {et~al.}(2025){Lothringer}, {Bennett}, {Sing},
  {Kehoe-Seamons}, {Rustamkulov}, {Reggiani}, {Schlaufman}, {McCreery},
  {Norris}, {Hauschildt}, {Cacho-Negrete}, {Gressier}, {Espinoza}, {Gapp},
  {Evans-Soma}, {Stevenson}, {Wakeford}, {Gibson}, {Wilson}, \&
  {Nikolov}}]{lothringer2025_wasp178}
{Lothringer}, J.~D., {Bennett}, K.~A., {Sing}, D.~K., {et~al.} 2025, arXiv
  e-prints, arXiv:2503.15472

\bibitem[{{Lothringer} {et~al.}(2022){Lothringer}, {Sing}, {Rustamkulov},
  {Wakeford}, {Stevenson}, {Nikolov}, {Lavvas}, {Spake}, \&
  {Winch}}]{lothringer2022_wasp178}
{Lothringer}, J.~D., {Sing}, D.~K., {Rustamkulov}, Z., {et~al.} 2022, \nat,
  604, 49

\bibitem[{{Lund} {et~al.}(2017){Lund}, {Rodriguez}, {Zhou}, {Gaudi}, {Stassun},
  {Johnson}, {Bieryla}, {Oelkers}, {Stevens}, {Collins}, {Penev}, {Quinn},
  {Latham}, {Villanueva}, {Eastman}, {Kielkopf}, {Oberst}, {Jensen}, {Cohen},
  {Joner}, {Stephens}, {Relles}, {Corfini}, {Gregorio}, {Zambelli}, {Esquerdo},
  {Calkins}, {Berlind}, {Ciardi}, {Dressing}, {Patel}, {Gagnon}, {Gonzales},
  {Beatty}, {Siverd}, {Labadie-Bartz}, {Kuhn}, {Col{\'o}n}, {James}, {Pepper},
  {Fulton}, {McLeod}, {Stockdale}, {Calchi Novati}, {DePoy}, {Gould},
  {Marshall}, {Trueblood}, {Trueblood}, {Johnson}, {Wright}, {McCrady},
  {Wittenmyer}, {Johnson}, {Sergi}, {Wilson}, \& {Sliski}}]{lund2017}
{Lund}, M.~B., {Rodriguez}, J.~E., {Zhou}, G., {et~al.} 2017, \aj, 154, 194

\bibitem[{{Malik} {et~al.}(2017){Malik}, {Grosheintz}, {Mendon{\c{c}}a},
  {Grimm}, {Lavie}, {Kitzmann}, {Tsai}, {Burrows}, {Kreidberg}, {Bedell},
  {Bean}, {Stevenson}, \& {Heng}}]{malik2017}
{Malik}, M., {Grosheintz}, L., {Mendon{\c{c}}a}, J.~M., {et~al.} 2017, \aj,
  153, 56

\bibitem[{{Malik} {et~al.}(2019){Malik}, {Kitzmann}, {Mendon{\c{c}}a}, {Grimm},
  {Marleau}, {Linder}, {Tsai}, \& {Heng}}]{malik2019}
{Malik}, M., {Kitzmann}, D., {Mendon{\c{c}}a}, J.~M., {et~al.} 2019, \aj, 157,
  170

\bibitem[{{Mandel} \& {Agol}(2002)}]{MandelAgol2002apjLightcurves}
{Mandel}, K. \& {Agol}, E. 2002, \apjl, 580, L171

\bibitem[{{Marigo} {et~al.}(2017){Marigo}, {Girardi}, {Bressan}, {Rosenfield},
  {Aringer}, {Chen}, {Dussin}, {Nanni}, {Pastorelli}, {Rodrigues}, {Trabucchi},
  {Bladh}, {Dalcanton}, {Groenewegen}, {Montalb{\'a}n}, \& {Wood}}]{marigo17}
{Marigo}, P., {Girardi}, L., {Bressan}, A., {et~al.} 2017, \apj, 835, 77

\bibitem[{{Michaud}(1970)}]{michaud1970}
{Michaud}, G. 1970, \apj, 160, 641

\bibitem[{{Moran} {et~al.}(2024){Moran}, {Marley}, \& {Crossley}}]{moran2024}
{Moran}, S.~E., {Marley}, M.~S., \& {Crossley}, S.~D. 2024, \apjl, 973, L3

\bibitem[{{Murphy} {et~al.}(2016){Murphy}, {Fossati}, {Bedding}, {Saio},
  {Kurtz}, {Grassitelli}, \& {Wang}}]{murphy2016_vsini}
{Murphy}, S.~J., {Fossati}, L., {Bedding}, T.~R., {et~al.} 2016, \mnras, 459,
  1201

\bibitem[{{Nugroho} {et~al.}(2020){Nugroho}, {Gibson}, {de Mooij}, {Watson},
  {Kawahara}, \& {Merritt}}]{nugroho2020}
{Nugroho}, S.~K., {Gibson}, N.~P., {de Mooij}, E. J.~W., {et~al.} 2020, \mnras,
  496, 504

\bibitem[{{Owen} \& {Wu}(2017)}]{owen2017}
{Owen}, J.~E. \& {Wu}, Y. 2017, \apj, 847, 29

\bibitem[{{Pagano} {et~al.}(2024){Pagano}, {Scandariato}, {Singh}, {Lendl},
  {Queloz}, {Simon}, {Sousa}, {Brandeker}, {Cameron}, {Sulis}, {Van Grootel},
  {Wilson}, {Alibert}, {Alonso}, {Anglada}, {B{\'a}rczy}, {Navascues},
  {Barros}, {Baumjohann}, {Beck}, {Beck}, {Benz}, {Billot}, {Bonfils},
  {Borsato}, {Broeg}, {Bruno}, {Carone}, {Charnoz}, {Corral van Damme},
  {Csizmadia}, {Cubillos}, {Davies}, {Deleuil}, {Deline}, {Delrez},
  {Demangeon}, {Demory}, {Ehrenreich}, {Erikson}, {Fortier}, {Fossati},
  {Fridlund}, {Gandolfi}, {Gillon}, {G{\"u}del}, {G{\"u}nther}, {Helling},
  {Hoyer}, {Isaak}, {Kiss}, {Kopp}, {Lam}, {Laskar}, {Lecavelier des Etangs},
  {Magrin}, {Maxted}, {Mordasini}, {Munari}, {Nascimbeni}, {Olofsson},
  {Ottensamer}, {Pall{\'e}}, {Peter}, {Piotto}, {Pollacco}, {Ragazzoni},
  {Rando}, {Rauer}, {Reimers}, {Ribas}, {Rieder}, {Santos}, {S{\'e}gransan},
  {Smith}, {Stalport}, {Steller}, {Szab{\'o}}, {Thomas}, {Udry}, {Venturini},
  \& {Walton}}]{pagano2024}
{Pagano}, I., {Scandariato}, G., {Singh}, V., {et~al.} 2024, \aap, 682, A102

\bibitem[{{Parmentier} {et~al.}(2018){Parmentier}, {Line}, {Bean}, {Mansfield},
  {Kreidberg}, {Lupu}, {Visscher}, {D{\'e}sert}, {Fortney}, {Deleuil},
  {Arcangeli}, {Showman}, \& {Marley}}]{parmentier2018_UHJs}
{Parmentier}, V., {Line}, M.~R., {Bean}, J.~L., {et~al.} 2018, \aap, 617, A110

\bibitem[{{Pepe} {et~al.}(2021){Pepe}, {Cristiani}, {Rebolo}, {Santos},
  {Dekker}, {Cabral}, {Di Marcantonio}, {Figueira}, {Lo Curto}, {Lovis},
  {Mayor}, {M{\'e}gevand}, {Molaro}, {Riva}, {Zapatero Osorio}, {Amate},
  {Manescau}, {Pasquini}, {Zerbi}, {Adibekyan}, {Abreu}, {Affolter}, {Alibert},
  {Aliverti}, {Allart}, {Allende Prieto}, {{\'A}lvarez}, {Alves}, {Avila},
  {Baldini}, {Bandy}, {Barros}, {Benz}, {Bianco}, {Borsa}, {Bourrier},
  {Bouchy}, {Broeg}, {Calderone}, {Cirami}, {Coelho}, {Conconi}, {Coretti},
  {Cumani}, {Cupani}, {D'Odorico}, {Damasso}, {Deiries}, {Delabre},
  {Demangeon}, {Dumusque}, {Ehrenreich}, {Faria}, {Fragoso}, {Genolet},
  {Genoni}, {G{\'e}nova Santos}, {Gonz{\'a}lez Hern{\'a}ndez}, {Hughes},
  {Iwert}, {Kerber}, {Knudstrup}, {Landoni}, {Lavie}, {Lillo-Box}, {Lizon},
  {Maire}, {Martins}, {Mehner}, {Micela}, {Modigliani}, {Monteiro}, {Monteiro},
  {Moschetti}, {Murphy}, {Nunes}, {Oggioni}, {Oliveira}, {Oshagh}, {Pall{\'e}},
  {Pariani}, {Poretti}, {Rasilla}, {Rebord{\~a}o}, {Redaelli}, {Santana
  Tschudi}, {Santin}, {Santos}, {S{\'e}gransan}, {Schmidt}, {Segovia},
  {Sosnowska}, {Sozzetti}, {Sousa}, {Span{\`o}}, {Su{\'a}rez Mascare{\~n}o},
  {Tabernero}, {Tenegi}, {Udry}, \& {Zanutta}}]{pepe2021_espresso}
{Pepe}, F., {Cristiani}, S., {Rebolo}, R., {et~al.} 2021, \aap, 645, A96

\bibitem[{{Pino} {et~al.}(2020){Pino}, {D{\'e}sert}, {Brogi}, {Malavolta},
  {Wyttenbach}, {Line}, {Hoeijmakers}, {Fossati}, {Bonomo}, {Nascimbeni},
  {Panwar}, {Affer}, {Benatti}, {Biazzo}, {Bignamini}, {Borsa}, {Carleo},
  {Claudi}, {Cosentino}, {Covino}, {Damasso}, {Desidera}, {Giacobbe},
  {Harutyunyan}, {Lanza}, {Leto}, {Maggio}, {Maldonado}, {Mancini}, {Micela},
  {Molinari}, {Pagano}, {Piotto}, {Poretti}, {Rainer}, {Scandariato},
  {Sozzetti}, {Allart}, {Borsato}, {Bruno}, {Di Fabrizio}, {Ehrenreich},
  {Fiorenzano}, {Frustagli}, {Lavie}, {Lovis}, {Magazz{\`u}}, {Nardiello},
  {Pedani}, \& {Smareglia}}]{pino2020}
{Pino}, L., {D{\'e}sert}, J.-M., {Brogi}, M., {et~al.} 2020, \apjl, 894, L27

\bibitem[{{Piskunov} {et~al.}(1995){Piskunov}, {Kupka}, {Ryabchikova}, {Weiss},
  \& {Jeffery}}]{piskunov1995_vald}
{Piskunov}, N.~E., {Kupka}, F., {Ryabchikova}, T.~A., {Weiss}, W.~W., \&
  {Jeffery}, C.~S. 1995, \aaps, 112, 525

\bibitem[{{Ricker} {et~al.}(2015){Ricker}, {Winn}, {Vanderspek}, {Latham},
  {Bakos}, {Bean}, {Berta-Thompson}, {Brown}, {Buchhave}, {Butler}, {Butler},
  {Chaplin}, {Charbonneau}, {Christensen-Dalsgaard}, {Clampin}, {Deming},
  {Doty}, {De Lee}, {Dressing}, {Dunham}, {Endl}, {Fressin}, {Ge}, {Henning},
  {Holman}, {Howard}, {Ida}, {Jenkins}, {Jernigan}, {Johnson}, {Kaltenegger},
  {Kawai}, {Kjeldsen}, {Laughlin}, {Levine}, {Lin}, {Lissauer}, {MacQueen},
  {Marcy}, {McCullough}, {Morton}, {Narita}, {Paegert}, {Palle}, {Pepe},
  {Pepper}, {Quirrenbach}, {Rinehart}, {Sasselov}, {Sato}, {Seager},
  {Sozzetti}, {Stassun}, {Sullivan}, {Szentgyorgyi}, {Torres}, {Udry}, \&
  {Villasenor}}]{ricker2015_tess}
{Ricker}, G.~R., {Winn}, J.~N., {Vanderspek}, R., {et~al.} 2015, Journal of
  Astronomical Telescopes, Instruments, and Systems, 1, 014003

\bibitem[{{Rodr{\'\i}guez Mart{\'\i}nez} {et~al.}(2020){Rodr{\'\i}guez
  Mart{\'\i}nez}, {Gaudi}, {Rodriguez}, {Zhou}, {Labadie-Bartz}, {Quinn},
  {Penev}, {Tan}, {Latham}, {Paredes}, {Kielkopf}, {Addison}, {Wright},
  {Teske}, {Howell}, {Ciardi}, {Ziegler}, {Stassun}, {Johnson}, {Eastman},
  {Siverd}, {Beatty}, {Bouma}, {Bedding}, {Pepper}, {Winn}, {Lund},
  {Villanueva}, {Stevens}, {Jensen}, {Kilby}, {Crane}, {Tokovinin}, {Everett},
  {Tinney}, {Fausnaugh}, {Cohen}, {Bayliss}, {Bieryla}, {Cargile}, {Collins},
  {Conti}, {Col{\'o}n}, {Curtis}, {Depoy}, {Evans}, {Feliz}, {Gregorio},
  {Rothenberg}, {James}, {Joner}, {Kuhn}, {Manner}, {Khakpash}, {Marshall},
  {McLeod}, {Penny}, {Reed}, {Relles}, {Stephens}, {Stockdale}, {Trueblood},
  {Trueblood}, {Yao}, {Zambelli}, {Vanderspek}, {Seager}, {Jenkins}, {Henry},
  {James}, {Jao}, {Wang}, {Butler}, {Thompson}, {Shectman}, {Wittenmyer},
  {Bowler}, {Horner}, {Kane}, {Mengel}, {Morton}, {Okumura}, {Plavchan},
  {Zhang}, {Scott}, {Matson}, {Mann}, {Dragomir}, {G{\"u}nther}, {Ting},
  {Glidden}, \& {Quintana}}]{rodriguez-martinez2020}
{Rodr{\'\i}guez Mart{\'\i}nez}, R., {Gaudi}, B.~S., {Rodriguez}, J.~E.,
  {et~al.} 2020, \aj, 160, 111

\bibitem[{{Rothman} {et~al.}(2010){Rothman}, {Gordon}, {Barber}, {Dothe},
  {Gamache}, {Goldman}, {Perevalov}, {Tashkun}, \&
  {Tennyson}}]{rothman2010_hitemp}
{Rothman}, L.~S., {Gordon}, I.~E., {Barber}, R.~J., {et~al.} 2010, \jqsrt, 111,
  2139

\bibitem[{{Royer} {et~al.}(2007){Royer}, {Zorec}, \&
  {G{\'o}mez}}]{royer2007_rotationAstars}
{Royer}, F., {Zorec}, J., \& {G{\'o}mez}, A.~E. 2007, \aap, 463, 671

\bibitem[{{Ryabchikova} {et~al.}(2009){Ryabchikova}, {Fossati}, \&
  {Shulyak}}]{ryabchikova2009_hd49933}
{Ryabchikova}, T., {Fossati}, L., \& {Shulyak}, D. 2009, \aap, 506, 203

\bibitem[{{Ryabchikova} {et~al.}(2015){Ryabchikova}, {Piskunov}, {Kurucz},
  {Stempels}, {Heiter}, {Pakhomov}, \& {Barklem}}]{ryabchikova2015_vald}
{Ryabchikova}, T., {Piskunov}, N., {Kurucz}, R.~L., {et~al.} 2015, \physscr,
  90, 054005

\bibitem[{{S{\'a}nchez-L{\'o}pez} {et~al.}(2022){S{\'a}nchez-L{\'o}pez}, {Lin},
  {Snellen}, {Casasayas-Barris}, {Garc{\'\i}a Mu{\~n}oz}, {Lamp{\'o}n}, \&
  {L{\'o}pez-Puertas}}]{sanchez2022_paschenBeta_kelt9b}
{S{\'a}nchez-L{\'o}pez}, A., {Lin}, L., {Snellen}, I.~A.~G., {et~al.} 2022,
  \aap, 666, L1

\bibitem[{{Shulyak} {et~al.}(2009){Shulyak}, {Ryabchikova}, {Mashonkina}, \&
  {Kochukhov}}]{shulyak2009}
{Shulyak}, D., {Ryabchikova}, T., {Mashonkina}, L., \& {Kochukhov}, O. 2009,
  \aap, 499, 879

\bibitem[{{Shulyak} {et~al.}(2004){Shulyak}, {Tsymbal}, {Ryabchikova},
  {St{\"u}tz}, \& {Weiss}}]{Shulyak2004}
{Shulyak}, D., {Tsymbal}, V., {Ryabchikova}, T., {St{\"u}tz}, C., \& {Weiss},
  W.~W. 2004, \aap, 428, 993

\bibitem[{{Sing} {et~al.}(2019){Sing}, {Lavvas}, {Ballester}, {Lecavelier des
  Etangs}, {Marley}, {Nikolov}, {Ben-Jaffel}, {Bourrier}, {Buchhave}, {Deming},
  {Ehrenreich}, {Mikal-Evans}, {Kataria}, {Lewis}, {L{\'o}pez-Morales},
  {Garc{\'\i}a Mu{\~n}oz}, {Henry}, {Sanz-Forcada}, {Spake}, {Wakeford}, \&
  {PanCET Collaboration}}]{sing2019}
{Sing}, D.~K., {Lavvas}, P., {Ballester}, G.~E., {et~al.} 2019, \aj, 158, 91

\bibitem[{{Speagle}(2020)}]{speagle2020_dynesty}
{Speagle}, J.~S. 2020, \mnras, 493, 3132

\bibitem[{{Sreejith} {et~al.}(2023){Sreejith}, {France}, {Fossati}, {Koskinen},
  {Egan}, {Cauley}, {Cubillos}, {Ambily}, {Huang}, {Lavvas}, {Fleming},
  {Desert}, {Nell}, {Petit}, \& {Vidotto}}]{sreejith2023_cute_wasp189b}
{Sreejith}, A.~G., {France}, K., {Fossati}, L., {et~al.} 2023, \apjl, 954, L23

\bibitem[{{Stangret} {et~al.}(2024){Stangret}, {Fossati}, {D'Arpa}, {Borsa},
  {Nascimbeni}, {Malavolta}, {Sicilia}, {Pino}, {Biassoni}, {Bonomo}, {Brogi},
  {Claudi}, {Damasso}, {Di Maio}, {Giacobbe}, {Guilluy}, {Harutyunyan},
  {Lanza}, {Mart{\'\i}nez Fiorenzano}, {Mancini}, {Nardiello}, {Scandariato},
  {Sozzetti}, \& {Zingales}}]{stangret2024}
{Stangret}, M., {Fossati}, L., {D'Arpa}, M.~C., {et~al.} 2024, \aap, 692, A76

\bibitem[{{Sun} \& {Chiappini}(2024)}]{sun2024_rotationEvolution}
{Sun}, W. \& {Chiappini}, C. 2024, arXiv e-prints, arXiv:2406.18268

\bibitem[{{Talens} {et~al.}(2018){Talens}, {Justesen}, {Albrecht}, {McCormac},
  {Van Eylen}, {Otten}, {Murgas}, {Palle}, {Pollacco}, {Stuik}, {Spronck},
  {Lesage}, {Grundahl}, {Fredslund Andersen}, {Antoci}, \&
  {Snellen}}]{talens2018}
{Talens}, G.~J.~J., {Justesen}, A.~B., {Albrecht}, S., {et~al.} 2018, \aap,
  612, A57

\bibitem[{{Tennyson} {et~al.}(2016){Tennyson}, {Yurchenko}, {Al-Refaie},
  {Barton}, {Chubb}, {Coles}, {Diamantopoulou}, {Gorman}, {Hill}, {Lam},
  {Lodi}, {McKemmish}, {Na}, {Owens}, {Polyansky}, {Rivlin}, {Sousa-Silva},
  {Underwood}, {Yachmenev}, \& {Zak}}]{tennyson2016_exomol}
{Tennyson}, J., {Yurchenko}, S.~N., {Al-Refaie}, A.~F., {et~al.} 2016, Journal
  of Molecular Spectroscopy, 327, 73

\bibitem[{{Tsymbal}(1996)}]{tsymbal1996_width9}
{Tsymbal}, V. 1996, in Astronomical Society of the Pacific Conference Series,
  Vol. 108, M.A.S.S., Model Atmospheres and Spectrum Synthesis, ed. S.~J.
  {Adelman}, F.~{Kupka}, \& W.~W. {Weiss}, 198

\bibitem[{{Turner} {et~al.}(2020){Turner}, {de Mooij}, {Jayawardhana}, {Young},
  {Fossati}, {Koskinen}, {Lothringer}, {Karjalainen}, \&
  {Karjalainen}}]{turner2020_kelt9b}
{Turner}, J.~D., {de Mooij}, E. J.~W., {Jayawardhana}, R., {et~al.} 2020,
  \apjl, 888, L13

\bibitem[{{Volkov} {et~al.}(2011){Volkov}, {Johnson}, {Tucker}, \&
  {Erwin}}]{volkov2011}
{Volkov}, A.~N., {Johnson}, R.~E., {Tucker}, O.~J., \& {Erwin}, J.~T. 2011,
  \apjl, 729, L24

\bibitem[{{von Zeipel}(1924)}]{vonZeipel1924}
{von Zeipel}, H. 1924, \mnras, 84, 665

\bibitem[{{Wade, G. A.} {et~al.}(2001){Wade, G. A.}, {Bagnulo, S.}, {Kochukhov,
  O.}, {Landstreet, J. D.}, {Piskunov, N.}, \& {Stift, M. J.}}]{2001Wadeetal}
{Wade, G. A.}, {Bagnulo, S.}, {Kochukhov, O.}, {et~al.} 2001, \aap, 374, 265

\bibitem[{{Wong} {et~al.}(2020){Wong}, {Shporer}, {Kitzmann}, {Morris}, {Heng},
  {Hoeijmakers}, {Demory}, {Ahlers}, {Mansfield}, {Bean}, {Daylan},
  {Fetherolf}, {Rodriguez}, {Benneke}, {Ricker}, {Latham}, {Vanderspek},
  {Seager}, {Winn}, {Jenkins}, {Burke}, {Christiansen}, {Essack}, {Rose},
  {Smith}, {Tenenbaum}, \& {Yahalomi}}]{wong2020_dynamics_kelt9b}
{Wong}, I., {Shporer}, A., {Kitzmann}, D., {et~al.} 2020, \aj, 160, 88

\bibitem[{{Wyttenbach} {et~al.}(2020){Wyttenbach}, {Molli{\`e}re},
  {Ehrenreich}, {Cegla}, {Bourrier}, {Lovis}, {Pino}, {Allart}, {Seidel},
  {Hoeijmakers}, {Nielsen}, {Lavie}, {Pepe}, {Bonfils}, \&
  {Snellen}}]{Wyttenbach_2020}
{Wyttenbach}, A., {Molli{\`e}re}, P., {Ehrenreich}, D., {et~al.} 2020, \aap,
  638, A87

\bibitem[{{Yan} \& {Henning}(2018)}]{yan_and_henning2018_HalphaKELT9b}
{Yan}, F. \& {Henning}, T. 2018, Nature Astronomy, 2, 714

\bibitem[{{Yan} {et~al.}(2022){Yan}, {Reiners}, {Pall{\'e}}, {Shulyak},
  {Stangret}, {Molaverdikhani}, {Nortmann}, {Molli{\`e}re}, {Henning},
  {Casasayas-Barris}, {Cont}, {Chen}, {Czesla}, {S{\'a}nchez-L{\'o}pez},
  {L{\'o}pez-Puertas}, {Ribas}, {Quirrenbach}, {Caballero}, {Amado},
  {Galad{\'\i}-Enr{\'\i}quez}, {Khalafinejad}, {Lara}, {Montes}, {Morello},
  {Nagel}, {Sedaghati}, {Zapatero Osorio}, \&
  {Zechmeister}}]{yan2022_mascara2b}
{Yan}, F., {Reiners}, A., {Pall{\'e}}, E., {et~al.} 2022, \aap, 659, A7

\bibitem[{{Yelle}(2004)}]{yelle2004}
{Yelle}, R.~V. 2004, \icarus, 170, 167

\bibitem[{{Young} {et~al.}(2020){Young}, {Fossati}, {Koskinen}, {Salz},
  {Cubillos}, \& {France}}]{young2020}
{Young}, M.~E., {Fossati}, L., {Koskinen}, T.~T., {et~al.} 2020, \aap, 641, A47

\bibitem[{{Young} {et~al.}(2024){Young}, {Spring}, \&
  {Birkby}}]{young2024_wasp121b}
{Young}, M.~E., {Spring}, E.~F., \& {Birkby}, J.~L. 2024, \mnras, 530, 4356

\end{thebibliography}
\begin{appendix}
\FloatBarrier
\section{Additional figures and tables}\label{appendix:figures_tables}
%
%-------------------------------------
\begin{table}[]
\renewcommand{\arraystretch}{1.1}
\caption{Derived stellar atmospheric abundances.}
\begin{tabular}{l|cccc}
\hline
\hline
Ion & $\log (N/N_{\rm tot})$ & $\sigma\log (N/N_{\rm tot})$ & n & [$N/N_{\rm tot}$] \\
\hline
C{\sc i}	& $-$3.79 & 0.08 &  18 & $-$0.18 \\
N{\sc i}	& $-$4.23 & 0.02 &   2 & $-$0.02 \\
O{\sc i}	& $-$3.29 & 0.01 &   1 &    0.06 \\
Na{\sc i}	& $-$5.58 & 0.09 &   2 &    0.22 \\
Mg{\sc i}	& $-$4.34 & 0.07 &   3 &    0.10 \\
Mg{\sc ii}	& $-$4.34 & 0.02 &   2 &    0.10 \\
Al{\sc ii}	& $-$5.35 & 0.07 &   4 &    0.24 \\
Si{\sc i}	& $-$4.47 & 0.44 &  11 &    0.06 \\
Si{\sc ii}	& $-$4.45 & 0.04 &   3 &    0.08 \\
S{\sc i}	& $-$4.53 & 0.09 &   6 &    0.39 \\
K{\sc i}	& $-$6.78 & 0.09 &   1 &    0.23 \\
Ca{\sc i}	& $-$5.72 & 0.12 &  15 & $-$0.02 \\
Ca{\sc ii}	& $-$5.72 & 0.04 &   2 & $-$0.02 \\
Sc{\sc ii}	& $-$9.22 & 0.09 &   5 & $-$0.33 \\
Ti{\sc i}	& $-$6.87 & 0.09 &   1 &    0.22 \\
Ti{\sc ii}	& $-$6.85 & 0.19 &  21 &    0.24 \\
V{\sc ii}	& $-$7.55 & 0.04 &   1 &    0.56 \\
Cr{\sc i}	& $-$6.02 & 0.15 &  18 &    0.38 \\
Cr{\sc ii}	& $-$5.92 & 0.07 &  36 &    0.48 \\
Mn{\sc i}	& $-$6.54 & 0.09 &   1 &    0.07 \\
Mn{\sc ii}	& $-$6.54 & 0.07 &   6 &    0.07 \\
Fe{\sc i}	& $-$4.31 & 0.15 & 198 &    0.23 \\
Fe{\sc ii}	& $-$4.31 & 0.10 & 216 &    0.23 \\
Co{\sc i}	& $-$6.76 & 0.08 &   1 &    0.29 \\
Co{\sc ii}	& $-$6.74 & 0.03 &   1 &    0.31 \\
Ni{\sc i}	& $-$5.48 & 0.13 &  39 &    0.34 \\
Ni{\sc ii}	& $-$5.40 & 0.02 &   2 &    0.42 \\
Cu{\sc i}	& $-$7.38 & 0.09 &   1 &    0.47 \\
Zn{\sc i}	& $-$7.03 & 0.10 &   3 &    0.45 \\
Sr{\sc ii}	& $-$8.59 & 0.18 &   1 &    0.58 \\
Y{\sc ii}	& $-$9.48 & 0.11 &   3 &    0.35 \\
Zr{\sc ii}	& $-$8.71 & 0.06 &   1 &    0.75 \\
Ba{\sc ii}	& $-$8.93 & 0.16 &   3 &    0.93 \\
La{\sc ii}	& $-$9.70 & 0.09 &   1 &    1.24 \\
Ce{\sc ii}	& $-$9.52 & 0.13 &   5 &    0.94 \\
Pr{\sc iii}	& $-$9.93 & 0.05 &   1 &    1.39 \\
Nd{\sc iii}	& $-$9.94 & 0.04 &   1 &    0.68 \\
\hline
\end{tabular}
\tablefoot{The second and third columns give the abundance obtained for each ion and the corresponding total uncertainty (see text). The fourth column lists the number of lines measured to derive the abundance, while the last column gives the abundance relative to solar \citep{asplund2009}.}
\label{tab:abundances}
\end{table}
%-------------------------------------
%-------------------------------------
\begin{figure}[h!]
                \centering
                \includegraphics[width=9cm]{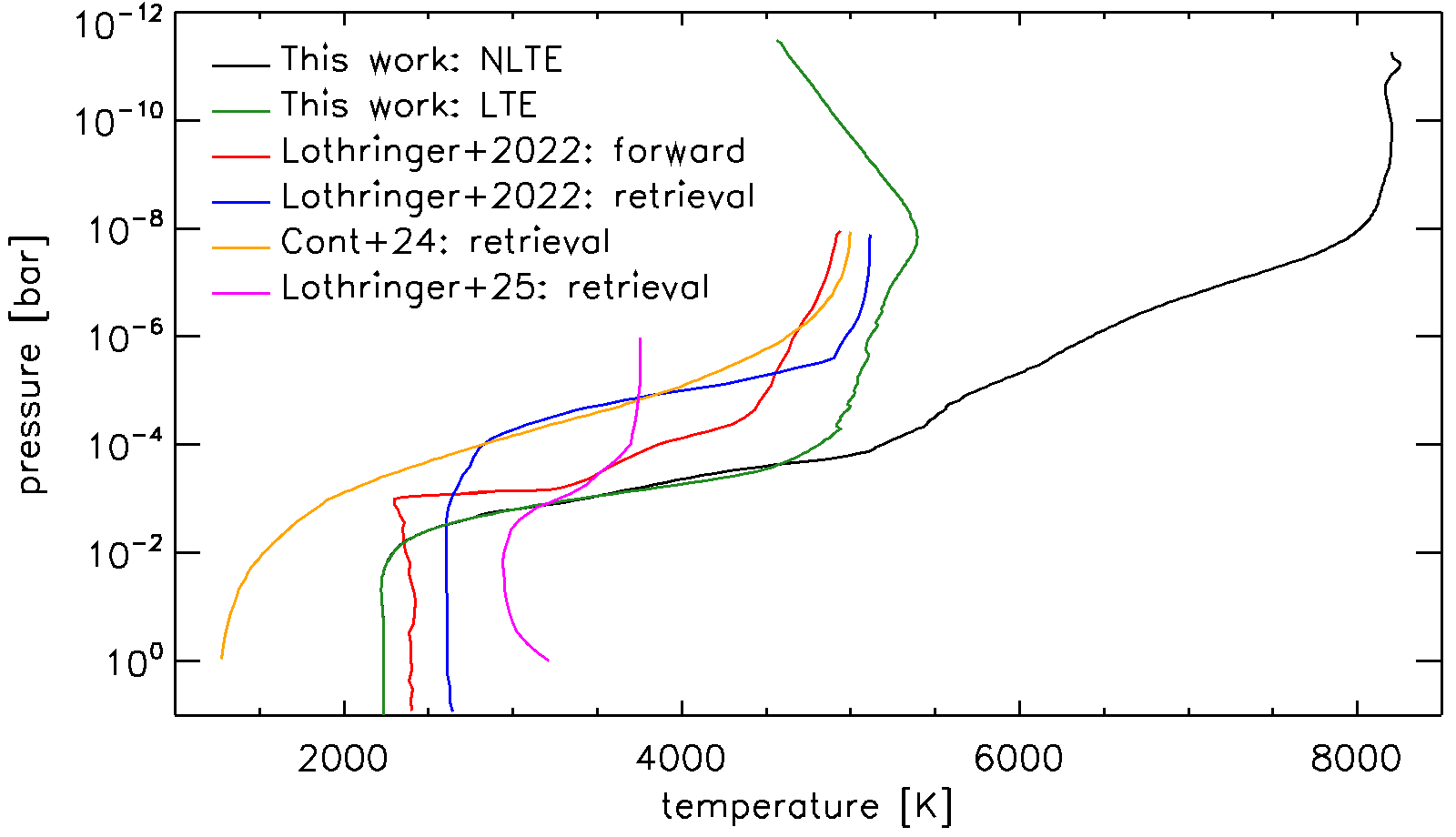}
                \caption{Comparison between the composite TP profiles computed in NLTE (black) and LTE (green) with those obtained by \citet[][LTE]{lothringer2022_wasp178} employing forward modelling (red) and retrieval (blue), \citet{cont2024} employing retrieval (yellow), and \citet{lothringer_and_barman2019} employing retrieval (magenta).} 
                \label{fig:TPliteratureCompare}
\end{figure}
%-------------------------------------
%-------------------------------------
\begin{figure*}[h!]
                \centering
                \includegraphics[width=18cm]{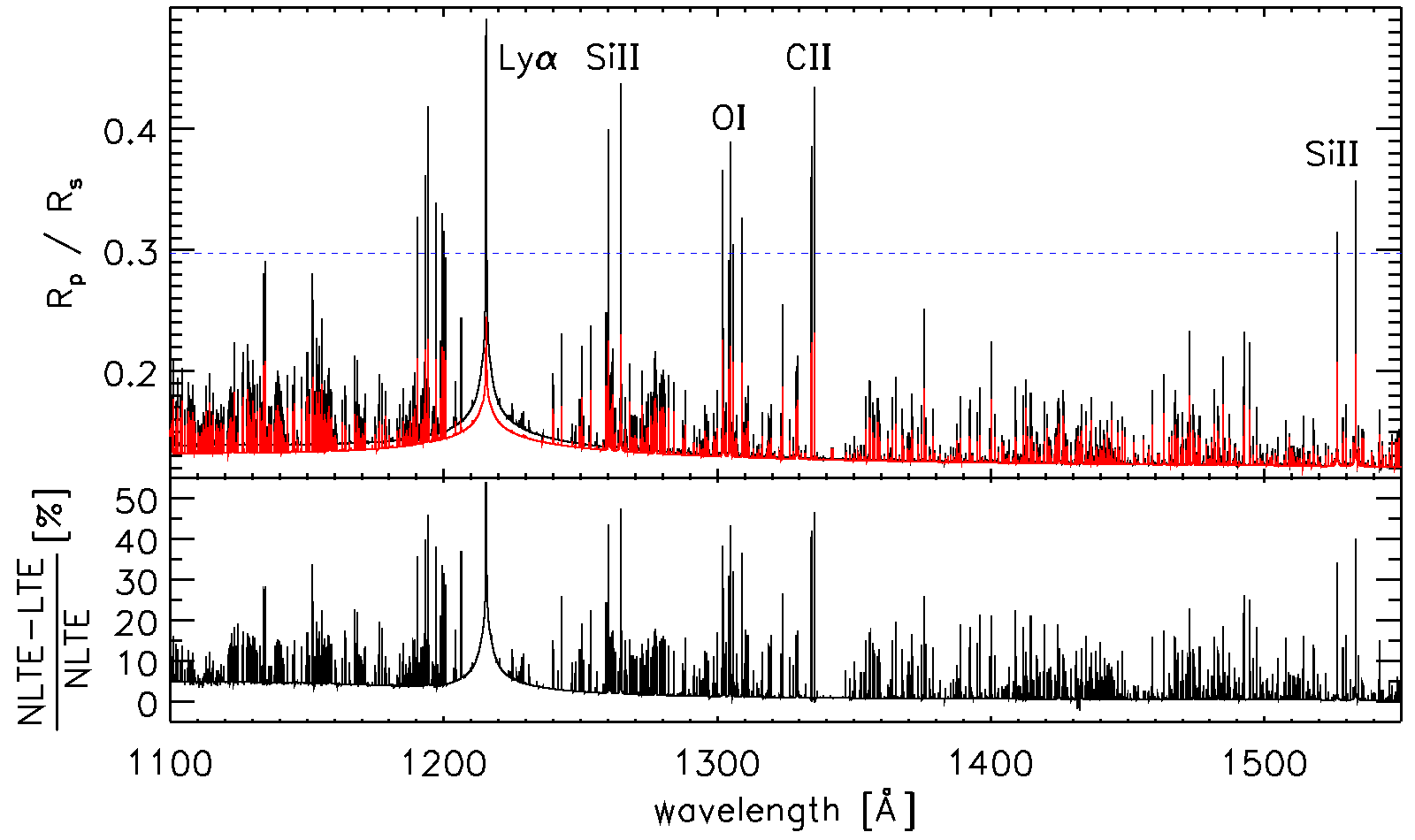}
                \caption{Same as Figure~\ref{fig:transmission_spectra}, but for the 1100--1550\AA\ band.} 
                \label{fig:transmission_spectra_1100-1550}
\end{figure*}
%-------------------------------------
%-------------------------------------
\begin{figure*}[h!]
                \centering
                \includegraphics[width=18cm]{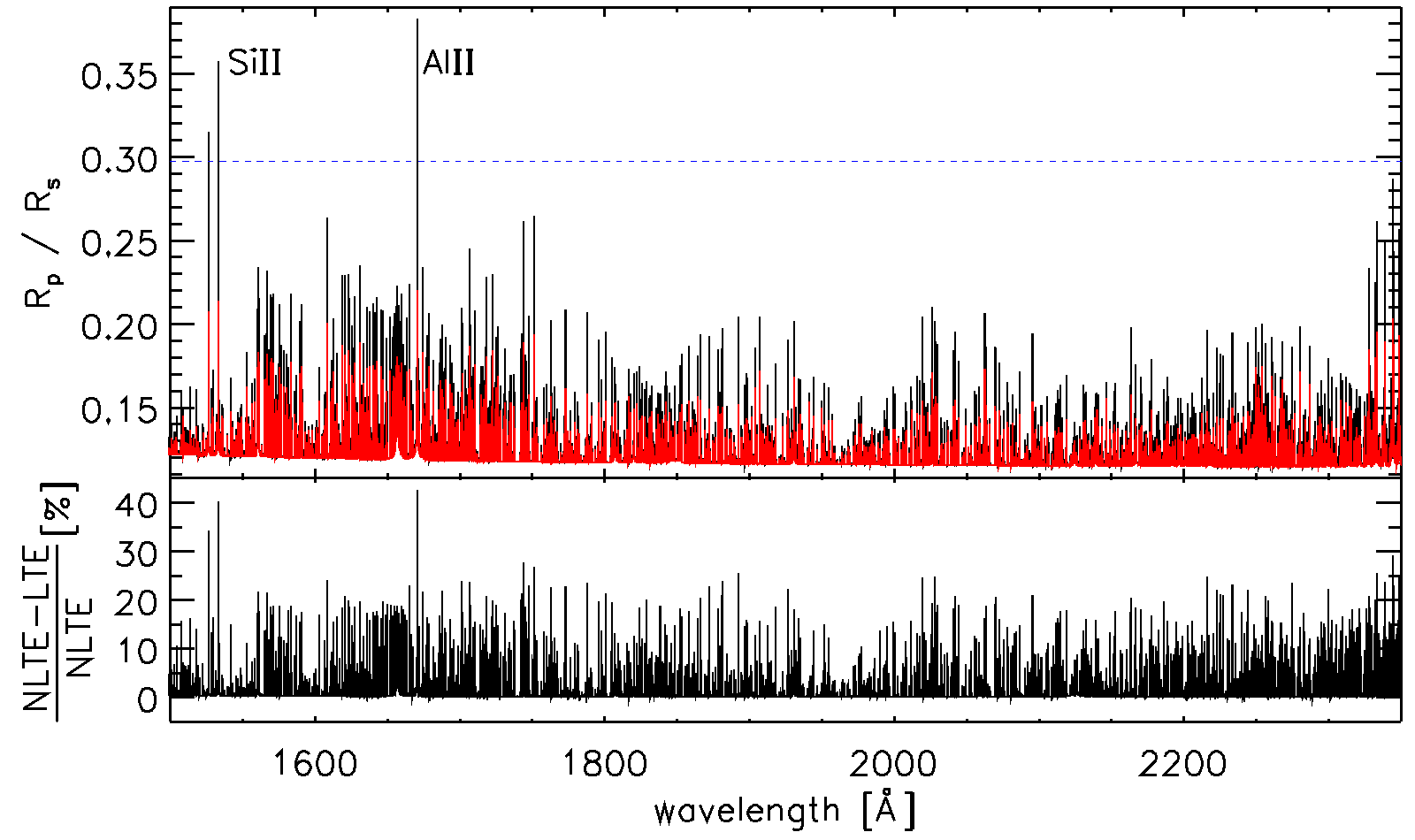}
                \caption{Same as Figure~\ref{fig:transmission_spectra}, but for the 1500--2350\AA\ band.} 
                \label{fig:transmission_spectra_1500-2350}
\end{figure*}
%-------------------------------------
%-------------------------------------
\begin{figure*}[h!]
                \centering
                \includegraphics[width=18cm]{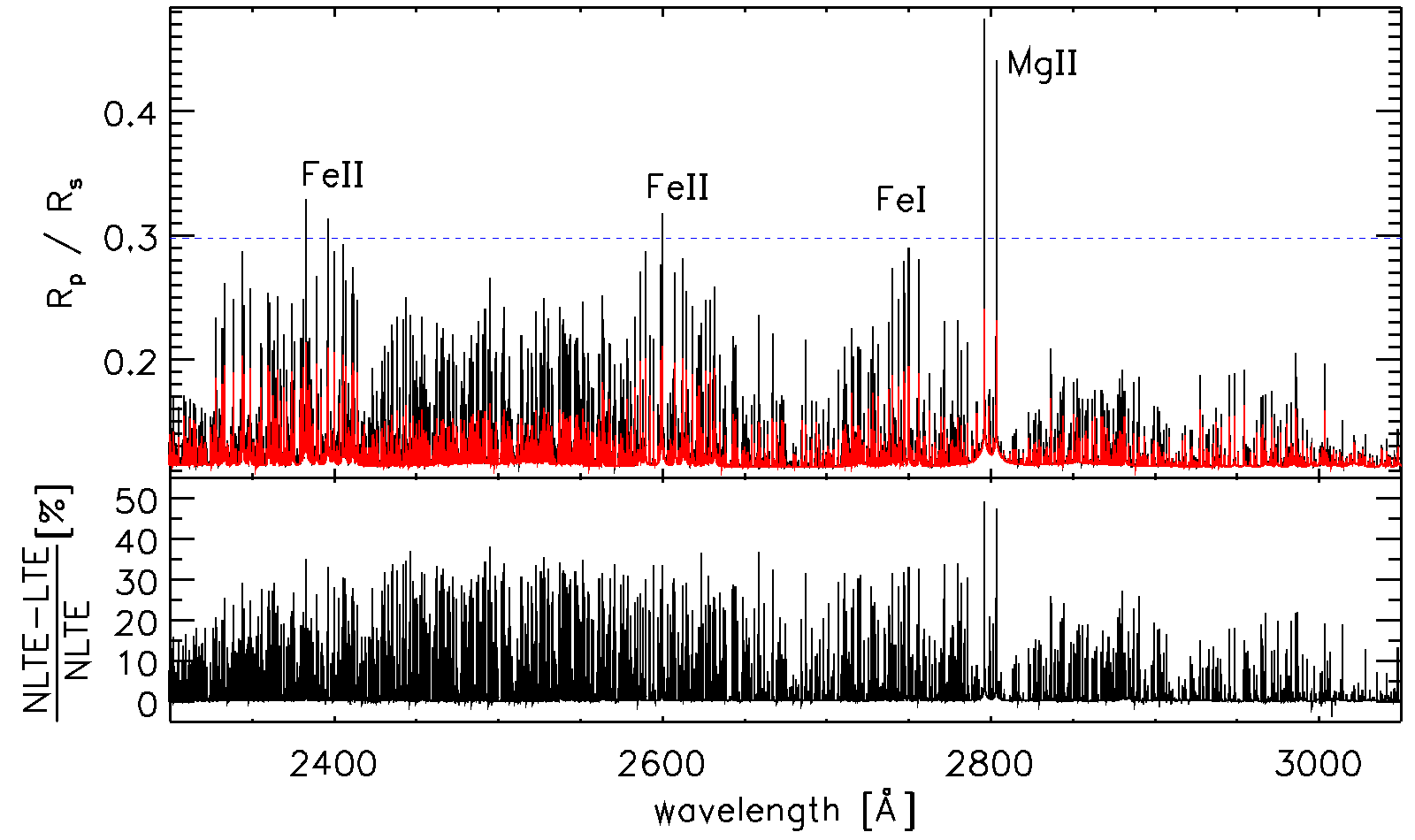}
                \caption{Same as Figure~\ref{fig:transmission_spectra}, but for the 2300--3050\AA\ band.} 
                \label{fig:transmission_spectra_2300-3050}
\end{figure*}
%-------------------------------------
%-------------------------------------
\begin{figure*}[h!]
                \centering
                \includegraphics[width=18cm]{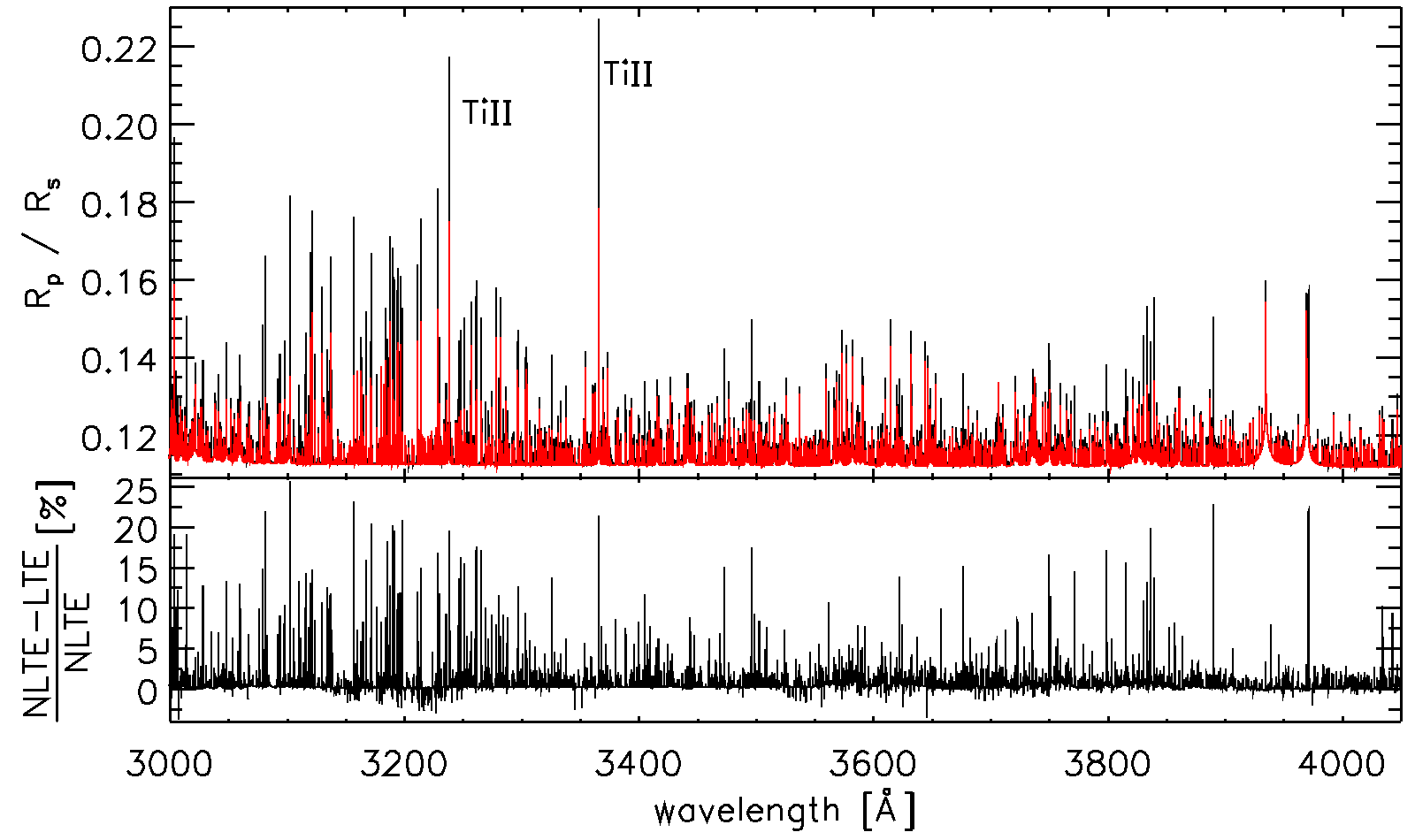}
                \caption{Same as Figure~\ref{fig:transmission_spectra}, but for the 3000--4050\AA\ band.} 
                \label{fig:transmission_spectra_3000-4050}
\end{figure*}
%-------------------------------------
%-------------------------------------
\begin{figure*}[h!]
                \centering
                \includegraphics[width=18cm]{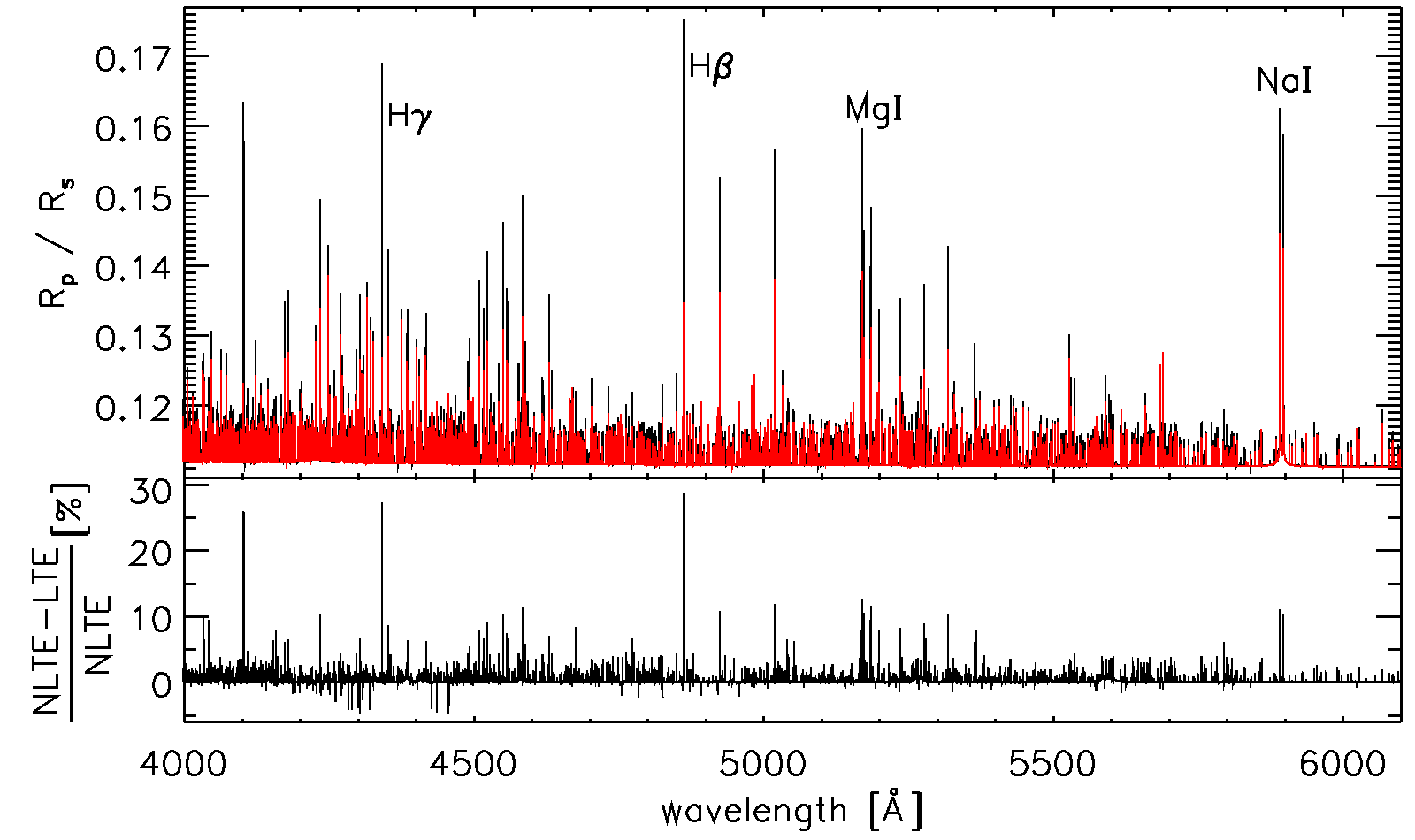}
                \caption{Same as Figure~\ref{fig:transmission_spectra}, but for the 4000--6100\AA\ band.} 
                \label{fig:transmission_spectra_4000-6100}
\end{figure*}
%-------------------------------------
%-------------------------------------
\begin{figure*}[h!]
                \centering
                \includegraphics[width=18cm]{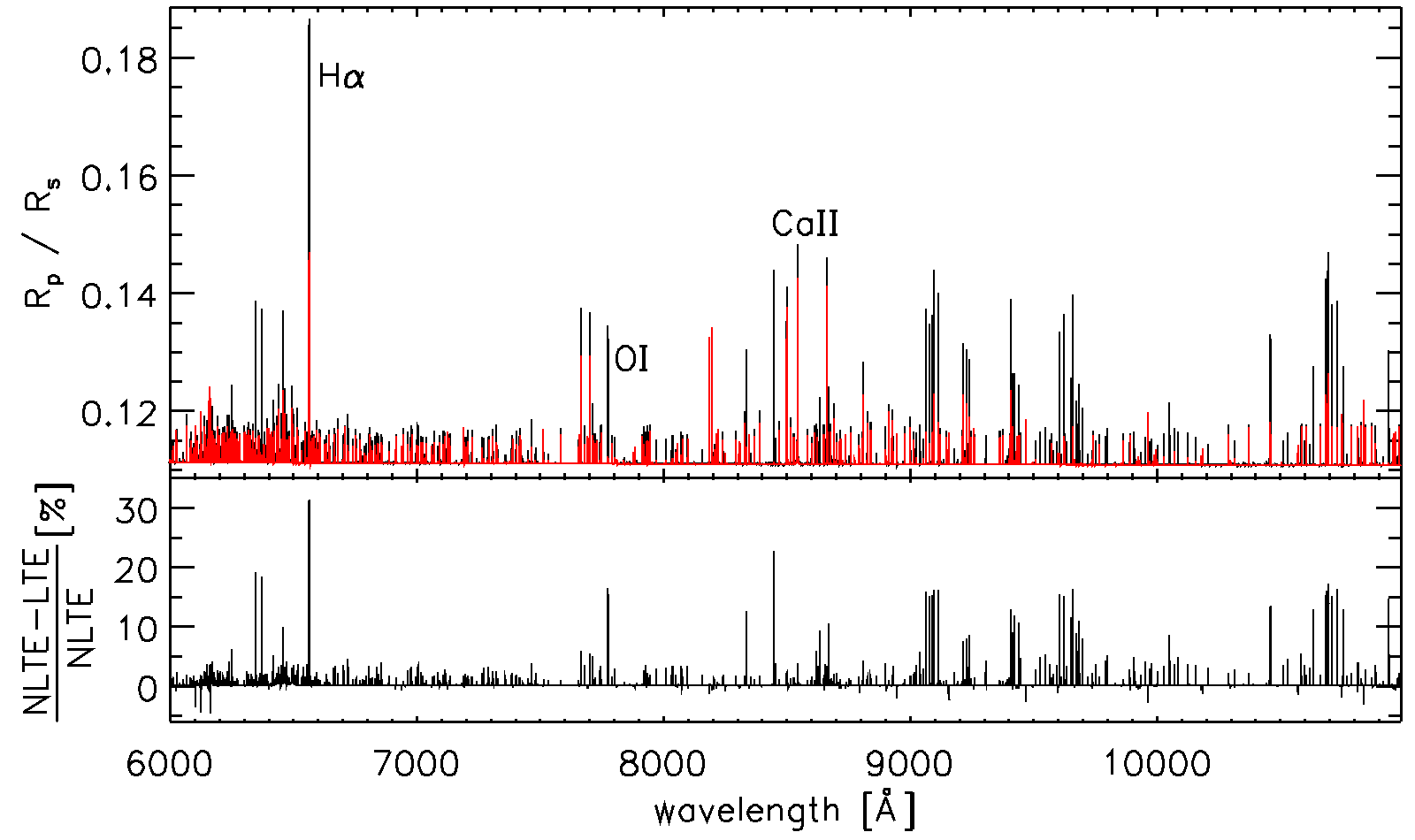}
                \caption{Same as Figure~\ref{fig:transmission_spectra}, but for the 6000--11000\AA\ band.} 
                \label{fig:transmission_spectra_6000-11000}
\end{figure*}
%-------------------------------------
%
\FloatBarrier
%
%-------------------------------------
\begin{figure*}[h!]
                \centering
                \includegraphics[width=18cm]{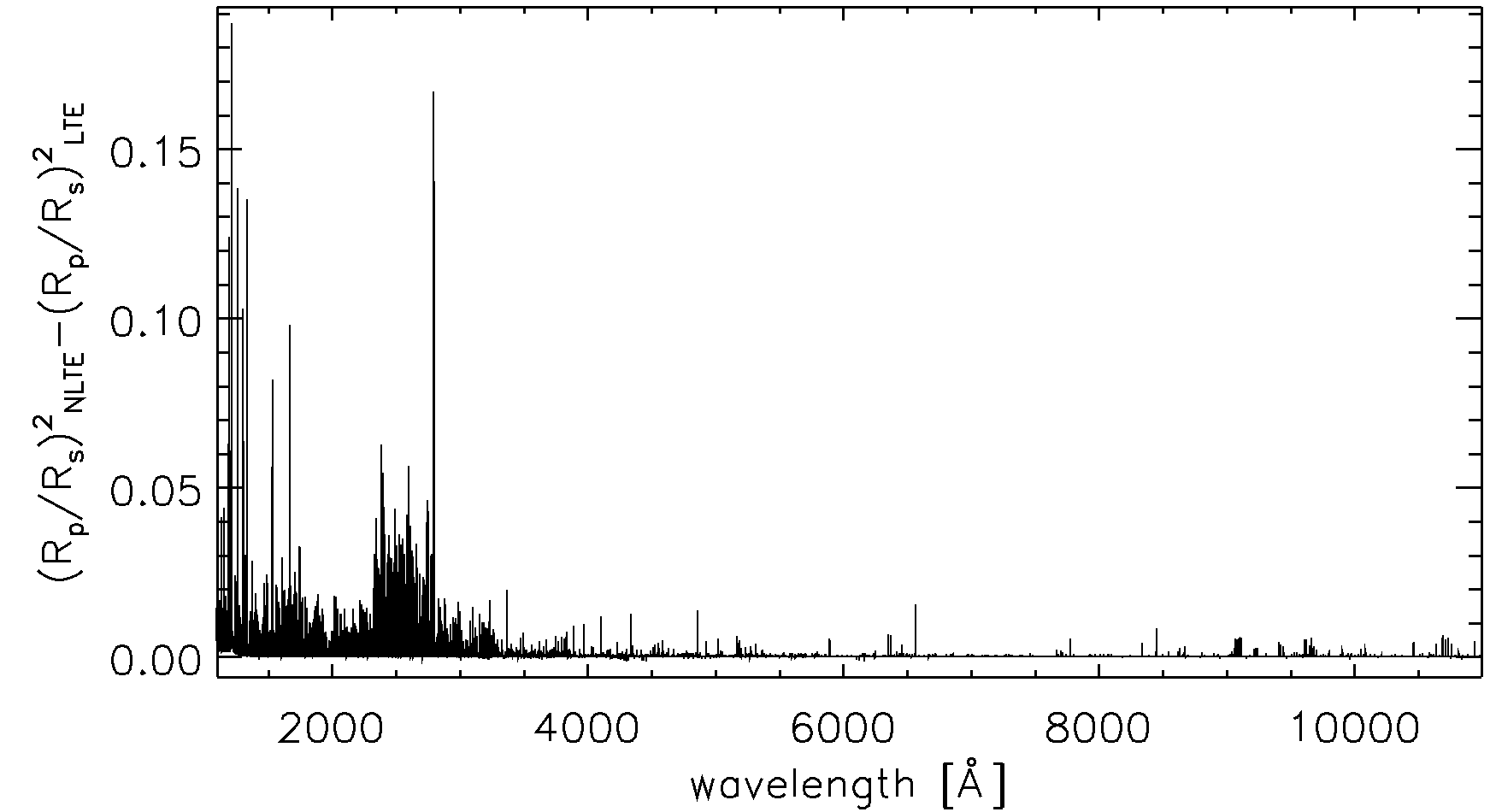}
                \includegraphics[width=18cm]{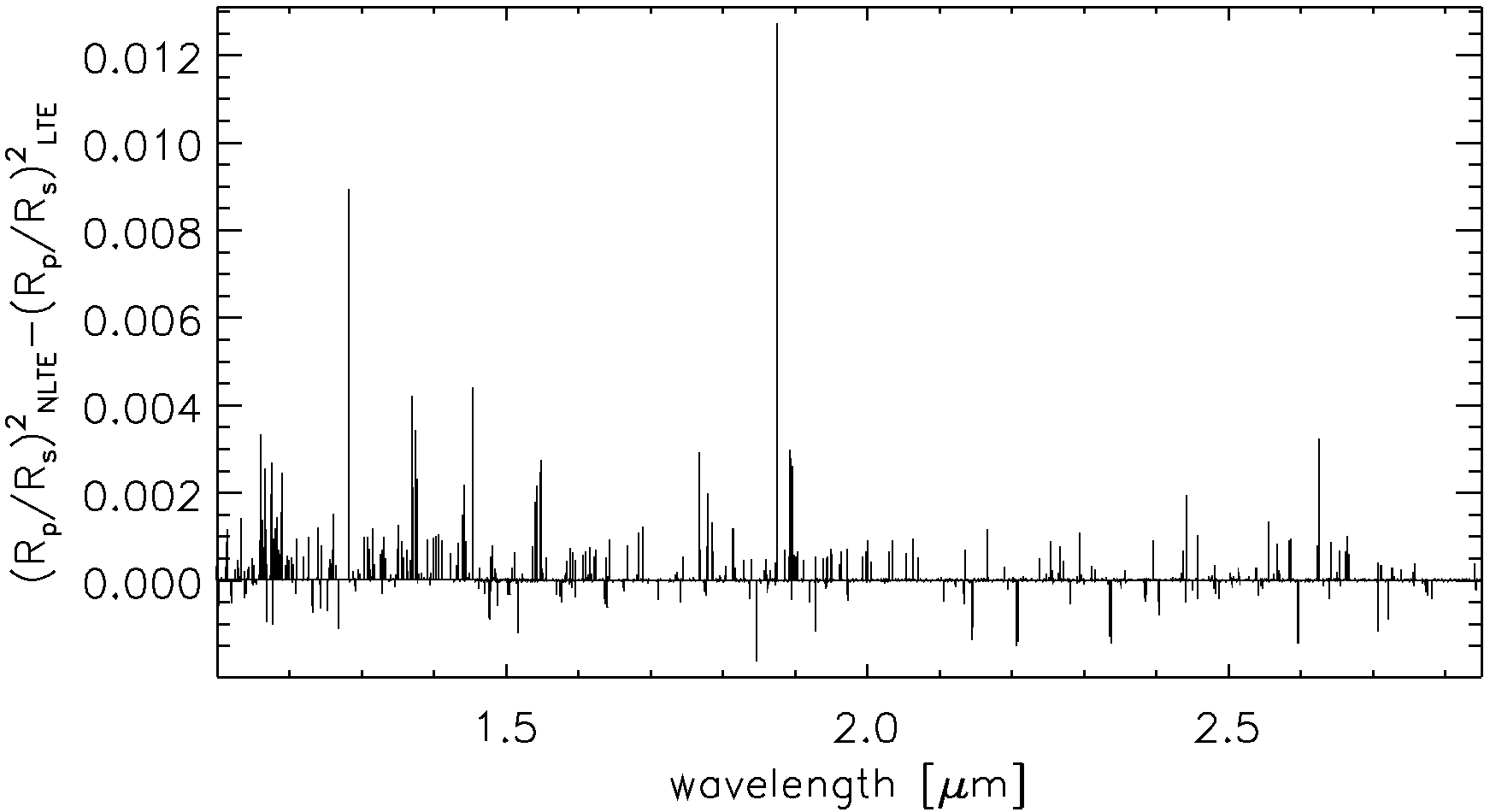}
                \caption{Transit depth difference between the NLTE and LTE transmission spectra shown in Figure~\ref{fig:transmission_spectra}. The top plot covers the UV and optical range, while the bottom plot covers the infrared band.} 
                \label{fig:NLTEvsLTE_transit_depth_difference}
\end{figure*}
%-------------------------------------
%
\FloatBarrier
\section{HST STIS data analysis}\label{appendix:hst_stis_analysis}
Before proceeding with the re-analysis of the STIS observations, we derived precise transit ephemerides by analysing the available TESS photometry, which has been obtained during sectors 11 (2019), 38 (2021), and 65 (2023). We modelled all transit events present in the data using {\sc allesfitter} \citep{guenther2021_allesfitter}, while simultaneously detrending the light curves with splines. We computed the posterior distributions of the parameters of interest using the Dynamic Nested Sampling inference algorithm \citep[see e.g.][]{feroz2008_nested_sampling,feroz2019_nested_sampling} that is integrated in {\sc allesfitter} via the {\sc dynesty} package \citep{speagle2020_dynesty}. Finally, we obtained an orbital period $P= 3.34483532_{-7.7\times10^{-7}}^{+7.6\times10^{-7}}$\,days and a reference mid-transit time T$_{0,{\rm BJD}}=2459348.73196\pm$0.00013\,days, which agree within 1$\sigma$ to the values adopted by \citet{lothringer2022_wasp178}.

The STIS data are affected by significant systematics. We tackled their removal using two techniques, both based on polynomial detrending. The first technique is described in \citet{cubillos2020_nuv_stis_hd209458b}, \citet{cubillos2023_nuv_stis_hd189733b}, and \citet{sreejith2023_cute_wasp189b} and builds up on the jitter detrending scheme introduced by \cite{sing2019}. We corrected for systematic effects the order-by-order white light curves using a function $S_{\lambda}(t, \phi, j_i)$ based on time $t$, telescope phase $\phi$, and the jitter parameter that exhibited the highest degree of correlation with the data $j_i$. The transit model $T_{\lambda}(t)$ has been implemented following \citet{MandelAgol2002apjLightcurves}; the transit ephemerides were derived from the TESS data as described above and the limb-darkening coefficients were calculated through the open-source routines of \citet{EspinozaJordan2015mnrasLimbDarkeningI}. In detail, the model $F_{\lambda}(t)$ applied to the raw light curves is
\begin{equation}
F_\lambda(t) = T_{\lambda}(t)\,S_{\lambda}(t, \phi, j_i)\,.
\end{equation}
This approach, which we call `one jitter parameter' method, assumes that within each spectral order the systematics are wavelength independent. It turns out that the optimal model of the instrumental systematics takes the form
\begin{align}
\nonumber
& S_{\lambda}(t, \phi, j_i) =  1 +\ a_{0}(t-t_{0}) + a_{1}(t-t_{0})^{2} + a_{2}(t-t_{0})^{3}  \\
\nonumber
& \hspace{1.0cm} + b_{0}(\phi-\phi_{0}) + b_{1}(\phi-\phi_{0})^{2} 
  +\ b_{2}(\phi-\phi_{0})^{3} + b_{3}(\phi-\phi_{0})^{4} \\
& \hspace{1.0cm} + c_0(j_i-\langle{j_i}\rangle)\,,
\label{eq:systematics}
\end{align}
where $a_{k}$, $b_{k}$, and $c_{k}$ are the polynomial coefficients of the fit, and $t\sb{0}$ and $\phi\sb{0}$ are the reference values for time and phase, set as mid-transit time $t\sb{0}=T\sb{0}$ and the HST telescope mid-phase $\phi\sb{0}=0.2$, respectively. The term $\langle{j_i}\rangle$ denotes the mean value of the jitter parameter $j_i$ along the visit. The best fit in this case corresponds to the jitter parameter $LOS\_Zenith$.

To select the optimal systematic model, we used the Bayesian Information Criterion (BIC) and Akaike Information Criterion (AIC). Once the optimal systematic model has been identified, we extracted the transmission spectra using the `divide-white' spectral analysis method \citep[e.g.][]{cubillos2020_nuv_stis_hd209458b,cubillos2023_nuv_stis_hd189733b,sreejith2023_cute_wasp189b}, which corrects instrumental systematics by dividing white-light curves by the best-fit transit model, assuming weak wavelength dependence. Systematics-corrected light curves are constructed by propagating uncertainties through all steps and dividing the data into wavelength bins. These wavelength-specific light curves are modelled with a \citet{MandelAgol2002apjLightcurves} transit model, fixing the orbital parameters and the limb-darkening coefficients. The free parameters for each fit are transit depth, mid-transit time, and visit-specific out-of-transit flux. Remarkably, the BIC criterion led us using a 3rd order polynomial in time (BIC\,=\,3313.4), while for example when using a 1st order polynomial in time we obtained a higher BIC of 3637.5. This is further addressed later in this section.

\begin{figure}
                \centering
                \includegraphics[width=9cm]{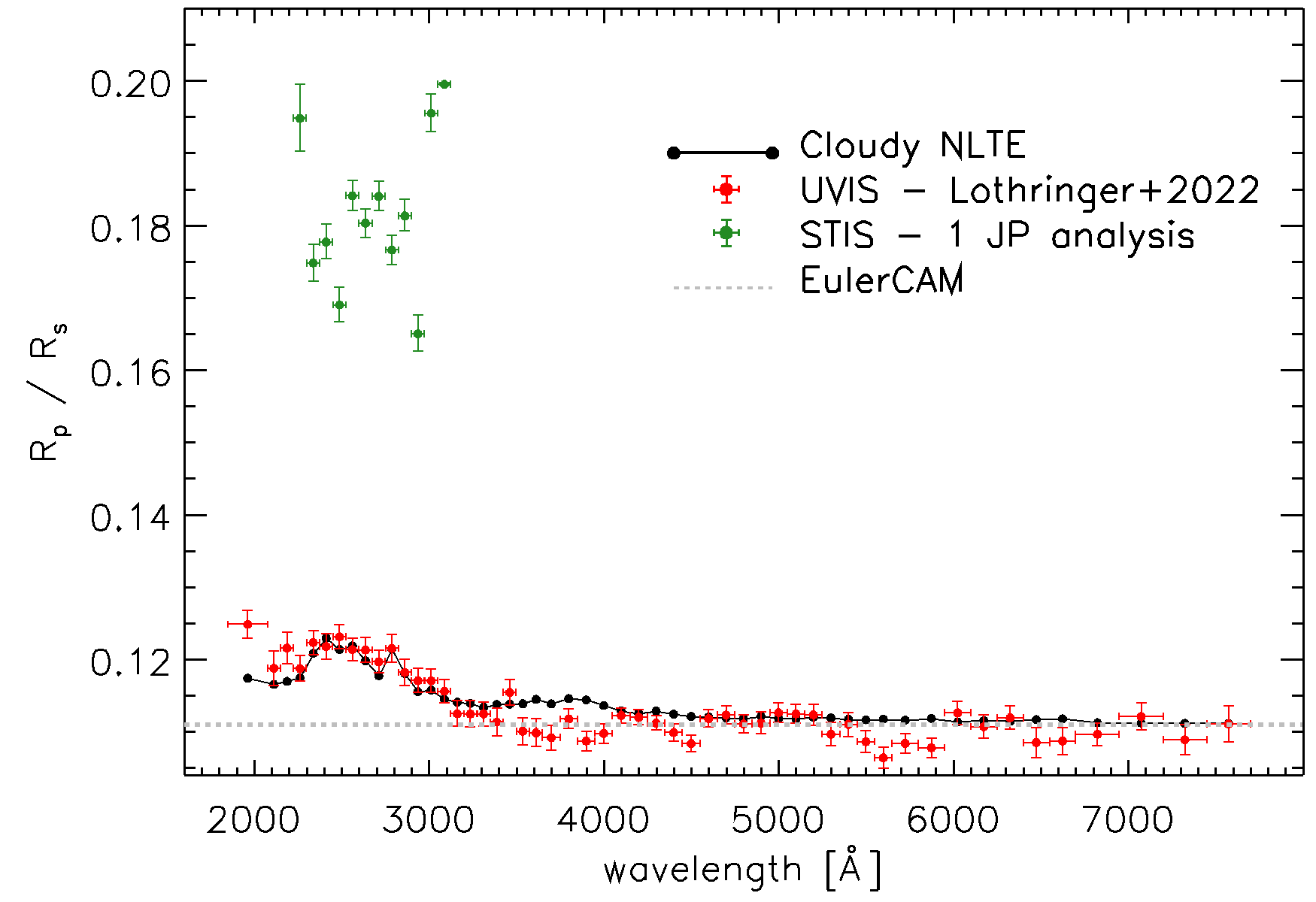}
                \caption{Comparison of the NLTE (black) theoretical transmission spectrum with the HST WFC3/UVIS observations published by \citet[][ red]{lothringer2022_wasp178} and the STIS transmission spectrum (green) re-analysed employing a detrending based on one jitter parameter and 3rd order polynomial in time (see text), binned to the same binning of the UVIS spectrum. The horizontal error bars shown for the red and green points correspond to the size of the wavelength bins. The gray horizontal dashed line indicates the broad-band $R_{\rm p}/R_{\rm s}$ obtained from EulerCAM \citep{hellier2019}.}
                \label{fig:comparison_with_STIS_1JitterParam}
\end{figure}

\begin{figure*}[h!]
                \centering
                \includegraphics[width=18cm,trim={2cm 0 2cm 0},clip]{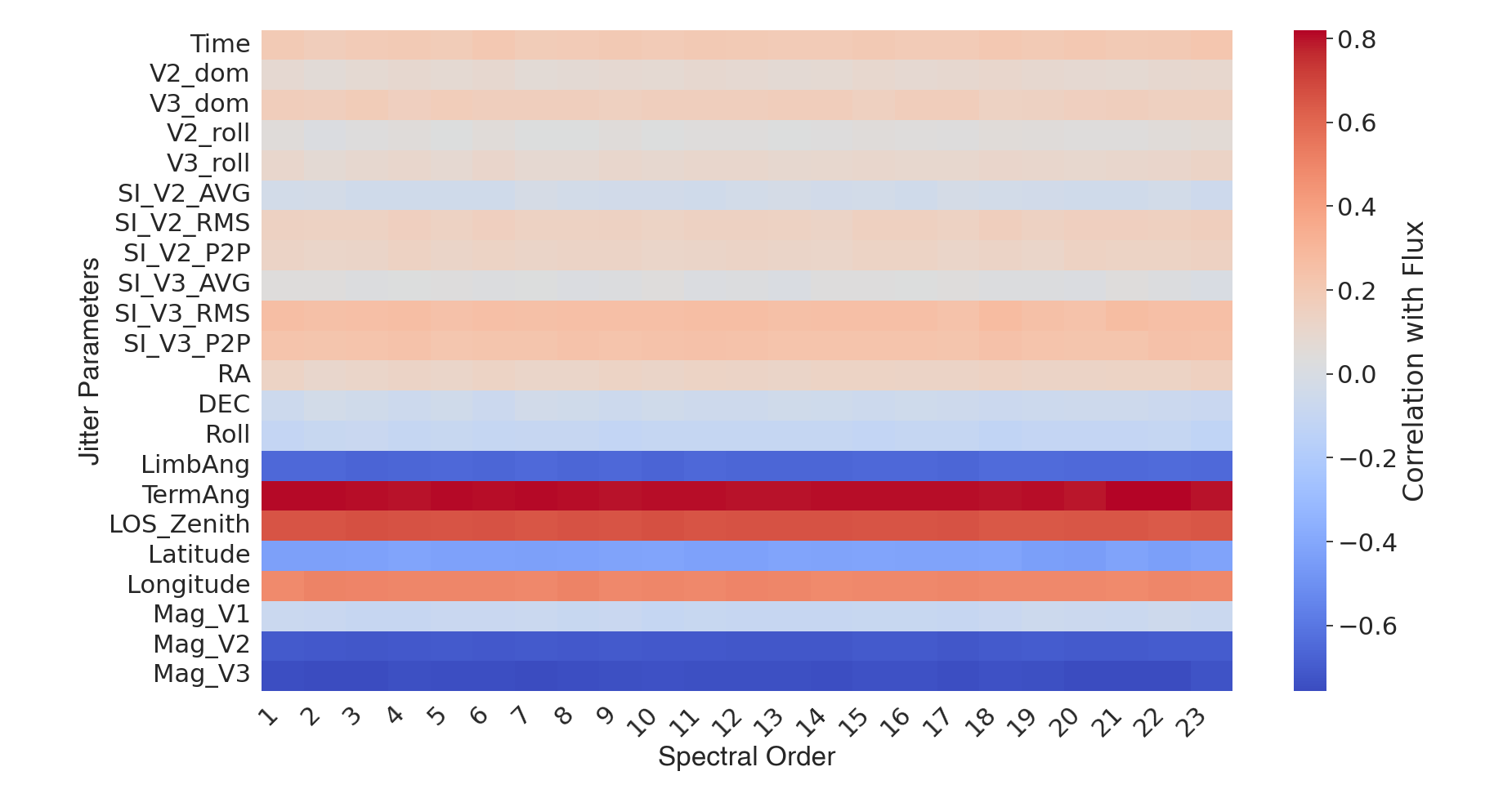}
                \caption{Heat map showing the correlation of jitter parameters and time with flux across each spectral order. The x-axis represents the spectral order, while the y-axis lists the parameters. The colour indicates the strength of the correlation, with red and blue representing positive and negative correlations, respectively, as indicated by the colour bar on the right hand side.}
                \label{fig:jittercorr}
\end{figure*}

However, this approach led to a transmission spectrum that was incompatible with both the UVIS data and the broad-band EulerCAM photometry (Figure~\ref{fig:comparison_with_STIS_1JitterParam}), indicating that the model did not adequately capture the underlying systematics influencing the data. Therefore, additional factors not accounted for in our initial model (Equation~\ref{eq:systematics}) contribute significantly to the observations. To better understand the root cause of these discrepancies, we calculated the correlation between the observed data and a wide range of parameters, including time and the HST jitter parameters available through the data distributed by the HST archive. The result of the correlation analysis are summarised in Figure~\ref{fig:jittercorr}, which identifies $Termang$, $LOS\_Zenith$, $Latitude$, $Longitude$, $MagV2$, $MagV3$ as the parameters with the highest correlation. As we found the correlation to be uniform across the covered wavelength range, we chose to use the same correlation parameters across all orders. Notably, these parameters show a degree of correlation with the data stronger than that shown by time. 
%-------------------------------------
%-------------------------------------
%-------------------------------------
%-------------------------------------

To account for these findings, the second technique we employed to analyse the data, which we call `multiple jitter parameter' method, enables one to account for several parameters simultaneously against which to decorrelate. Therefore, we considered the parameters showing the strongest degree of correlation mentioned above as input to the decorrelation model based on the MCMCI code\footnote{{\tt https://github.com/Bonfanti88/MCMCI}} \citep{2020AndreaMCMCI} that we briefly describe below. The MCMCI is a Fortran-based fitting code that integrates a Markov Chain Monte Carlo (MCMC) algorithm with an isochrone placement algorithm to fully characterise exoplanetary systems and supports analysis of light curves and radial velocity (RV) time series by modelling transits, occultations, and the Rossiter-McLaughlin effect. It samples the parameter space using the Metropolis-Hastings algorithm and Gibbs sampling, to infer system properties such as planetary radius, mass, orbital eccentricity, and host star characteristics. The code is capable of correcting for observational trends and systematics in the data. Bayesian priors are incorporated and the code allows for flexible handling of orbital eccentricity to reduce biases. The code selects optimal models based on the BIC, enabling precise characterisation of single and multi-planet systems. This tool can take up to six parameters ($J_i$) at the same time against which to decorrelate and employs the following function of time $t$ to model the transit light curve
\begin{equation}
F_\lambda(t) = T_{\lambda}(t)\;S_{\lambda}(t, J_1, J_2, J_3, J_4, J_x, J_y)\,,
\end{equation}
where $T_{\lambda}(t)$ is the
\citet{MandelAgol2002apjLightcurves} transit model. The $S_{\lambda}()$ term is a polynomial model (with order not greater than 4) of the instrumental systematics taking the form
\begin{align}
\nonumber
S_{\lambda} = & \, \sum_{i=1}^{n_t} a_i \cdot {t}^i + \sum_{i=1}^{n_{J_1}} b_i \cdot {J_1}^i + \sum_{i=1}^{n_{J_2}} c_i \cdot {J_2}^i + \sum_{i=1}^{n_{J_3}} d_i \cdot {J_3}^i +\\
&  \sum_{i=1}^{n_{J_4}} e_i \cdot {J_4}^i +     \sum_{i=0}^{n_{J_{xy}}}\,\sum_{j=0}^{n_{J_{xy}}-i} f_{ij} \cdot J_x^i \cdot J_y^j\,, 
\end{align}
where $a_{i}$, $b_{i}$, $c_{i}$, $d_{i}$, and $e_{i}$ are the polynomial coefficients of the fit for time and the jitter parameters ($J_{1}$--$J_{4}$; i.e. $TermAng$, $LOS\_Zenith$, $MagV2$, $MagV3$), while the terms $n_t$ and $n_{J_i}$ indicate the order of the polynomials used for each parameter. The $f_{ij}$ are the polynomial coefficients associated with the two jitter parameters $J_{x}$ and $J_{y}$, which can have mutual correlation (in our case $Latitude$ and $Longitude$).

We generated order-specific white light curves and conducted model fitting using MCMCI for various polynomial combinations of the jitter parameters. To ensure consistency across spectral orders, the same polynomial configuration was applied for all orders during the analysis. The optimal model (i.e. the sequence of polynomial orders $n_t$, $n_{J_1}$, $n_{J_2}$, $n_{J_3}$, $n_{J_4}$, and $n_{J_{xy}}$) was identified by evaluating the BIC, with the combination (3rd order for ${J_1}$ and ${J_2}$, 1st order for ${J_4}$, and 4th order for ${J_{xy}}$ ) yielding the lowest BIC selected as the best-fit model (BIC\,=\,5667.84 as compared to 68570.7 for an all-zero polynomial baseline). Based on this optimal solution, we constructed a systematics model using the best-fit parameters to account for instrumental and observational trends. Then, we applied the `divide-white' approach, as described earlier, to extract the transmission spectrum for the wavelength bins of interest.

\begin{figure}[h!]
                \centering
                \includegraphics[width=9cm]{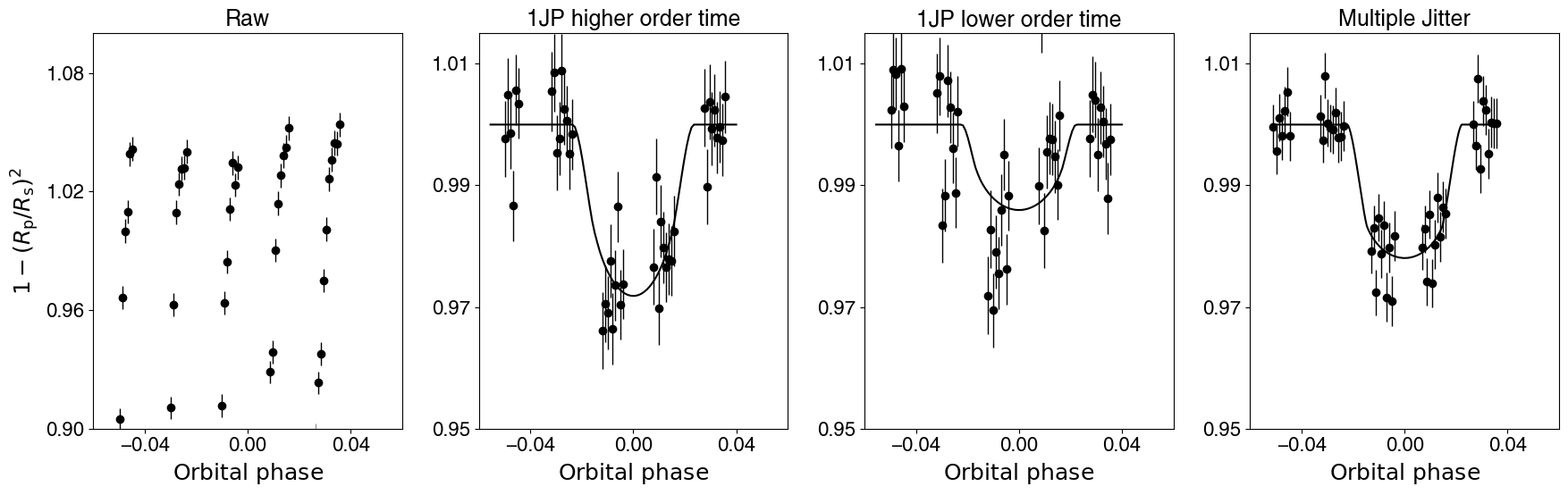}
                \includegraphics[width=9cm]{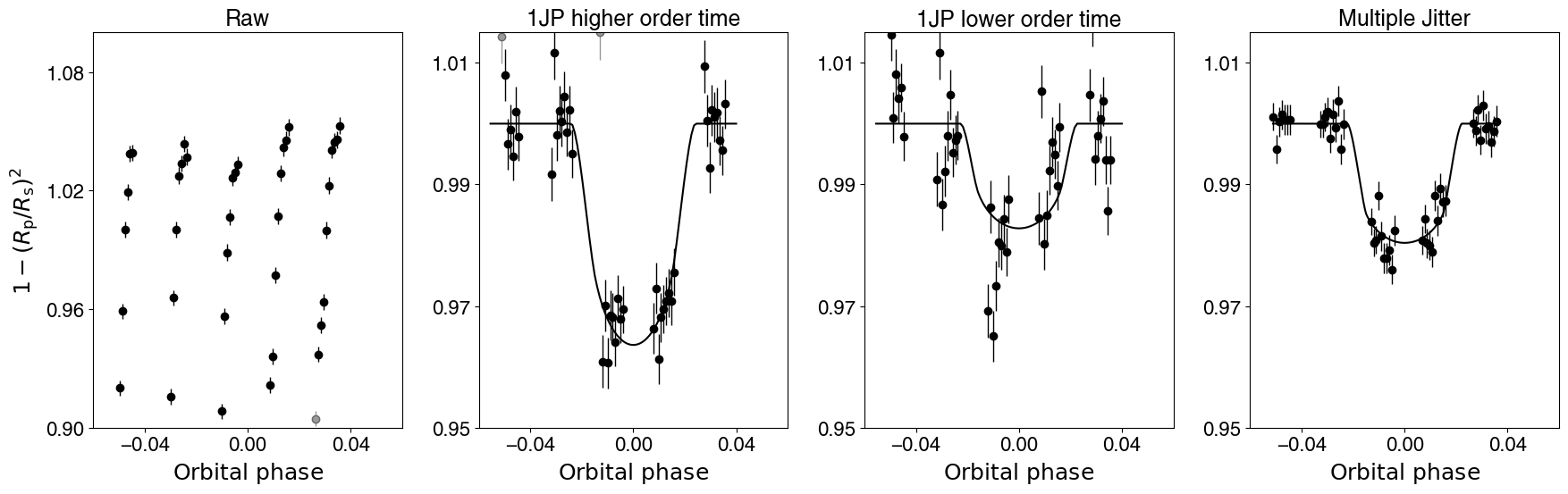}
                \includegraphics[width=9cm]{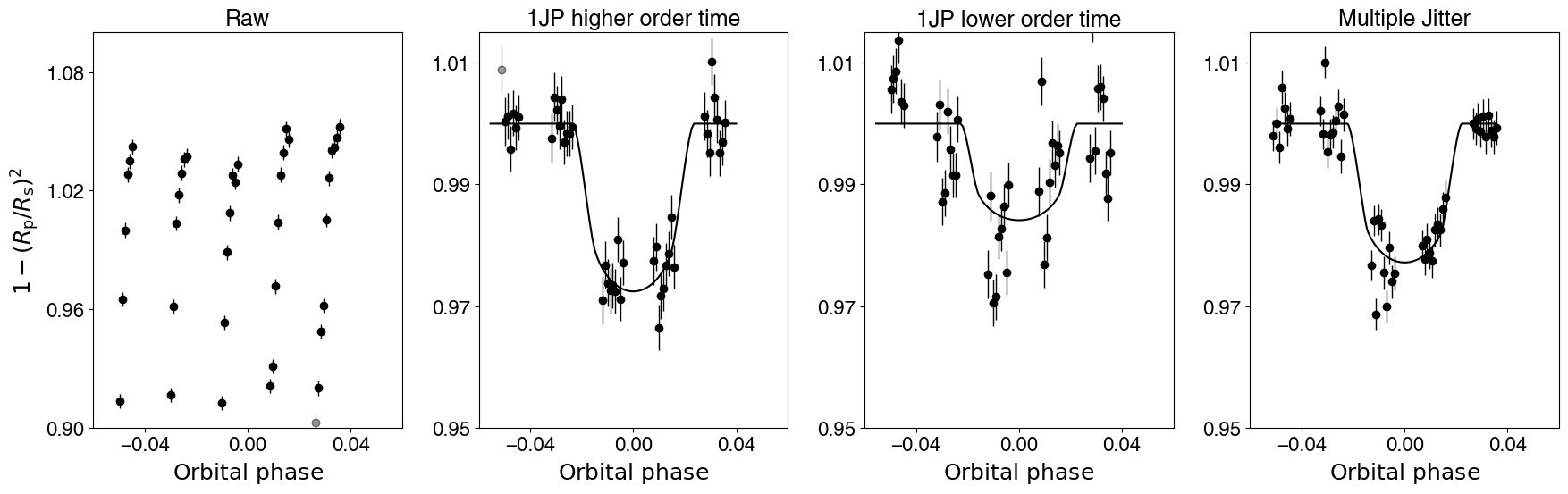}
                \caption{Light curves extracted from the STIS data for three spectral orders located around the beginning (21st spectral order; 2326.7--2365.8\,\AA; top), middle (11th spectral order; 2629.1--2673.3\,\AA; middle), and end (2nd spectral order; 2977.4--3027.3\,\AA; bottom) of the range covered by the data. Within each plot, the left panel is for the raw light curve, the two middle panels are for the data analysed using the `one jitter parameter' (high and low polynomial order in time) analysis method, and the right panel is for the data analysed using the `multiple jitter parameter' analysis method.} 
                \label{fig:lightcurves}
\end{figure}

Figure~\ref{fig:lightcurves} shows example light curves for three spectral orders located around the beginning, middle, and end of the wavelength range covered by the STIS observations. In particular, this plot compares the raw light curves (left column), which clearly show the presence of strong systematic noise, with the light curves obtained using the `one jitter parameter' (high and low polynomial order in time; second and third columns, respectively) and `multiple jitter parameter' (last column) analysis methods described above. The light curves obtained using the `one jitter parameter' analysis method return either an unrealistic transmission spectrum (i.e. for the high polynomial order in time; see Figure~\ref{fig:comparison_with_STIS_1JitterParam}) or light curves that appear to be still affected by residual systematic noise (i.e. for the low polynomial order in time), in addition to be disfavoured by the BIC. Instead, the light curves obtained applying the `multiple jitter parameter' analysis method present significantly less systematic noise. This demonstrates the higher quality of the `multiple jitter parameter' analysis method with respect to the `one jitter parameter' analysis method.
%-------------------------------------
%-------------------------------------

\begin{figure*}[h!]
                \centering
                \includegraphics[width=18cm,trim={3cm 0 4cm 0},clip]{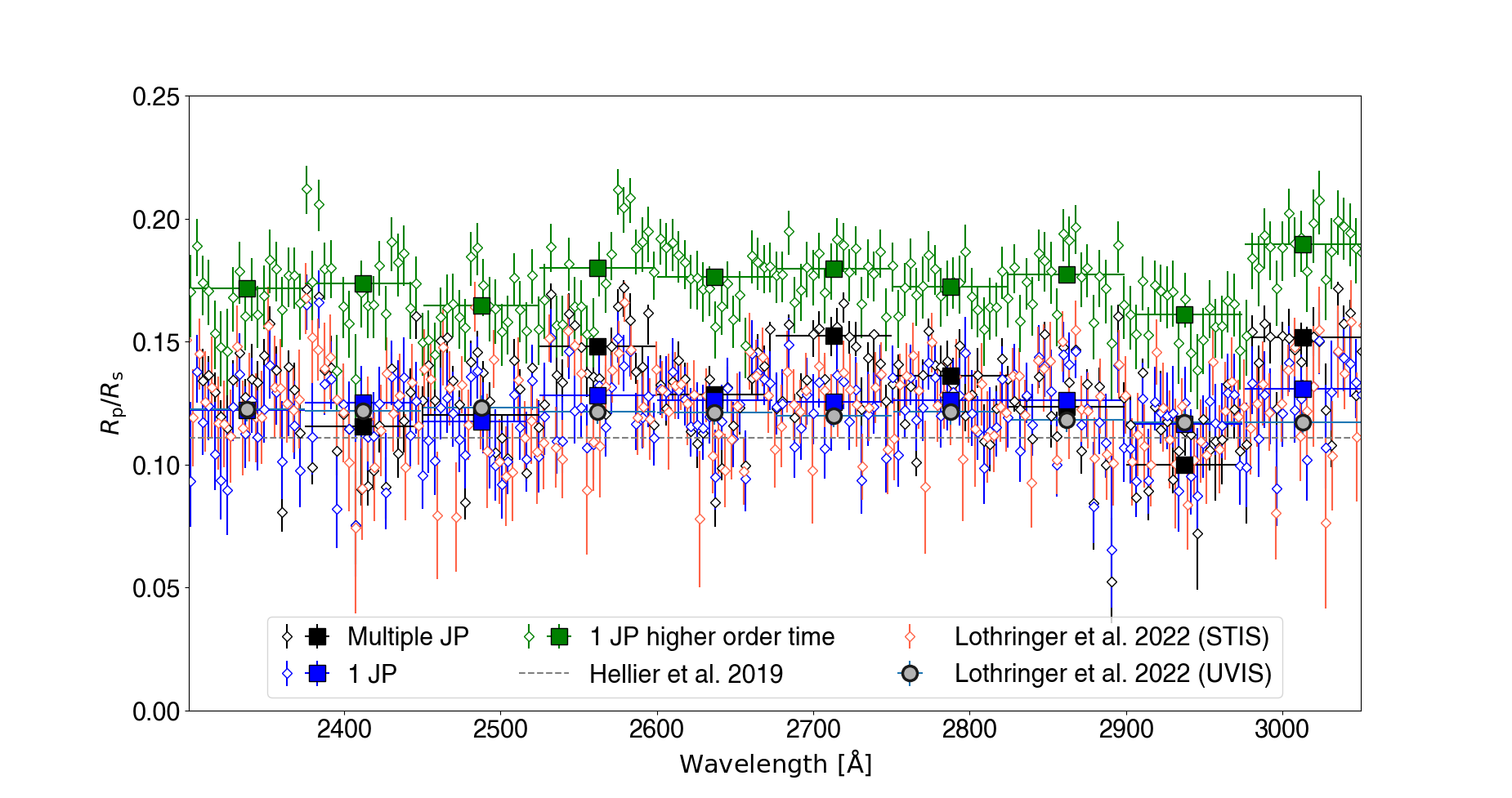}
                \caption{Comparison of STIS transmission spectra of WASP-178b obtained following different data analysis procedures. The black empty diamonds show the transmission spectrum obtained using the `multiple jitter parameter' method (`Multiple JP'). The blue and green empty diamonds show, respectively, the results obtained from the `one jitter parameter' analysis methods considering a lower (`1 JP lower order time') and higher (`1 JP higher order time') polynomial order in time, with the latter being favoured by the BIC. For each of these, the filled squares show the transmission spectra obtained binning at UVIS resolution. For comparison, the red empty diamonds show the transmission spectrum presented by \citet[][priv. comm. from the authors]{lothringer2022_wasp178}. The grey circles show the UVIS transmission spectrum from \citet{lothringer2022_wasp178}. The horizontal grey dashed line indicates the broad-band $R_{\rm p}/R_{\rm s}$ obtained from EulerCAM \citep{hellier2019}.} 
                \label{fig:tp_comparison}
\end{figure*}

Figure~\ref{fig:tp_comparison} compares the STIS transmission spectra of WASP-178b obtained using the different systematics correction procedures described above, with that presented by \citet[][priv. comm. from the authors]{lothringer2022_wasp178}. Figure~\ref{fig:tp_comparison} further shows the re-analysed STIS transmission spectra extracted binning at the resolution of UVIS and the observed UVIS transmission spectrum. This plot clearly shows that using just one jitter parameter with the higher polynomial order in time introduces a bias in the transmission spectrum resulting in much deeper transits, possibly due to over-fitting of the data (see also Figure~\ref{fig:comparison_with_STIS_1JitterParam}). Artificially removing the correlation with time or limiting it to a first order polynomial, either considering one or multiple jitter parameters results in a transmission spectrum comparable to that presented by \citet{lothringer2022_wasp178} and more in line with what expected following the broad-band transit radius. Remarkably, the transmission spectrum obtained arbitrarily using a low polynomial order in time within the `one jitter parameter' analysis method matches well the UVIS transmission spectrum, though this specific analysis method is disfavoured by the BIC (compared to the one obtained considering a high polynomial order in time) and by the presence of residual systematic noise in the light curves.  

Following the comparisons shown in Figure~\ref{fig:tp_comparison} and the higher degree of sophistication of the second technique in removing systematic noise (Figure~\ref{fig:lightcurves}), we consider the transmission spectrum obtained using the `multiple jitter parameter' analysis method as the final one. However, the analysis and comparisons presented here clearly indicate that the results obtained from this data set are significantly model-dependent.
%-------------------------------------
%-------------------------------------

%
\end{appendix}  
\end{document}